\documentclass[fleqn,usenatbib]{mnras}

\usepackage{newtxtext,newtxmath}
\usepackage[T1]{fontenc}

\DeclareRobustCommand{\VAN}[3]{#2}
\let\VANthebibliography\thebibliography
\def\thebibliography{\DeclareRobustCommand{\VAN}[3]{##3}\VANthebibliography}

\usepackage{graphicx}	% Including figure files
\usepackage{amsmath}	% Advanced maths commands
\usepackage{pdflscape}   % Landscape table

\newcommand{\edit}[1]{#1}
\newcommand{\edittwo}[1]{#1}
\newcommand{\editthree}[1]{#1}
\newcommand{\editfour}[1]{#1}
\newcommand{\editfive}[1]{#1}

\title[Finding Class II YSOs with a Bayes Classifier]{A Naive Bayes Classifier for identifying Class II YSOs}

\author[A. J. Wilson et al.]{
	Andrew J. Wilson\thanks{E-mail: aw648@exeter.ac.uk, andyjwilson\_uk@hotmail.com},
	Ben S. Lakeland,
	Tom J. Wilson and
	Tim Naylor
\\
Department of Physics and Astronomy, University of Exeter, Stocker Road, Exeter EX4 4QL, UK\\
}

\date{Accepted XXX. Received YYY; in original form ZZZ}

\pubyear{2022}

\begin{document}
\label{firstpage}
\pagerange{\pageref{firstpage}--\pageref{lastpage}}
\maketitle

\begin{abstract}
	A naive Bayes classifier for identifying Class II YSOs has been constructed and applied to a region of the Northern Galactic Plane containing 8 million sources with good quality \textit{Gaia} EDR3 parallaxes. The classifier uses the five features: \textit{Gaia} $G$-band variability, \textit{WISE} mid-infrared excess, UKIDSS and 2MASS near-infrared excess, IGAPS H$\alpha$ excess and overluminosity with respect to the main sequence. A list of candidate Class II YSOs is obtained by choosing a posterior threshold appropriate to the task at hand, balancing the competing demands of completeness and purity. At a threshold posterior greater than 0.5 our classifier identifies 6\,504 candidate Class II YSOs. \edit{At this threshold we find a false positive rate around 0.02 per cent and a true positive rate of approximately 87 per cent for identifying Class II YSOs.} The ROC curve rises rapidly to almost one with an area under the curve around 0.998 or better, indicating the classifier is efficient at identifying candidate Class II YSOs. Our map of these candidates shows what are potentially three previously undiscovered clusters or associations. When comparing our results to published catalogues from other young star classifiers, we find between one \edit{quarter} and \edit{three quarters} of high probability candidates are unique to each classifier, telling us no single classifier is finding all young stars.
\end{abstract}


%%%%%%%%%%%%%%%%%%%%%%%%%%%%%%%%%%%%%%%%%%%%%%%%%%

%%%%%%%%%%%%%%%%% BODY OF PAPER %%%%%%%%%%%%%%%%%%

\section{Introduction}

A variety of machine learning techniques have already been used to identify candidate Young Stellar Objects (YSOs). Naive Bayes classifiers have advantages that may bring benefit to this task. The aim of this paper is to assess how well a carefully constructed naive Bayes classifier can identify YSOs. Our approach focuses on Class II YSOs, using their known properties as inputs.

\subsection{YSOs}
\label{sec:intro_ysos}

Stars form in clouds of cold dust and gas, where gravitational collapse creates knots of over density that grow with time. Protostars form in these knots, starting cool and emitting in the far infrared. The peak of their emission moves to shorter wavelengths as their cores grow and heat. The spectral index \citep{Lada1987,Adams1987,Kenyon1990,Andre1993,Andre2000} encapsulates the location of the peak emission in a single number and is used to divide YSOs into evolutionary classes. Class 0/I YSOs are nearly impossible to identify in the optical, requiring infrared observations to detect them within their prenatal clouds of dust and gas. Class III sources are easily observed but have lost some of the identifiable characteristics of YSOs. Class IIs on the other hand can be observed in the optical as well as the infrared, with their discs and accretion giving rise to the well known features of Class II YSOs. The disc creates an infrared excess \citep{Calvet1991} and can cause variable obscuration of the protostar giving rise to photometric variability \citep{Bouvier1999,Froebrich2018b}. Emission lines such as H$\alpha$ are generated in accretion columns and by chromospheric activity \citep{Herbig1958,Herbig1962b,Edwards1994,Hartmann1994,2013ApJ...767..112I}. Variable accretion leads to aperiodic variability while starspots and accretion hotspots lead to periodic variability at the rotation period of the protostar \citep{Herbst1994,Sergison2020}. Protostars have a larger physical size than Main Sequence (MS) stars of the same temperature, resulting in a greater luminosity than MS stars of the same colour. These identifiable observational properties of Class II YSOs are ideally suited to be the input features of machine learning classifiers.

\subsection{Machine learning for YSOs}
\label{sec:machine_learning_for_ysos}

There are two main branches of machine learning (ML). Supervised ML classifies data into predefined labels, while unsupervised ML is free to choose the data classifications. Hence, supervised machine learning is appropriate for identifying YSOs and \edit{it} has already been successfully applied to this task.

\cite{Miettinen2018} tested eight machine learning algorithms for classifying 319 known YSOs. Interestingly their standard R implementation of a naive Bayes classifier gave the worst performance of the eight classifiers. The assumption of Gaussian feature distributions in the R implementation is a potential cause of this weak result, as the probability distributions used to define the likelihoods in Bayes classifiers have a big impact on their performance. They also make no mention of their choice of prior or posterior threshold, both of which have significant impacts on the measured performance.

\cite{Marton2019} compared a range of approaches for identifying candidate YSOs, including support vector machines, neural networks, naive Bayes, $k$-NN and random forests. They found a random forest of 500 trees gave the best performance. In common with the work we present in this paper, they base their input data on \textit{Gaia}, though DR2 rather than EDR3, combined with \textit{WISE}, 2MASS and \textit{Planck}.

\cite{Vioque2020} created a neural network to search for new Herbig Ae/Be and PMS stars. They use similar features to our classifier, based on \textit{Gaia} DR2, 2MASS, \textit{WISE}, IPHAS and VPHAS+. The approach to creating their PMS classification means it is similar to our Class II YSO (\textsc{CII}) classification.

The SCAO (Spectrum Classifier of Astronomical Objects) developed by \cite{Chiu2021} uses a fully connected neural network to identify YSOs. They also tried extreme gradient boost, random forest, support vector machine and k-nearest neighbour machine learning techniques. The classifier uses spectral energy distributions (SEDs) constructed from \textit{Spitzer}, UKIDSS and 2MASS photometry. The results of SCAO run against the SEIP (\textit{Spitzer} Enhanced Imaging Products) catalogue are available online.

\cite{Kuhn2021} employ a random forest to identify YSOs for their SPICY catalogue using \textit{Spitzer} with 2MASS, UKIDSS and VVV photometry. Again this has similarities to our work.

The neural network named Sagitta created by \cite{McBride2021} identifies pre-main sequence (PMS) stars and assigns them an age. This is trained on the results of a hierarchical clustering analysis and a neural network named Auriga \citep{Kounkel2020}. They use \textit{Gaia} DR2 and 2MASS, so this has some similarities to our classifier, though we additionally incorporate data from UKIDSS, \textit{WISE} and IGAPS. Also, we focus on Class II YSOs while they assessed a wide range of PMS ages.

\edit{\cite{2021AA...647A.116C} constructed a neural network for identifying Class I and Class II YSOs. Their labelled data was created using a modified version of \cite{Gutermuth2009a} based purely on Sptizer photometry. The classifier was trained and tested on data from Orion, NGC 2264 and a sample of clouds within 1\,kpc.}

With the exception of \cite{Miettinen2018}, published catalogues are available for these classifiers. We compare them with the results from our classifier in Section~\ref{sec:comparisons_to_other_catalogues}.

\subsection{Possible advantages of a naive Bayes classifier}
\label{sec:intro_naive_bayes_classifier}

Our naive Bayes classifier is based on \cite{Broos2011}, a classifier of X-ray sources. They expanded the technique in \cite{Broos2013}, making use of infrared as well as X-ray observations to identify YSOs for the MYStIX project. Naive Bayes classifiers have also been used in other areas of astronomy, for example \cite{Li2020} built one to separate Type I and Type II Gamma-ray bursts. \editfour{While \cite{2022A&A...657A.138T} created a naive Bayes classifier for separating X-ray sources into AGN (Active Galactic Nuclei), stars, X-ray binaries and cataclysmic variables.}

Our classifier is applied across a broad region of the Northern Galactic Plane defined $20<l<220$ and $|b|<4$, we shall refer to this region as the NGPn. This region was chosen to overlap with the footprint of WEAVE \citep{2012SPIE.8446E..0PD}, a spectrograph commissioned for the WHT (William Herschel Telescope). \edittwo{In future we intend to test our classifier on the WEAVE SCIP (Stellar, Circumstellar and Interstellar Physics) survey \citep{2022arXiv221203981J}.} We impose a magnitude limit of $r<18$ to ensure the candidates are sufficiently bright for spectroscopic confirmation by WEAVE. Our NGPn data set is restricted to sources with \textit{Gaia} EDR3 parallaxes consistent with a distance of 2\,kpc or less, as beyond this limit we found the parallax measurements were dominated by their uncertainties. The results are posterior probabilities, distinguishing between the binary classifications of Class II YSO (\textsc{CII}) and all other types of object (\textsc{Other}).

The inputs to our classifier are the known features of Class II YSOs \edit{(described in Section~\ref{sec:features})}. Our decision to use known features avoids the risk of overfitting the classifier using non-physical properties. The simple algorithm of naive Bayes classifiers allows these features to be traced through the workings to the results. This provides astrophysical intuition into \edit{the} individual source classifications, not possible with black box classifiers. The feature data also make troubleshooting and refinement of the classifier straightforward.

A critical step in building a classifier is training \edit{it} to distinguish between the classes based on the feature values. Training can use purely theoretical models, or real data labelled by class using established techniques. We use labelled data to train our classifier \edit{(see Section~\ref{sec:training_data})}. With some types of classifier, the training sets need to be large and match the ratio of the classes in the population. With Bayes classifiers, the ratio of the classes is encoded by the priors \edit{(see Section~\ref{sec:naive_bayes_classifier})}, meaning we can use unbalanced training sets. Bayes classifiers also cope well with small training sets, a benefit for our relatively small training set of Class II YSOs \edit{(see Section~\ref{sec:training_data})}.

When combining data from multiple astronomical surveys, a source may be present in one survey and absent or difficult to reliably cross-match to another. This can present a significant problem for machine learning techniques that require all feature data to be present for all sources. In such cases, one possible solution is to limit the input data to the subset of sources common to all surveys, the approach employed by \cite{Vioque2020}. Alternatively, missing data can be populated with estimates, the approach taken by \cite{Kuhn2021} and \cite{McBride2021}. In contrast, naive Bayes classifiers do not require all feature data to be present for every source. They are able to simply skip over features without data, giving a result based on the available features for each individual source \edit{(see Section~\ref{sec:missing_data})}.

\subsection{Outline of this paper}
\label{sec:paper_outline}

The remainder of this paper is organised as follows. Section~\ref{sec:features} reviews the features of Class II YSOs. Section~\ref{sec:naive_bayes_classifier} introduces our naive Bayes classifier and explains how the features are incorporated. The results of the classifier are reviewed and compared to other published works in Section~\ref{sec:results}. We end by considering bias and how the classifier might be improved in Section~\ref{sec:improvements_and_bias}. Our catalogue is published to allow others to independently test our results, see the data availability statement at the end of this paper. \edit{We provide a table of the symbols used in this paper and a data dictionary for the online catalogue in the supplementary material.}

\section{Features of Class II YSOs}
\label{sec:features}

For the classifier to identify candidate Class II YSOs \edit{required} a set of observable features \edit{that were} able to separate them from other classes of object. We derived these features from publicly available surveys. To be useful for our work these surveys needed to provide good coverage of our NGPn footprint. Within these limitations we identified a set of five features with surveys to derive them: mid-infrared excess using \textit{WISE} \citep{Wright2010}, near-infrared using \editfour{UKIDSS} \citep{Lucas2008} and 2MASS \citep{Skrutskie2006}, H$\alpha$ excess and overluminosity with respect to the MS both using IGAPS \citep{Drew2005,Monguio2020}, and photometric variability from \textit{Gaia} EDR3 \citep{2016A&A...595A...1G,2021A&A...649A...1G}.

As this is an experimental classifier, we chose to prioritise quality at the expense of completeness when selecting data from the surveys. Otherwise it would have been difficult to separate poor classifier performance from incorrect classifications caused by low quality data. Our NGPn catalogue contains 8\,088\,045 sources, chosen for their good quality parallax measurements in \textit{Gaia} EDR3 then cross-matched against the other surveys. \edit{Table~\ref{tab:cii_features} gives a summary of the features, showing the percentage of our NGPn catalogue and the number of training sources for each feature.} As naive Bayes classifiers can cope with missing data (\edit{see} Section~\ref{sec:missing_data}), it was not necessary for the feature surveys to have data for all sources. Full details of the survey selection criteria and how they were combined are given in Appendix~\ref{sec:surveys}.

\begin{table*}
	\centering
	\caption{\edit{A summary of the features used by the classifier.}}
	\label{tab:cii_features}
	\begin{tabular}{llcrr}
		\hline
		Feature & Measure & Percent of NGPn & \# \textsc{CII} & \#\textsc{Other} \\
		\hline
		\textit{WISE} & $(W1-W2)$ excess & 35\% & 110 & 91\,673 \\
		$(H-K)$ excess total & Distance in $(H-K)$ from linear region of 1\,Myr isochrone in (J-H,H-K) & 80\%  &  &  \\
		$(H-K)$ UKIDSS & & 61\%  & 84 & 109\,623 \\
		$(H-K)$ 2MASS & & 19\% & 223 & 26\,639 \\
		H$\alpha$ excess & Distance in $(r-H\alpha)$ from empirical MS in $(r-i,r-H\alpha)$ & 76\% & 274 & 136\,782 \\
		Isochronal age & Location in $(r,r-i)$ CMD relative to young stellar isochrones & 74\% & \ 268 & 134\,279 \\
		Variability & \textit{Gaia} $G$-band variability from observed fractional standard deviation of the flux $\hat{\sigma}_{\rm O}$ & 97\% & 355 & 145\,769 \\
		\hline
	\end{tabular}
\end{table*}

\subsection{Training and test data sets}
\label{sec:training_data}

Training data sets were created to teach the classifier how to discriminate between Class II YSOs (\textsc{CII}) and all other types of source (\textsc{Other}). These were crucial in selecting and tuning the features, since the strength of a feature is determined by its ability to separate the \textsc{CII} from \textsc{Other}. This also means the quality of the classifier results depends to a large extent on the purity of the training sets, as contamination weakens the training. To allow an independent check of the classifier performance, we needed a separate test set of sources not used in the training. Hence, we created a master list of known sources of each class, and split them into training and test subsets.

To create our known \textsc{CII} master list, we performed a literature search for papers providing catalogues of robustly identified Class II YSOs. We only included catalogues where we would expect all or a significant proportion of sources to lie within 2\,kpc, to match the distance limit of our base data set. The full list of papers used to compile our master list of known \textsc{CII}s is given in Table~\ref{tab:classiiyso_trainingset}. Many of these are based on variations of the methods of \cite{Gutermuth2005,Gutermuth2008,Gutermuth2009a}, accepting this may give the classifier a bias towards identifying \textsc{CII} with the properties used by this approach. There are 2\,853 distinct \textsc{CII} from these catalogues within our NGPn footprint with matches to \textit{Gaia} EDR3 using a 1\arcsec search radius. This reduces to 439 \textsc{CII} in our NGPn data set, with the losses due to our parallax and photometric quality cuts \edit{(see Appendix~\ref{sec:base_data_set})}. The count of \textsc{CII} from the separate papers is 539 as some of the same sources appear in more than one paper. Although this small quantity of labelled data limits the granularity that can be achieved in training, it is adequate for creating the probability distributions used by the Bayes classifier.

The \textsc{Other} data set was more difficult to define as ideally it should contain a representative population of all types of object apart from Class II YSOs. In reality this is impossible since the populations of different types of star vary with location in the Galaxy, and there could be unidentified \textsc{CII}s in any region. Hence, we decided to use a region that was devoid of known star formation for our \textsc{Other} data set. We settled on the region bound by the Galactic coordinates $160<l<166$ and $-4<b<4$, which we shall refer to as the \textsc{Golden Rectangle}. It is devoid of major known star formation, has low extinction and is bland in all feature input catalogues. These very properties risk introducing bias (\edit{see} Section~\ref{sec:improvements_and_bias}), but we accept this for the benefits the approach offers. We can be confident there are at most a few Class II YSOs in this region. Also, the low extinction is beneficial since our reddening values are only approximate and unidentified bad reddening can lead to spurious feature results. The master data set of \textsc{Other} contains 189\,556 sources.

Our approach to selecting the \textsc{Other} data set means it may be contaminated by YSOs of any type including \textsc{CII}. We should expect this to be at a very low level and if there is significant contamination this would show up as a high false positive rate for the \textsc{Other} sources when validating the classifier. Ironically, this lack of YSOs in the \textsc{Golden Rectangle} leads to a training problem. It means the classifier will not be trained to place other types of YSOs into the \textsc{Other} classification. This is exacerbated as other types of YSO share similar feature properties to \textsc{CII}s. So we should reasonably expect the classifier to identify other types of YSO as \textsc{CII}, though in general they should be at a lower posterior \textsc{CII} than genuine Class II YSOs. This situation could be improved by expanding the classifier to more than two classes, adding the other types of YSO with training sets. For now, we simply acknowledge and accept this weakness.

We split the master data sets $80:20$ for training and testing, using the \textit{Gaia} EDR3 random index to split the data into random selections. This gave 365 sources in the \textsc{CII} training set and 91 sources in the \textsc{CII} test set. The \textsc{Other} master set was split 151\,645 for training and 37\,911 for testing. This imbalance in the training sets would be a significant problem for some types of ML. It is not a problem for Bayes classifiers as the expected ratio of the classes is encoded by the priors, not the relative sizes of the training sets.

\begin{table*}
	\centering
	\caption{The source catalogues used to compile the Class II YSO training and test sets.}
	\label{tab:classiiyso_trainingset}
	\begin{tabular}{lrccll}
		\hline
		Source & \# \textsc{CII} & Region & Distance kpc & Method & Table and criteria \\
		\hline
		\cite{Allen2008} & 7 & Lynds 988e & 0.7 &\cite{Gutermuth2005} & 3 \\
		\cite{Allen2012} & 86 & Cep OB3b & 0.7 & \cite{Gutermuth2009a} & 2 \\
		\cite{Billot2010} & 35 & Vul OB1 & 2.3 & \cite{Gutermuth2008} & 2 Class II \\
		\cite{Cody2014} (CSI 2264) & 73 & NGC 2264 & 0.76 & SED via \cite{Evans2009a} & 3 Class II \\
		\cite{Damiani2017} & 32 & North America \& Pelican Nebulae & 0.6 & NIR-excess & 5 and 3+4 IR excess \\
		\cite{Dewangan2011} & 5 & S235 & 2.10 & Slope of SED from \textit{Spitzer} & C2 \\
		\cite{Fang2020} & 135 & North America \& Pelican Nebulae & 0.795 & \cite{Fang2020} & 4 IRE=Y \\
		\cite{Gutermuth2009a} & 40 & 36 young clusters & $\leq1.7$ & \cite{Gutermuth2009a} & 4 Class II \\
		\cite{Jose2016} & 3 & W3-AFGL & $\approx2.0$ & \cite{Gutermuth2009a} & 3 Class II \\
		\cite{Jose2017} & 7 & Stock 8 & $\approx2.1$ & \cite{Gutermuth2009a} & 2 Class II \\
		\cite{Rapson2014} & 116 & Mon OB1 East / NGC 2264 & 0.76 & \cite{Gutermuth2009a} & 1 Class II \\
		\hline
	\end{tabular}
\end{table*}

\subsection{WISE W1-W2}
\label{sec:wise_feature}

Excess emission in the mid-infrared is a strong indicator of stellar youth \citep{Lada1984}. In Class II YSOs this is due to dust in the circumstellar disc reprocessing radiation from the YSO into the infrared \citep{Calvet1991}. Other works have successfully used photometry from IRAS \citep{Kenyon1990}, \textit{Spitzer} \citep{Robitaille2008} \edit{and \textit{WISE} \citep{Koenig2012,Koenig2014,Fischer2016,Marton2019}} to identify YSOs. For this work we considered \textit{Spitzer} and \textit{WISE}. \textit{Spitzer} provides better spatially resolved photometry though only covers patches of our NGPn footprint, while \textit{WISE} provides full coverage at lower resolution. Since coverage is important to us we settled on \textit{WISE} for our mid-infrared feature.

A proven technique for identifying YSOs is defining areas of Colour-Colour Diagrams (CCDs) associated with stellar youth. \edit{This has been successfully demonstrated by \cite{Koenig2012} using \textit{WISE} $(W1-W2,W2-W3)$, then refined by \cite{Koenig2014} and \cite{Fischer2016}. However, these approaches} have an inherent weakness due to large uncertainties in the \textit{WISE} photometry. These uncertainties get progressively worse to longer wavelength bands. In the ALLWISE catalogue 69 per cent of sources have an uncertainty better than 0.1 magnitudes in $W1$, 33 per cent in $W2$, and just 2 per cent in $W3$. Hence requiring good quality photometry in $W3$ drastically reduces the available data.

A key benefit of employing the $W3$ photometry is it allows CCDs to distinguish YSOs from Asymptotic Giant Branch (AGB) stars, star-forming galaxies, shock emission and PAH (Polycyclic Aromatic Hydrocarbon) knots. It can also separate Class I from Class II YSOs. The sources in our NGPn catalogue have been selected to ensure they have good quality \textit{Gaia} parallaxes consistent with a distance 2\,kpc or less. This excludes galaxies and the majority of AGB stars. Further, we do not expect our selection of the \textit{Gaia} optical data with $G_{\rm{RP}}<18.0$ to contain significant numbers of sources that are shock emission, PAH knots or Class I YSOs. These factors enable an alternative approach with our NGPn catalogue, using a simple $(W1-W2)$ excess to pick out Class II YSOs, retaining a greater proportion of the \textit{WISE} photometry as there is no need for a $W3$ quality cut.

\subsection{(H-K) Excess}
\label{sec:hk_feature}

Class II YSOs exhibit an excess in the near-infrared \citep{Meyer1997}. We derive an $(H-K)$ excess from $(J-H,H-K)$ CCDs using UKIDSS and 2MASS photometry, defining our excess as the distance in $(H-K)$ of dereddened photometry from a straight line fit to the linear region of a 1\,Myr isochrone.

\begin{figure}
	\includegraphics[width=\columnwidth]{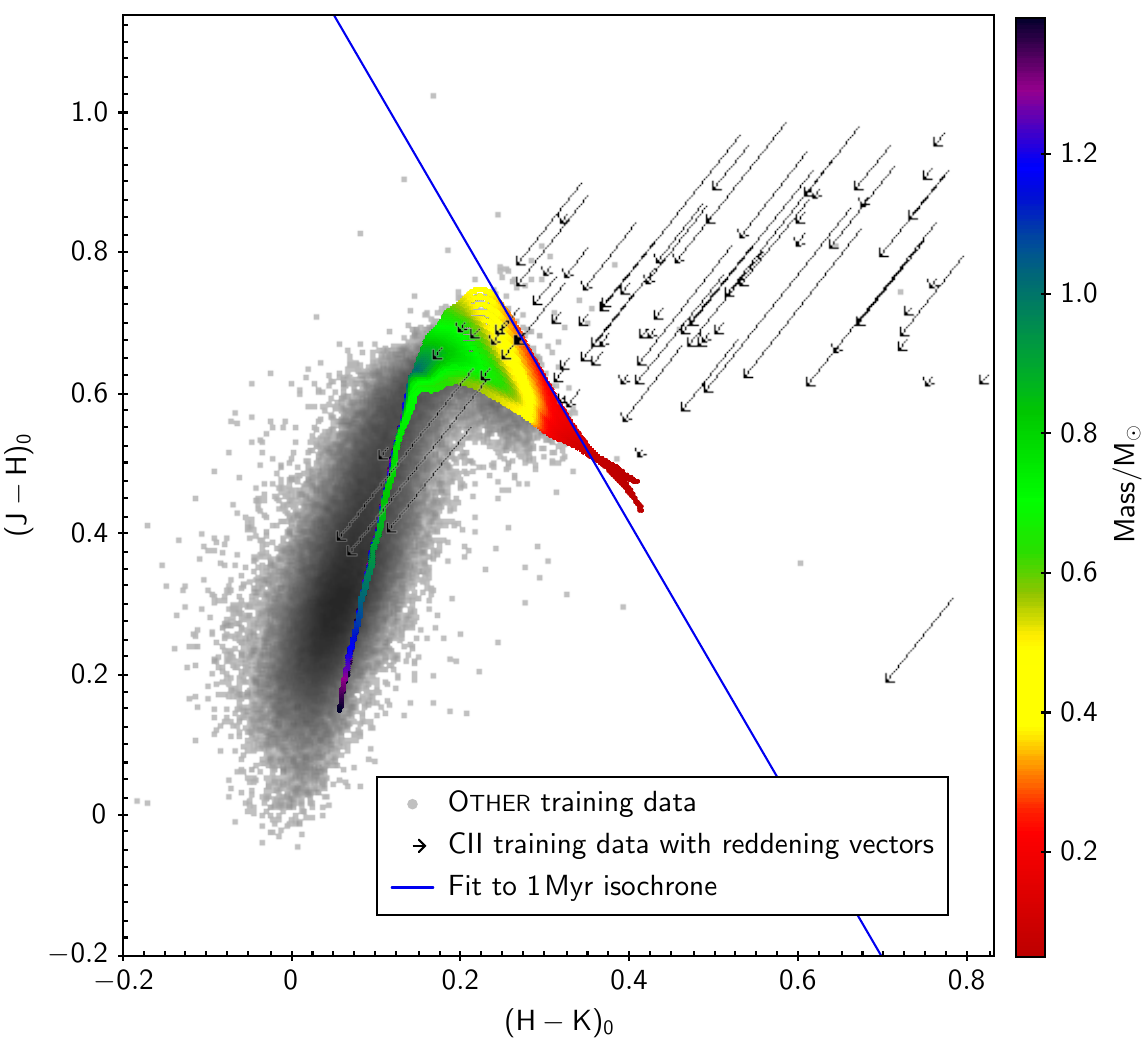}
	\caption{UKIDSS $(J-H,H-K)$ isochrones \edit{(see  Appendix~\ref{sec:isochrones})} for ages from 1\,Myr (top isochrone) to 100\,Myr (bottom isochrone) in steps of 0.1 dex. The blue line is the straight line fit to a 1\,Myr isochrone bewteen 0.013 and 0.35$\,\rm M_{\sun}$. The dereddened \textsc{Other} training set are grey dots. The black arrows are \textsc{CII} training set with reddening vectors, the start of each arrow is the observed location.}
	\label{fig:JHK_UKIDSS_iso_fig}
\end{figure}

The observations are dereddened in voxels of half a degree to each side in Galactic coordinates and 5\,pc in depth using transformed reddening values from \edittwo{STILISM \citep[][Appendix~\ref{sec:intersetllar_extinction}]{Lallement2014,Capitanio2017,2018A&A...616A.132L}.} We examined young stellar isochrones \edit{(described in  Appendix~\ref{sec:isochrones})} for ages from 1 to 100\,Myr in UKIDSS and 2MASS photometric systems \edit{(see Fig.~\ref{fig:JHK_UKIDSS_iso_fig})}. The stars tend \edit{to} descend in $(J-H)$ with age, while there is a more complex behaviour in $(H-K)$. For example a YSO of mass $0.4\,\rm M_{\sun}$ shows a slight increase in $(H-K)$ up to around 40\,Myr before decreasing. When compared to actual observations this effect is negligible, as can be seen in dereddened training data of  Fig.~\ref{fig:JHK_UKIDSS_iso_fig}, black reddening vectors for the \textsc{CII} and grey dots for the \textsc{Other}. The majority of the \textsc{CII} lie to higher $(J-H)$ and $(H-K)$ than the young stellar isochrones, while the \textsc{Other} tend to be closer to the older isochrones. The \textsc{CII} occupy the top right region of the CCD primarily due to infrared emission from their discs \citep{2000prpl.conf..377C}.

To have confidence in our approach, we briefly explore the reasons why many of the \textsc{CII} lie far from the young stellar isochrones in Fig.~\ref{fig:JHK_UKIDSS_iso_fig}. Variability in $J$, $H$ and $K_S$ was observed by \cite{Carpenter2001} for PMS stars in the Orion Nebula Cluster. Although much of the variability was colourless, some stars exhibited colour changes of 0.5 or more in $(H-K_S)$ and $(J-H)$. Their modelling could not fully explain these colour changes, though they considered a combination of cool and hot spots, variable extinction and changes in the circumstellar disc to be plausible explanations. The scatter in our \textsc{CII} training set is consistent with a snapshot from a population of varying young stars. There may also be a component caused by underestimation of the reddening, as our approach to dereddening will not capture the small scale variations in dust obscuration near to young stars. While we cannot give a detailed explanation for individual displacements of \textsc{CII} from the isochrones, these probable causes would not invalidate our approach.

A 1\,Myr \cite{Baraffe2015} interior with BT-Settl atmosphere $(J-H,H-K)$ isochrone has a linear region between 0.13 to 0.35 $\,\rm M_{\sun}$. We fitted a straight line to this region using an unweighted non-linear least-squares fit, with all data points lying within $\pm0.1$ magnitudes of the line \edit{(see Fig.~\ref{fig:JHK_UKIDSS_iso_fig})}. Separate fits were performed for UKIDSS
\begin{equation}
	(H-K) = -0.4835 (J-H) + 0.6003
	\label{eq:JHK_UKIDSS_ISOfit}
\end{equation}
and 2MASS
\begin{equation}
	(H-K) = -0.3733 (J-H) + 0.5188.
	\label{eq:JHK_2MASS_ISOfit}
\end{equation}

The UKIDSS and 2MASS photometry were dereddened as described in Appendix~\ref{sec:intersetllar_extinction}. The dereddened $(J-H)$ was used to calculate an expected $(H-K)$ value from the straight line fit. The excess was then calculated as the difference between the dereddened $(H-K)$ and the expected value. As the UKIDSS photometry has better spatial resolution, we took the UKIDSS results in preference to 2MASS where the photometry was available.

Away from the linear region our approach may not provide reliable results. In particular, the isochrones indicate that stars with $M \goa 0.8\,\rm M_{\sun}$ pile up on a line nearly perpendicular to the lower mass linear region. This means their observed location will be degenerate to age and mass. We do not consider this to be a serious problem, as these stars have both a $(J-H)$ and $(H-K)$ that is smaller than the majority of our \textsc{CII} training data. So we may miss some higher mass young stars rather than incorrectly classify older stars as \textsc{CII}. The effects of this degenerate region will be captured during training, when it can be expected to slightly weaken the feature.

\subsection{H$\alpha$ excess}
\label{sec:ha_feature}

It has long been known that young stars frequently display H$\alpha$ emission \citep{Herbig1958,Herbig1962b}, with an H$\alpha$ equivalent width $>10$\,\r{A} a defining characteristic of Classical T-Tauri stars. This emission is understood to originate in matter accreting from the disc on to the forming star via magnetically confined columns \citep{Edwards1994,Hartmann1994}. While this property is not unique to YSOs, an H$\alpha$ excess determined with a narrow band filter is nonetheless a good indicator of stellar youth. We use the IGAPS survey, an updated version of the INT Photometric H$\alpha$ Survey (IPHAS) used by \cite{Witham2008} and \cite{Barentsen2011} to successfully identify candidate YSOs. The IGAPS $i$-band AB photometry was transformed to the INT photometric system by the method described in Appendix~\ref{sec:igaps}.

\begin{figure}
	\includegraphics[width=\columnwidth]{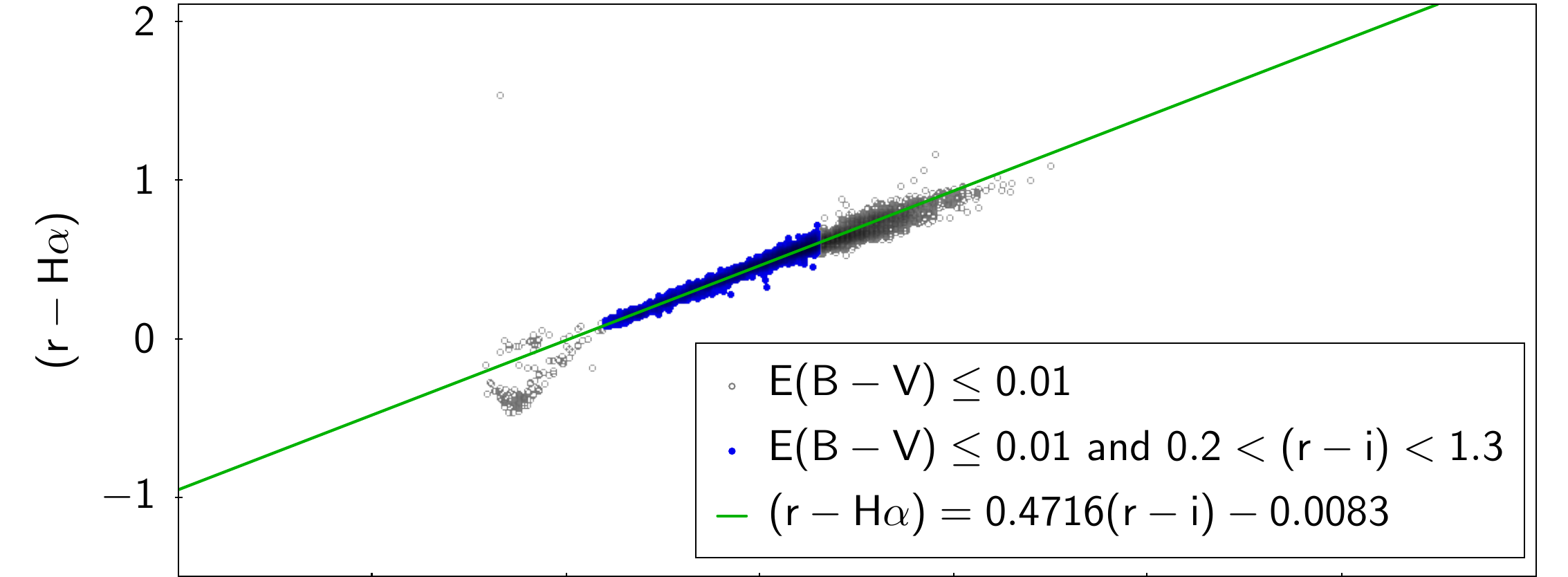}
	\includegraphics[width=\columnwidth]{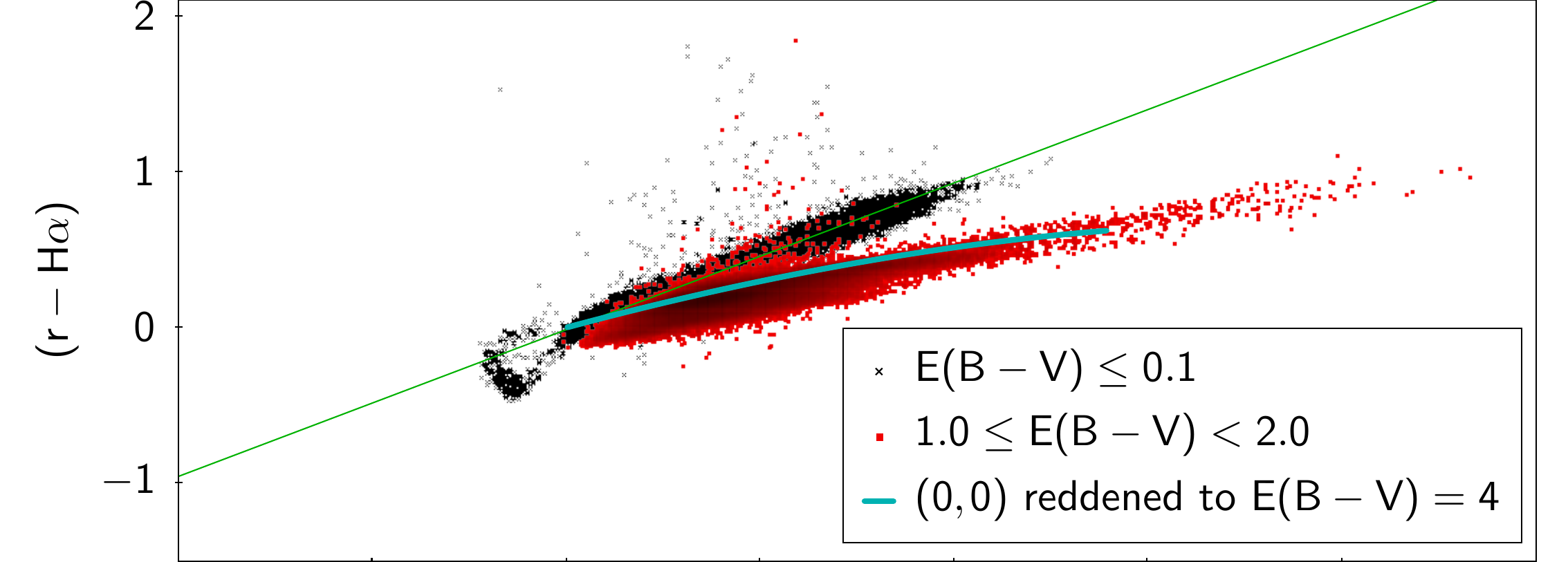}
	\includegraphics[width=\columnwidth]{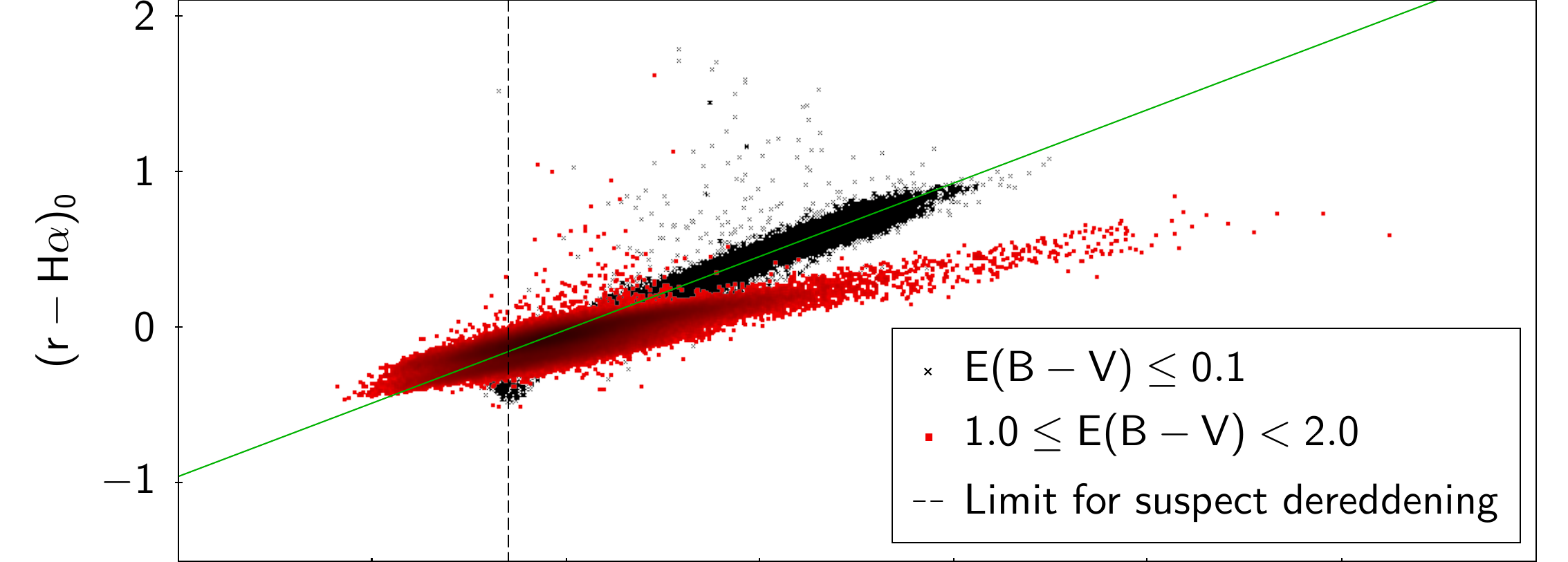}
	\includegraphics[width=\columnwidth]{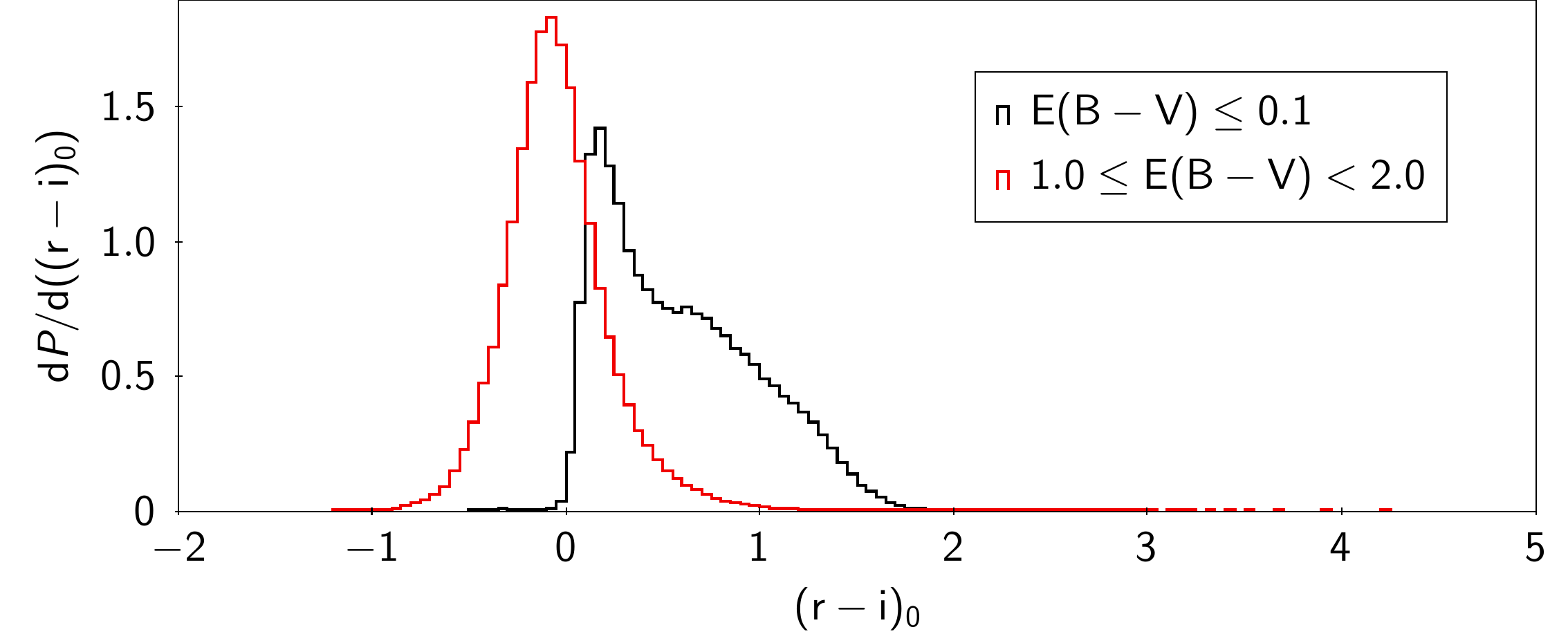}
	\caption{\edit{All four panels show subsets of IGAPS sources in the NGPn data. The empirical MS is the green line. In the top panel, the grey open circles are sources with $E(B-V) \leq 0.01$, and the blue filled circles have the additional restriction $0.2<(r-i)<1.3$ used for the empirical MS fit. The second panel shows IGAPS sources with $E(B-V) \leq 0.1$ as black crosses and $1.0 \leq E(B-V) < 2.0$ as red filled squares. The cyan curve is a reddening line to $E(B-V)=4$ for unreddened sources at $(0,0)$. The third panel is the same data as the second, except both $(r-i)$ and $(r-H\alpha)$ have been dereddened. The black dashed line indicates the limit we impose for suspect reddening at $(r-i) \leq -0.3$. The bottom panel shows a histogram of  dereddened $(r-i)$ for the same data selections and colouring as the third panel.}}
	\label{fig:IGAPS_rHa_ri}
\end{figure}

We define our H$\alpha$ excess feature as the excess in $(r-H\alpha)$ from dereddened $(r-H\alpha,r-i)$ CCDs \edit{(see Fig.~\ref{fig:IGAPS_rHa_ri})}. The MS in these CCDs can be approximated by a straight line. Thus the $(r-H\alpha)$ excess is simply calculated as the distance in $(r-H\alpha)$ between the dereddened observed value and an empirical MS.

The empirical MS was derived \edit{using} an unweighted non-linear least-squares fit to the IGAPS sources \edit{(see the top panel of Fig.~\ref{fig:IGAPS_rHa_ri})}. To minimize the \edit{effect} of reddening, only sources with an $E(B-V) \leq 0.01$ were selected from our NGPn data set. To prevent sources outside the linear region from skewing the fit, the data were restricted to $0.2<(r-i)<1.3$. \edit{The little hook of sources blueward of $(r-i)=0$ correspond to the higher mass stars seen in the isochrones of \citet[figure 4]{Drew2005}.} Outliers were removed by iteratively excluding sources greater than 0.1 magnitudes in $(r-H\alpha)$ from the fit. After two iterations this gave the stable straight line fit
\begin{equation}
	(r - H\alpha) = 0.4716 (r - i) - 0.0083.
	\label{eq:ri_rHa_MSfit}
\end{equation}

\edit{Although there is no simple uniform reddening vector \citep{Drew2005}, the effect of reddening in $(r-H\alpha,r-i)$ is to shift sources along the approximate direction of the MS. This can been seen in the second panel of Fig.~\ref{fig:IGAPS_rHa_ri}, where reddened stars selected by $1.0 \leq E(B-V) <2.0$ follow the cyan reddening line (illustrated for $E(B-V)=4$). It can be seen that the low reddening sources with $E(B-V) \leq 0.1$ lie along the MS while the reddened sources tend to lie slightly beneath it. The reddened sources can be traced back along the reddening line to an $(r-i)$ and $(r-H\alpha)$ of around zero, corresponding to the higher mass stars. The third panel of Fig.~\ref{fig:IGAPS_rHa_ri} shows the same data, now dereddened by our method described in Appendix~\ref{sec:intersetllar_extinction}. The scattering of sources along the reddening line even after dereddening illustrates that our dereddening is approximate. We discuss this in Appendix~\ref{sec:suspect_reddening} where we explain how uncertainties in the \textit{Gaia} parallaxes can lead to incorrect reddening values. The bottom panel of Fig.~\ref{fig:IGAPS_rHa_ri} is a histogram of the same low reddening and dereddened data from the panel above. The peak of the dereddened data lies at an $(r-i)$ of about -0.1 with nearly all sources within $\pm1.0$. Hence the long tail to $(r-i)>1.0$ of poorly dereddened data in the third panel contains very few sources. This incorrect dereddening} is not a serious problem where $(r-i)>0$ as they fall below the empirical MS line, making their $(r-H\alpha)$ excess more negative. \edit{This could result in some excess sources being missed.} It is more of a problem for $(r-i)<0$ as these sources are incorrectly given a positive $(r-H\alpha)$ excess. \edit{As described in Appendix~\ref{sec:suspect_reddening}, we flag sources with $(r-i) \leq -0.3$ as suspect reddening and we do not include their H$\alpha$ excess feature in the classifier calculation.}

\subsection{Isochronal Age}
\label{sec:isochronal_age_feature}

A YSO of a given temperature will have a larger radius than a MS star, and hence a higher luminosity. Therefore YSOs occupy a region above the MS in Colour-Magnitude Diagrams (CMDs). To convert this into a feature usable by the classifier, we interpolated the ages of sources from their location on CMDs relative to young stellar isochrones.

\begin{figure}
	\includegraphics[width=\columnwidth]{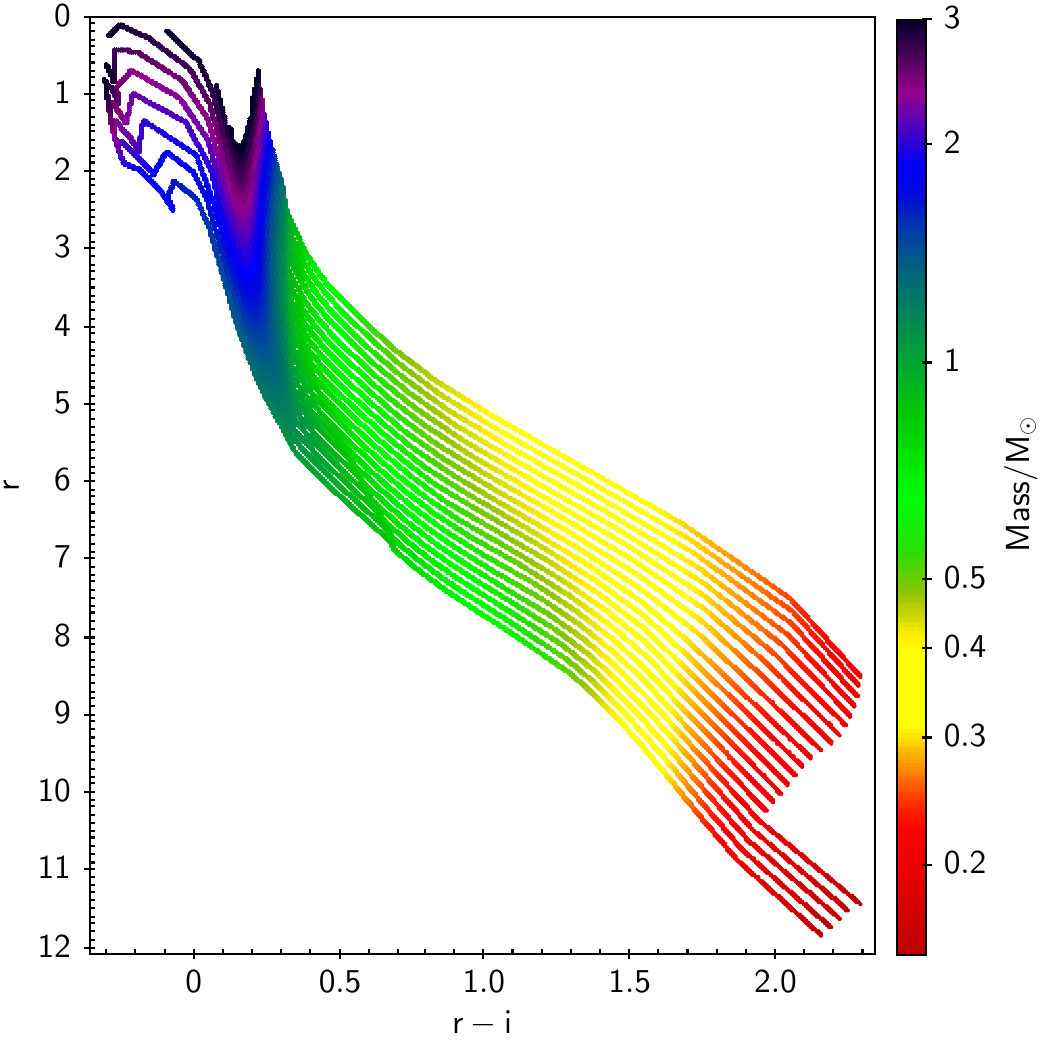}
	\caption{IGAPS $(r,r-i)$ \edit{isochrones (see  Appendix~\ref{sec:isochrones})} from 100\,kyr (top isochrone) to 10\,Myr (bottom isochrone) in steps of 0.1 dex. The isochrones cover the mass range from 0.05 to 3.00$\,\rm M_{\sun}$ and are coloured by mass on a logarithmic scale.}
	\label{fig:IGAPS_Dotter08_BTSettl_Bell}
\end{figure}

As YSOs emit strongly in the red and IR, we chose IGAPS $(r,r-i)$ CMDs for this feature. These CMDs have the benefit that the reddening vectors run approximately parallel to the line of the MS, \edit{translating primarily into an uncertainty on the mass with only a marginal uncertainty in age}. As with the H$\alpha$ excess feature, the IGAPS $i$-band AB photometry was transformed to the INT photometric system by the method of Appendix~\ref{sec:igaps}. The IGAPS $r$ magnitude was corrected for distance using the \textit{Gaia} EDR3 parallaxes. A set of isochrones \edit{(described in Appendix~\ref{sec:isochrones})} were created covering $\log_{10}(Age)$ from 5.0 to 7.0, masses from $0.05 \,\rm M_{\sun}$ to $3.00 \,\rm M_{\sun}$, and at reddenings from $E(B-V)=0.00$ to $E(B-V)=5.95$ in steps of 0.05. The unreddened isochrones are shown in Fig.~\ref{fig:IGAPS_Dotter08_BTSettl_Bell}. The photometry was matched to a set of reddened isochrones at or just below the reddening value assigned to the source using the STILISM reddening map \edit{(see Appendix~\ref{sec:intersetllar_extinction})}.

To keep the feature derivation straightforward, all sources were assumed to be single stars. While this assumption will be wrong for a large proportion of stars, to a limited extent the training will deal with the problem, since the training data should contain representative proportions of multiple stellar systems. The problem was also mitigated by the selection criteria of our NGPn data set. We required \edit{that} all sources have good quality \textit{Gaia} EDR3 parallax measurements. This selected sources that were consistent with a single star model in the \textit{Gaia} pipeline. This \edit{will} have removed a subset of multiple star systems where they were not resolvable into distinct sources by \textit{Gaia} and yet their orbital motion caused quality issues in the \textit{Gaia} pipeline.

While the isochrones gave good coverage of the YSO region in CMD space, the data points were at discrete values in $(r,r-i)$. To return an age and mass for a general point in $(r,r-i)$ we performed an interpolation for each set of reddened isochrones. The interpolation was calculated using a collection of Radial Basis Functions (RBFs) generated by the \textsc{scipy} interpolate library with a thin plate function ($r^2 \times \log(r)$ where $r$ is the Euclidian radius) and zero smoothing. The isochrones started to overlap above $2.00 \,\rm M_{\sun}$, so the interpolation was limited to $\leq 2.00 \,\rm M_{\sun}$.

As \textsc{scipy} interpolate can extrapolate outside of the isochrones, a series of quality flags were included in the results \edit{(see Table~\ref{tab:isoage_flags})}. These indicate whether the data fell outside the brightest, faintest, bluest and reddest of the isochrone bounds, and whether the interpolation result was above or below the age and mass range of the isochrones. Flags were also created for a backwards interpolation quality check, using another set of RBFs to interpolate the colour and magnitude from the interpolated age and mass. The backwards interpolation was only performed on data within the colour and magnitude bounds of the isochrones. When we checked the backwards interpolation using a regular grid of data points, most fell very close to their original $(r,r-i)$, though \edit{0.6 per cent of sources with IGAPS photometry landed over 0.1 magnitudes from their input $(r,r-i)$}. To allow the selection of good quality interpolations, a pair of flags indicate whether the back interpolated magnitude and colour fell within a tolerance of 0.1.

\begin{table*}
	\centering
	\caption{The flags returned by the classifier when performing the isochronal age interpolation. The final column states whether the result was used in the likelihood calculation described in Section~\ref{sec:isochronal_age_likelihoods}.}
	\label{tab:isoage_flags}
	\begin{tabular*}{\textwidth}{llc} % four columns, alignment for each
		\hline
		Flag & Description & Generate likelihood \\
		\hline
		brighter\_min\_iso\_mag & The source is >0.1 magnitudes brighter than the minimum of the isochrones. & No \\
		fainter\_max\_iso\_mag & The source is >0.1 magnitudes fainter than the maximum of the isochrones. & No \\
		above\_iso\_col & The source is >0.1 magnitudes redder than the red end of the isochrones. & No \\
		below\_iso\_col & The source is >0.1 magnitude bluer than the blue end of the isochones. & No \\
		good\_except\_back\_interp\_colour & No issues except the colour from back interpolating is >0.1 magnitudes different. & No \\
		good\_except\_back\_interp\_mag & No issues except the magnitude from back interpolating is >0.1 magnitudes different. & No \\
		above\_max\_iso\_red & The reddening of the source is above the maximum reddening of the isochrones. & No \\
		older\_max\_iso\_age & The derived age is above maximum age of the isochrones, \editthree{10\,Myr}. & Yes \\
		younger\_min\_iso\_age & The derived age is below the minimum age of the isochrones, \editthree{100\,kyr}. & No \\
		above\_max\_iso\_mass & The derived mass is above the maximum mass of 2 $\,\rm M_{\sun}$ of the isochrones. & No \\
		below\_min\_iso\_mass & The derived mass is below the minimum mass of 0.05 $\,\rm M_{\sun}$ of the isochrones. & No \\
		good\_classiiyso\_fit & The source is a good fit to the young stellar isochrones with no issues. & Yes \\
		\hline
	\end{tabular*}
\end{table*}

\subsection{Gaia G-band variability}
\label{sec:variability_feature}

A significant proportion of YSOs show some form of variability, and it is likely that all go through periods of variability as they evolve. This behaviour can be grouped into the two broad categories of periodic and aperiodic \citep{Carpenter2001}. Periodic variations are typically due to stellar rotation, caused by surface features such as starspots and accretion hot spots \citep{Herbst1994}. Aperiodic variability can be caused by variable accretion \citep{Herbst1994,Sergison2020} and variations in obscuration due to the circumstellar disc \citep{Bouvier1999,Froebrich2018b}.

Over the course of its mission \textit{Gaia} will make multiple observations (CCD transits) of each source, with new observations added in each release. This work uses data from EDR3, having an average of 376 $G$-band observations for each source in the NGPn data set. The number of observations per source varies across the sky due to the \textit{Gaia} scanning law. While the \textit{Gaia} archive does not provide a measure of variability, it provides the standard error and the number of observations contributing to the $G$-band photometry. These allow the reconstruction of the standard deviation of the $G$-band observations. We define our variability feature as the observed fractional standard deviation of the flux, identical to the variability amplitude of \cite{Deason2017}
\begin{equation}
	\hat{\sigma}_{\rm O} = \sqrt{N_{\rm{obs}}} \times \frac{S_{\rm O}}{\edit{\bar{F}}} = \frac{\sigma_{\rm O}}{\edit{\bar{F}}} ,
	\label{eq:GaiaVarAmp}
\end{equation}
where $N_{\rm{obs}}$ is the number observations (CCD transits) contributing to the $G$-band photometry, \edit{$\bar{F}$} is the observed $G$-band mean flux, $S_{\rm O}$ is the standard error of the $G$-band mean flux, and $\sigma_{\rm O}$ is the standard deviation of the $G$-band photometry.

\begin{figure}
	\includegraphics[width=\columnwidth]{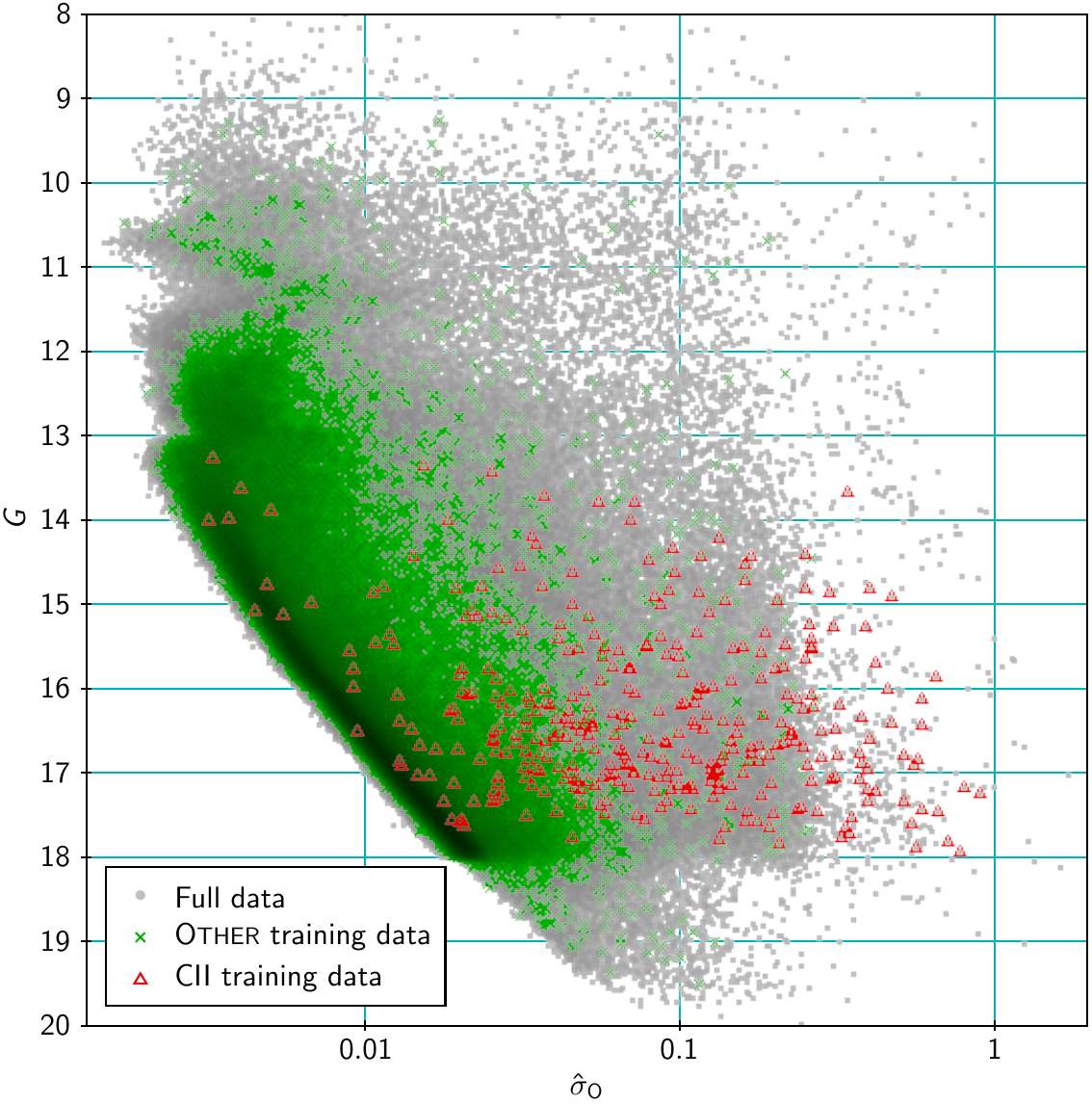}
	\caption{The relationship between \textit{Gaia} EDR3 $\hat{\sigma}_{\rm O}$ (fractional standard deviation of the flux) and $G$-band magnitude. The grey dots are the full NGPn dataset, green crosses are the \textsc{Other} training set, and red triangles are the \textsc{CII} training set.}
	\label{fig:Gvar_G_vs_FSDF}
\end{figure}

The $\hat{\sigma}_{\rm O}$ is a function of magnitude \edit{(see Fig.~\ref{fig:Gvar_G_vs_FSDF})}. There is a clear ridgeline in the plot where most sources lie. These are the stars with little or no intrinsic variability whose $\hat{\sigma}_{\rm O}$ is dominated by the instrumental noise. Those with larger $\hat{\sigma}_{\rm O}$ are the variable stars. The locations of the two training sets in Fig.~\ref{fig:Gvar_G_vs_FSDF} match our expectations. In general the \textsc{Other} training set will be composed of sources with low intrinsic variability. These tend to be on or near the ridgeline, with some scatter to larger $\hat{\sigma}_{\rm O}$ as the training set is expected to contain a variety of variable stars. While the \textsc{CII} sources are expected to be highly variable and are scattered across a range of $\hat{\sigma}_{\rm O}$ values.

The variations in $\hat{\sigma}_{\rm O}$ with magnitude are driven by changes in the standard error for a source given in the \textit{Gaia} catalogue. The general trend of the ridgeline to larger $\hat{\sigma}_{\rm O}$ with fainter magnitudes is due to the increase in photon noise. The other variations are rooted in the \textit{Gaia} gate and window class configurations that define how the telescope observes sources across different magnitude ranges \citep{Evans2017,Evans2018}. The changes in $\hat{\sigma}_{\rm O}$ for $G<12$ are due to gates reducing the exposure time for brighter sources. At around $G=13$ the window class changes from 2D for brighter sources to 1D for fainter sources. At $G=16$ there is a change in the window size, causing a subtle change in the slope of the ridgeline. These changes do not occur at precise magnitude boundaries as the star mapper CCDs estimate the magnitude of each source to decide the observation configuration \citep{Evans2017}. These star mapper estimates will naturally be less precise than the photometric measurements. Also, some of the intrinsically variable sources will move between gate and window classes as they vary. Since each source will be given a single mean $G$-band magnitude in the \textit{Gaia} archive, this will lead to blurring of the $\hat{\sigma}_{\rm O}$ behaviour at the boundaries.

\section{Naive Bayes Classifier}
\label{sec:naive_bayes_classifier}

Our goal is to create a classifier for producing a catalogue of candidate Class II YSOs. The catalogue needs to cover our NGPn footprint without gaps, unavoidable bias, and include sufficient data to allow us to understand the reasons behind individual probability assignments. To achieve these goals we chose to construct a naive Bayes classifier similar to \cite{Broos2011,Broos2013}, \cite{Li2020} and \cite{2022A&A...657A.138T}.

Naive Bayes classifiers are a kind of supervised machine learning. The results are a set of posterior probabilities for two or more classes. The inputs are a set of priors for these classes, and one or more features with likelihoods split by the classes. The features are measurements or derived properties, that are chosen for their ability to discriminate between the classes.

\edit{The learning process for Naive Bayes classifiers creates probability distributions by feature and class from the training data. These are referred to as the feature likelihoods. Our training approach for four of the five features defined the likelihoods from direct probability distributions of the training data. This gave robust likelihoods, even with relatively small data sets. The exception was the variability feature, where we created a model to generate probability density functions, with the model parameters set by the training data.}

The power of the classifier comes from combining different features. Where multiple features reinforce a particular class, this increases the posterior for that class. Where different features favour different classes or do not give a strong result, then the posterior tends to be dominated by the prior. The classifier is called naive as it assumes no correlation between features. While this assumption is usually false, the inclusion of correlations is unlikely to fundamentally change the classification results, and leaving out correlations makes for a significantly simpler classifier.

With naive Bayes classifiers, it is the ratio of the feature likelihoods between classes that drives the posterior probabilities. This means the training work grows with the square of the number of \edit{classes}. We settled on a simple binary classification of \textsc{CII} and \textsc{Other}.

As the likelihoods for a class are calculated from the training data for that class, the ratio of sources in the training sets does not need to match the ratio in the population. Instead the values of the priors encode the expected ratio of the classes in the population. Since the true number of Class II YSOs in the Galaxy is unknown, we chose the crude estimate for the priors of one Class II YSO in a thousand sources. We did not use a spatially varying prior with higher values in known star forming regions, as this would have biased our results to known regions.

\edit{As the true priors for our classifier are unknown, our choice of prior biases the posterior probabilities, meaning they are not literal probabilities. For example, a \textsc{CII} posterior of 0.8 does not mean a literal 80 per cent probability that the source is a Class II YSO. However, as we use the same priors for all sources, the rank of sources by posterior is identical to the rank by probability.}

\subsection{Mathematics of the classifier}

We use the nomenclature $P_{\rm s}$ for posterior probability, $P_{\rm r}$ for the prior probability, and $L$ for likelihood (the probability of a feature measurement for a given class from the training data). \edit{The priors satisfy
	\begin{equation}
		\sum_{i=1}^{x } P_{\rm r}(C_x) = 1
		\label{eq:prior}
	\end{equation}
	where $C_x$ is the class $x$. To keep nomenclature concise we define the shorthand posterior probability as}
\begin{equation}
	P_{\rm s}(C_x) = P(C_x|D_1,D_2,...,D_y) 
	\label{eq:posterior}
\end{equation}
where $D_y$ is the data for feature $y$. We define the likelihood of feature $f$ for class $x$
\begin{equation}
	L_f(C_x) = P(D_f|C_x) .
	\label{eq:likelihood}
\end{equation}

The naive Bayes classifier formula \citep{Duda2001,Broos2011,Broos2013} for the posterior probability of class $a$ for a binary classifier where the other class is $b$ and there are $m$ features is
\begin{equation}
	\begin{split}
		P_{\rm s}(C_a) &= \frac{P_{\rm r}(C_a) \prod_{f=1}^m L_f(C_a)}{ P_{\rm r}(C_a) \prod_{f=1}^m L_f(C_a) + P_{\rm r}(C_b) \prod_{f=1}^m L_f(C_b) } \\
		&= \frac{1}{1 + \frac{P_{\rm r}(C_b)}{P_{\rm r}(C_a)} \prod_{f=1}^m \frac{L_f(C_b)}{L_f(C_a)} } .
	\end{split}
	\label{eq:bayes_full}
\end{equation}
\edit{The second version of the formula provides insight into the calculation, since it demonstrates that the power of the classifier comes from the ratios of the likelihoods between the classes.}

A useful aspect of naive Bayes classifiers is the likelihoods may take the form of pure probabilities or probability densities. The fundamental restriction is the likelihoods at any specific value of a feature must not blend a pure probability with a probability density. They must be either one or the other to maintain the integrity of the likelihood ratio. For our feature likelihoods we use both pure probabilities and probability densities, though we never mix them in a ratio calculation \edit{(see Section~\ref{sec:likelihood_bins})}.

\subsection{Missing data}
\label{sec:missing_data}

Naive Bayes classifiers have a distinct advantage when it comes to missing data. They are able to simply bypass a feature if no data is available. Equation~\ref{eq:bayes_full} indicates the approach. As the ratio of the likelihoods drives the calculation, any result that gives a likelihood ratio of one will have no effect on the posterior. If all features were missing then the priors would be recovered as the posteriors. This is precisely the result we require in the absence of data. Hence, setting the likelihoods of a feature to the same value where the data is missing removes that feature from the calculation. For simplicity we chose likelihood values of one where a feature has no data, the approach taken by \cite{Broos2013}.

This has an important implication in how the features from multiple catalogues are combined. We do not need to limit our result set to the intersection of sources present in all the contributing catalogues. This gives us a much larger result set and allows us to add features without reducing the size of our result set. While it is beneficial to use as many features as possible, a posterior based on fewer than five features is valid. Our cap on the likelihood ratios prevents any single feature from overwhelming the prior. If only a single feature is present, this cap limits the posterior favouring \textsc{CII} to less than 0.5, regardless of the strength of result for that feature \edit{(see Section~\ref{sec:feature_likelihoods})}.

\subsection{Feature likelihoods}
\label{sec:feature_likelihoods}

The likelihoods are defined by probability distributions for each feature and class. We use either discrete distributions calculated by binning the training data, or continuous distributions from models whose parameters are set by the training data.

It is the ratio of the likelihoods between classes that defines the power of a feature to distinguish \textsc{CII} from \textsc{Other} sources. To avoid any single feature dominating the results from all other features, \edit{we set bounds to limit the ratio of individual feature likelihoods. We define these bounds as} the ratio of the priors and their reciprocal
\begin{equation}
	\frac{P_{\rm r}(\textsc{CII})}{P_{\rm r}(\textsc{Other})} \leq \frac{L(\textsc{CII})}{L(\textsc{Other})} \leq \frac{P_{\rm r}(\textsc{Other})}{P_{\rm r}(\textsc{CII})}.
	\label{eq:likelihood_ratio_prior_inequality}
\end{equation}
Were the Bayes classifier to be based on a single feature, this would lead the \textsc{CII} posterior being capped to
\begin{equation}
	\frac{1}{1 + \frac{P_{\rm r}(\textsc{Other})}{P_{\rm r}(\textsc{CII})} \times \frac{P_{\rm r}(\textsc{Other})}{P_{\rm r}(\textsc{CII})}}  < P_{\rm s}(\textsc{CII}) < \frac{1}{1 + \frac{P_{\rm r}(\textsc{Other}) }{P_{\rm r}(\textsc{CII}) } \times \frac{P_{\rm r}(\textsc{CII})}{P_{\rm r}(\textsc{Other})}}.
	\label{eq:initial_lhoodratio_inequality}
\end{equation}
As there is a large difference in the value of the priors, this can be simplified to
\begin{equation}
	\left( \frac{P_{\rm r}(\textsc{CII})}{P_{\rm r}(\textsc{Other})} \right)^2 \loa P_{\rm s}(\textsc{CII}) \leq 0.5.
	\label{eq:final_lhoodratio_inequality}
\end{equation}

\subsubsection{Choice of likelihood bins}
\label{sec:likelihood_bins}

The likelihoods for all but the variability feature were derived by calculating the proportion of sources in the training sets across a set of carefully chosen bins. To minimize the impact of small number statistics, a minimum of 10 sources per bin was imposed. The probability of a source having a value within any given bin was defined as the proportion of the training set within the bin. These were converted to probability densities using the bin widths, except at the end bins. The ends of the distributions are ill defined and sensitive to outliers, making it difficult to specify bin widths. Hence, the likelihoods in the end bins were left as pure probabilities. The same inner edges of the end bins were used for both classes, to ensure calculated likelihood ratios were always composed solely from probability densities or pure probabilities. This approach ensured likelihoods for any value of the features, even beyond the range of the training data.

The inner edge of the end bins were defined by the tails of the distributions and our requirement for bins to contain at least 10 sources. Hence, the likelihood ratios in these end bins usually cover a wide range of the feature space with a single value for the ratio. This is not a serious problem, since they will also have the most extreme value for the ratio favouring \textsc{CII} or \textsc{Other}, the exact behaviour that is required away from the overlap region. These are also the regions where the ratio may be capped, and hence it is not a great loss that the distributions do not capture a gradual change in the ratio away from the transition region.

The overlap region between the training sets is where the feature changes from favouring one classification to the other. If a feature is good at distinguishing between the classes, then the bulk of the training data will inevitably occupy different ranges of the feature space for the different classes. This can lead to a small region of overlap with a small proportion of the training data. While the \textsc{Other} training set contains 151\,645 sources, the \textsc{CII} training set is relatively small with 411 sources. Hence, the \textsc{CII} training is prone to suffer\edit{ing} from a lack of sources in the overlap region. Since we impose a minimum of 10 sources per bin, this can lead to an overlap with very few bins or even a single \textsc{CII} bin. As the likelihoods in the transition region are probability densities, it is not necessary for the \textsc{Other} and \textsc{CII} to be split into the same bins. To an extent this mitigates the problem, as the \textsc{Other} distribution can be split into multiple bins even if there is only a single \textsc{CII} bin.

The likelihood ratios were manually reviewed. Where feasible, bin edges were tweaked to minimize large jumps in the ratio between neighbouring bins. Outliers could lead to regions where the likelihood ratio became unphysical. These were analysed and where necessary adjustments were made to the bins.

\subsubsection{WISE likelihoods}

\begin{figure}
	\includegraphics[width=\columnwidth]{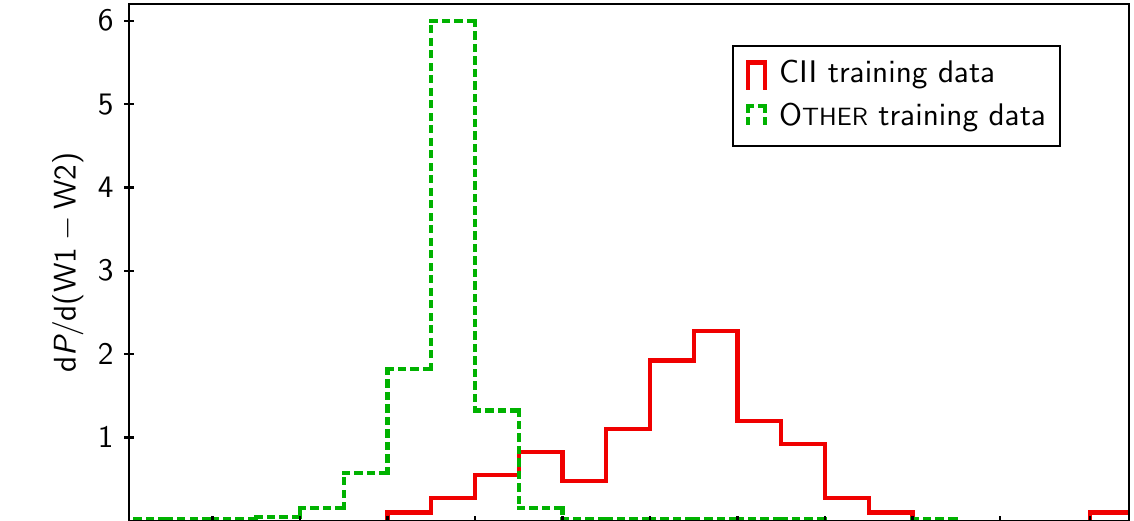}
	\includegraphics[width=\columnwidth]{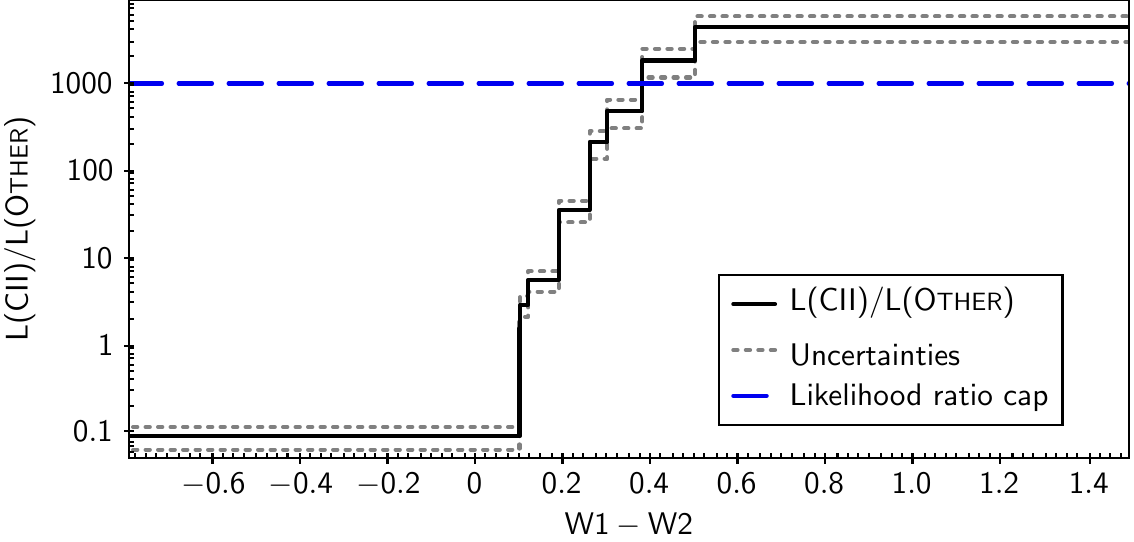}
	\caption{\edit{The top panel is a histogram of the \textit{WISE} $(W1-W2)$ training data with bins chosen to show the distribution of the training data. The \textsc{CII} are the solid red line, and \textsc{Other} sources are the green dotted line. The bottom panel is the ratio of the \textsc{CII} likelihoods divided by the \textsc{Other} likelihoods defined in Table~\ref{tab:wise_likelihoods}. The solid black line is the likelihood ratio, the grey dotted lines are the uncertainties, and the dashed blue line marks the cap on the ratio of the priors.}}
	\label{fig:WISE_Hist}
\end{figure}

\edit{A histogram of the \textit{WISE} training data is displayed in the top panel of Fig.~\ref{fig:WISE_Hist}. This illustrates that there is a good level of overlap between the two training sets. The likelihoods and their ratios are provided in Table~\ref{tab:wise_likelihoods} and a plot of the ratios with their uncertainties is given in the bottom panel of Fig.~\ref{fig:WISE_Hist}. Since we enforced a minimum of ten sources per bin, we approximated the counting uncertainties on the likelihoods to be the square root of the number of sources. To calculate the uncertainty on the likelihood ratio we added the fractional uncertainties from each class in a given bin in quadrature.}

When selecting the bins for the likelihood calculation, the requirement for a minimum of ten sources per bin resulted in the end bins occupying a significant proportion of the $(W1-W2)$ range. This left $0.1 < (W1-W2) \leq 0.5$ for the transition region, containing 1\,366 \textsc{Other} sources and 47 \textsc{CII}. The large number of \textsc{Other} sources enabled a gradual change in the likelihood ratio, with most \textsc{CII} bins covered by multiple \textsc{Other} bins.

Due to the power of the feature at large $(W1-W2)$, the likelihood ratios were capped from the second to last bin, $(W1-W2)>0.38$. This capped the likelihoods for 1\,568 sources and limited the true size of the transition region to $0.100 < (W1-W2) \leq 0.380$. \edit{The transition region contained 1.3 per cent of sources with the \textit{WISE} feature.}

\begin{table}
	\centering
	\caption{The \textit{WISE} feature likelihoods and their ratios. The end bins are pure probabilities while the inner bins are probability densities. Note that probability densities can attain a value greater than 1. Where the ratio is greater than the ratio of the priors, the likelihoods are capped to the ratio of the priors.}
	\label{tab:wise_likelihoods}
	\begin{tabular}{lcccc} % four columns, alignment for each
		\hline
		$(W1-W2)$ & \textsc{CII} & \textsc{Other} & Ratio \\
		range & likelihood & likelihood & L(\textsc{CII})/L(\textsc{Other}) \\
		\hline
		Below 0.100 & 0.0909 & 0.985 & 0.0932 \\
		0.100 to 0.101 & 0.682 & 0.425 & 1.60 \\
		0.101 to 0.120 & 0.682 & 0.231 & 2.95 \\
		0.120 to 0.190 & 0.682 & 0.120 & 5.67 \\
		0.190 to 0.260 & 0.682 & 0.0190 & 35.9 \\
		0.260 to 0.300 & 0.985 & 0.00464 & 212 \\
		0.300 to 0.380 & 0.985 & 0.00205 & 482 \\
		0.380 to 0.500 & 1.67 & 0.000909 & 1\,830 \\
		Above 0.500 & 0.482 & 0.000109 & 4\,420 \\
		\hline
	\end{tabular}
\end{table}

\subsubsection{(H-K) Excess  likelihoods}

The $(H-K)$ excess feature has two sets of likelihoods, one for UKIDSS and the other for 2MASS.  At first glance the number of \textsc{CII} with UKIDSS photometry \edit{quoted in Table~\ref{tab:cii_features}} appears anomalously low. This is explained as UKIDSS does not provide full coverage of our NGPn footprint, missing the region containing the young cluster Cep OB3b that contributes a significant proportion of the \textsc{CII} training set.

\begin{figure}
	\includegraphics[width=\columnwidth]{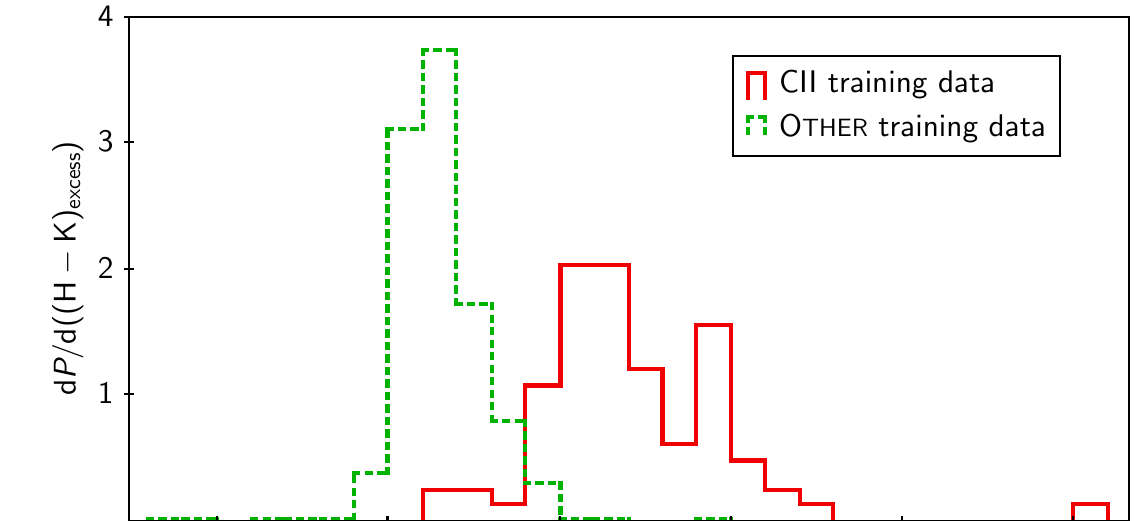}
	\includegraphics[width=\columnwidth]{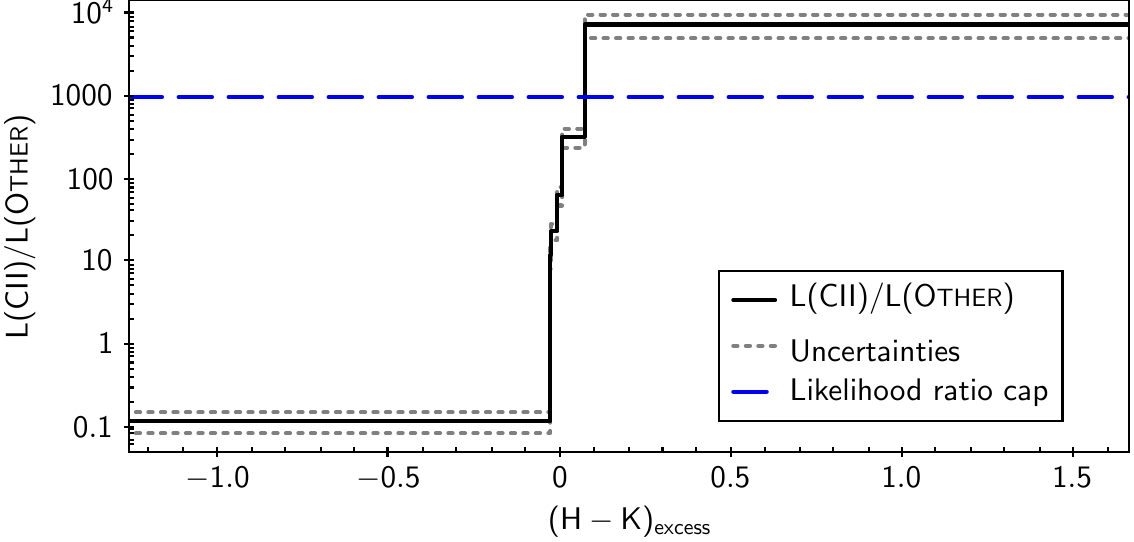}
	\caption{\edit{As Fig.~\ref{fig:WISE_Hist} but for the $(H-K)$ excess feature UKIDSS training data and likelihoods. The bottom panel likelihood ratios are defined in Table~\ref{tab:hk_ukidss_likelihoods}.}}
	\label{fig:UKIDSS_Hist}
\end{figure}

\begin{figure}
	\includegraphics[width=\columnwidth]{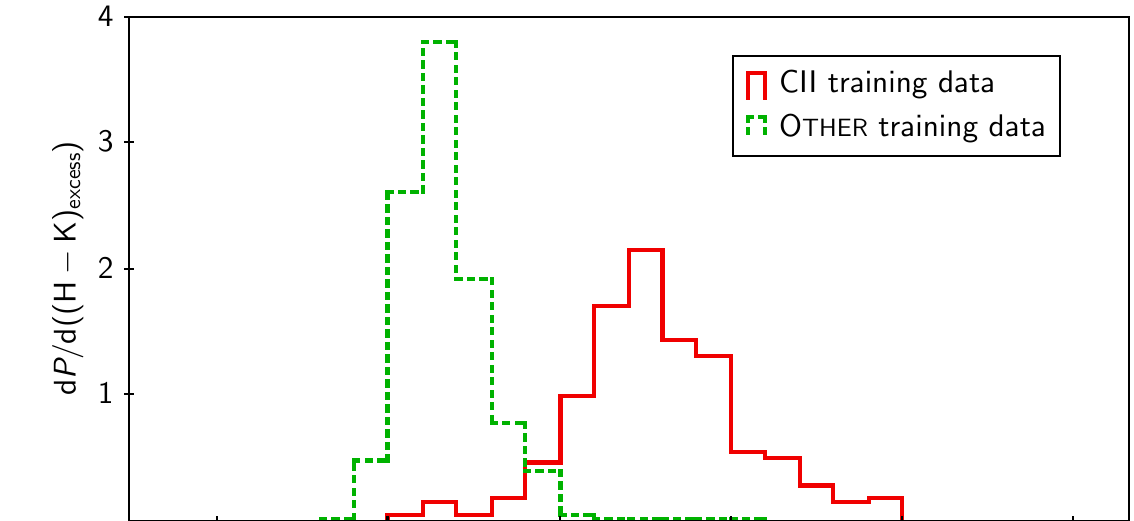}
	\includegraphics[width=\columnwidth]{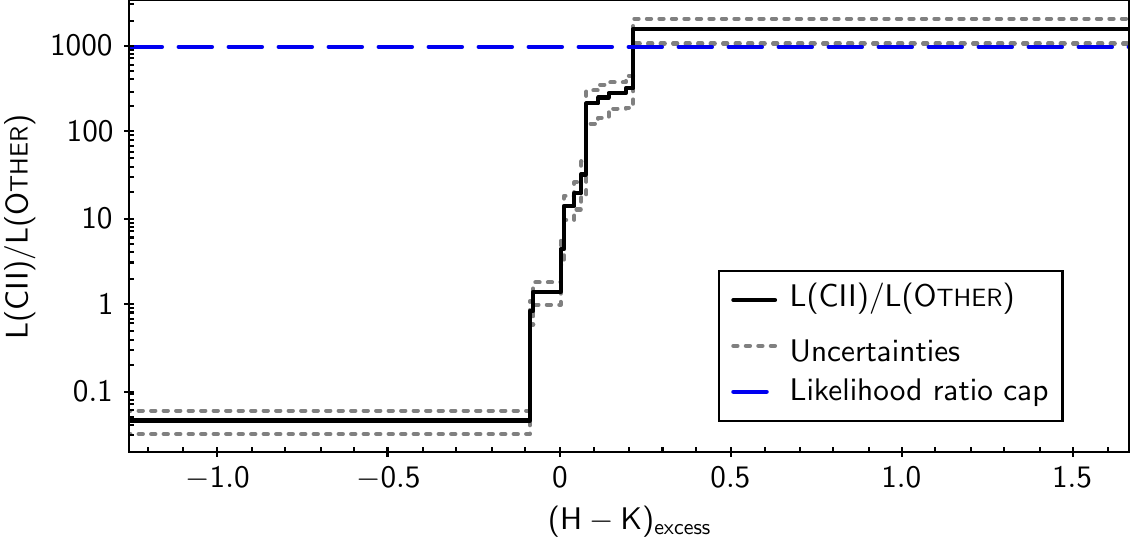}
	\caption{\edit{As Fig.~\ref{fig:WISE_Hist} but for the $(H-K)$ excess feature 2MASS training data and likelihoods. The bottom panel likelihood ratios are defined in Table~\ref{tab:hk_2mass_likelihoods}.}}
	\label{fig:2MASS_Hist}
\end{figure}

The training data from both surveys provide good overlap of the \textsc{CII} and the \textsc{Other} probability distributions \edit{as can be seen in the top panels of Fig.~\ref{fig:UKIDSS_Hist}~and~\ref{fig:2MASS_Hist}}. A small bump is present in the \textsc{CII} distributions of both surveys at a value corresponding to the peak of the \textsc{Other} training data. This could indicate a slight contamination of the \textsc{CII} training set, or it may be an artefact due to the inability of the feature to distinguish \textsc{CII} from \textsc{Other} for $M \goa 0.8\,\rm M_{\sun}$, where the isochrones deviate from the linear appoximation used for this feature.

The values of the UKIDSS likelihoods are given in Table~\ref{tab:hk_ukidss_likelihoods} and 2MASS in Table~\ref{tab:hk_2mass_likelihoods}. \edit{Their ratios with uncertainties are displayed in the bottom panels of Fig.~\ref{fig:UKIDSS_Hist}~and~\ref{fig:2MASS_Hist}.} Once again, our requirement for at least ten sources in each bin leads to the end bins occupying a significant proportion of the feature range. This results in a rapid change in the likelihood ratio across the remaining region $-0.030 < (H-K) \leq 0.070$ for UKIDSS and \edit{a more gradual change across} $-0.090 < (H-K) \leq 0.212$ for 2MASS. The steps in likelihood ratios between neighbouring bins are more coarse for UKIDSS than 2MASS. The \textsc{Other} data allow the single UKIDSS \textsc{CII} bin in the transition region to be split into four, though there is still a \edit{jump by a factor of 103 at the lower end and by 23} at the upper end of the feature values, compared to 19 and 5 for 2MASS.

Only the upper $(H-K)$ end bin had its likelihood ratio capped. This leads to a cap on 7\,491 sources, \edit{of which 1\,866 are from UKIDSS and 5\,625 are from 2MASS}. There were 102\,649 sources in the transition region, 1.6 per cent of sources with an $(H-K)$ likelihood.

\begin{table}
	\centering
	\caption{The UKIDSS $(H-K)$ feature likelihoods and their ratio. \edit{The notes and rules from Table~\ref{tab:wise_likelihoods} apply.}}
	\label{tab:hk_ukidss_likelihoods}
	\begin{tabular}{lcccc} % four columns, alignment for each
		\hline
		UKIDSS $(H-K)$ & \textsc{CII} & \textsc{Other} & Ratio \\
		excess range & likelihood & likelihood & $L(\textsc{CII})/L(\textsc{Other})$ \\
		\hline
		Below -0.030 & 0.119 & 0.997 & 0.119 \\
		-0.030 to -0.029 & 1.90 & 0.155 & 12.3 \\
		-0.029 to -0.010 & 1.90 & 0.0802 & 23.8 \\
		-0.010 to 0.005 & 1.90 & 0.0292 & 65.3 \\
		0.005 to 0.070 & 1.90 & 0.00575 & 331 \\
		Above 0.070 & 0.690 & 0.0000912 & 7\,570 \\
		\hline
	\end{tabular}
\end{table}

\begin{table}
	\centering
	\caption{The 2MASS $(H-K)$ feature likelihoods and their ratio. \edit{The notes and rules from Table~\ref{tab:wise_likelihoods} apply.}}
	\label{tab:hk_2mass_likelihoods}
	\begin{tabular}{lcccc} % four columns, alignment for each
		\hline
		2MASS $(H-K)$ & \textsc{CII} & \textsc{Other} & Ratio \\
		excess range & likelihood & likelihood & $L(\textsc{CII})/L(\textsc{Other})$ \\
		\hline
		Below -0.090 & 0.0446 & 0.962 & 0.0464 \\
		-0.090 to -0.080 & 0.493 & 0.571 & 0.864 \\
		-0.080 to 0.000 & 0.493 & 0.341 & 1.45 \\
		0.000 to 0.010 & 0.493 & 0.109 & 4.53 \\
		0.010 to 0.040 & 0.828 & 0.0576 & 14.4 \\
		0.040 to 0.060 & 0.828 & 0.0413 & 20.0 \\
		0.060 to 0.075 & 0.828 & 0.0250 & 33.1 \\
		0.075 to 0.110 & 1.41 & 0.00635 & 222 \\
		0.110 to 0.140 & 1.64 & 0.00635 & 259 \\
		0.140 to 0.190 & 1.97 & 0.00678 & 291 \\
		0.190 to 0.212 & 2.24 & 0.00678 & 331 \\
		Above 0.212 & 0.603 & 0.000375 & 1\,610 \\
		\hline
	\end{tabular}
\end{table}

\subsubsection{$H\alpha$ excess likelihoods}
\label{sec:halpha_excess_likelihoods}

\begin{figure}
	\includegraphics[width=\columnwidth]{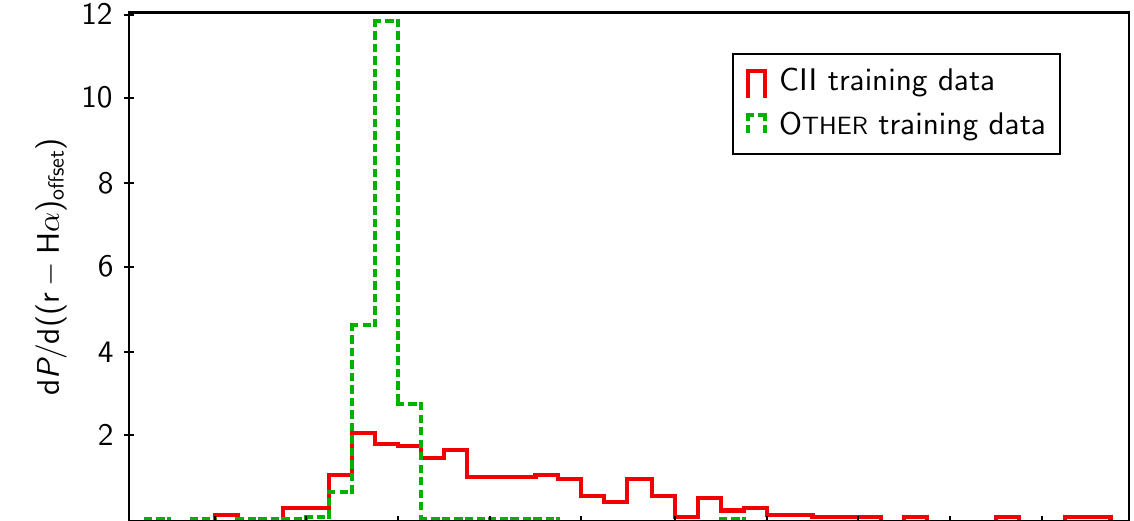}
	\includegraphics[width=\columnwidth]{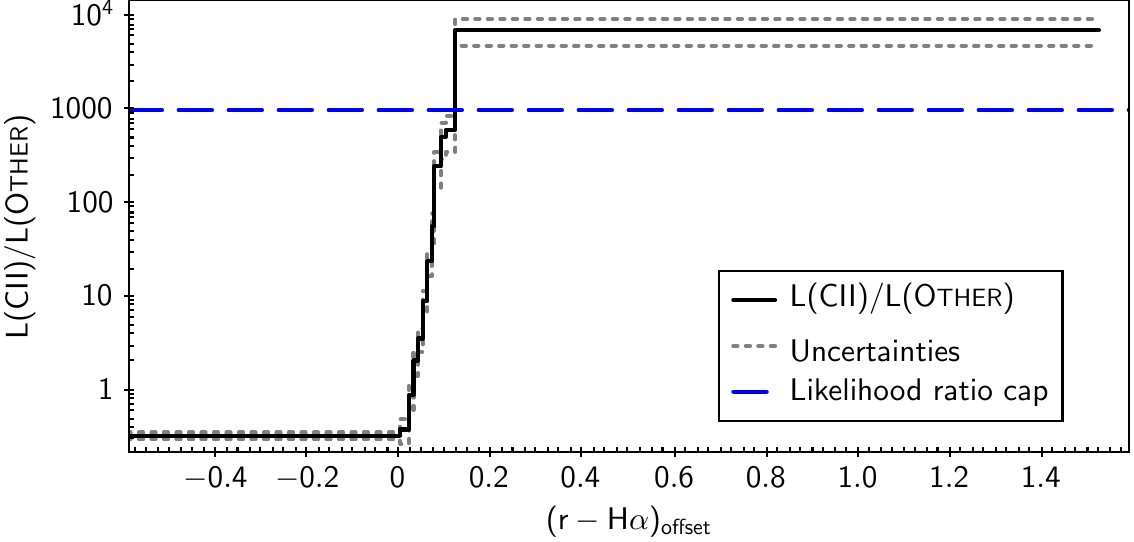}
	\caption{\edit{As Fig.~\ref{fig:WISE_Hist} but for the $H\alpha$ excess feature. The bottom panel likelihood ratios are defined in Table~\ref{tab:HaEx_likelihoods}.}}
	\label{fig:HaEx_Hist}
\end{figure}

\edit{The \textsc{CII} training set has a broad range in the $(r-H\alpha)$ offset feature space, completely covering the narrower range of the \textsc{Other} training set (see the top panel of Fig.~\ref{fig:HaEx_Hist}).} We would expect all \textsc{CII} to have a positive $(r-H\alpha)$ excess, though there are sources in the \textsc{CII} training set with negative $(r-H\alpha)$ excess. This is most likely caused by incorrect dereddening. Our $(r-i) \leq -0.3$ indicator for bad reddening is relatively crude and will not pick up all suspect values. We could reasonably expect incorrect reddening to affect \textsc{CII} more than \textsc{Other} sources as they form in dusty environments. Hence, we treat the negative $(r-H\alpha)$ excess with scepticism. To avoid unphysical likelihood ratios in bins splitting the negative $(r-H\alpha)$ excess values, we set the lower end bin to cover all sources with $(r-H\alpha) < 0$, giving a likelihood ratio consistent with the trend to positive $(r-H\alpha)$ excess.

The values of the likelihoods and their ratios are given in Table~\ref{tab:HaEx_likelihoods}. \edit{The ratios with uncertainties are displayed in the bottom panel of Fig.~\ref{fig:HaEx_Hist}.} The end bins occupy a significant proportion of the $H\alpha$ excess feature, resulting in a rapid change in the likelihood ratio across the remaining region $0.000 < (r-H\alpha) \leq 0.121$. This transition region contains 1\,064\,294 sources, 17 per cent of the sources with an $H\alpha$ excess likelihood. The training data allowed a good distribution of bins, avoiding large jumps in the likelihood ratio between neighbouring bins. The largest jump is a factor of 12 between the last two bins, though the capping of the likelihood ratio limited this to less than a factor of 2.

The likelihood ratio was only capped for the final bin, $(r-H\alpha)>0.121$. This lead to 1\,932 sources receiving capped likelihoods, 0.03 per cent of the sources with \edit{an} $H\alpha$ excess likelihood.

\begin{table}
	\centering
	\caption{The $H\alpha$ excess feature likelihoods \edit{and their ratio. The notes and rules from Table~\ref{tab:wise_likelihoods} apply.}}
	\label{tab:HaEx_likelihoods}
	\begin{tabular}{lcccc} % four columns, alignment for each
		\hline
		$(r-H\alpha)$ & \textsc{CII} & \textsc{Other} & Ratio \\
		offset range & likelihood & likelihood & $L(\textsc{CII})/L(\textsc{Other})$ \\
		\hline
		Below 0.000 & 0.285 & 0.861 & 0.331 \\
		0.000 to 0.020 & 2.01 & 5.22 & 0.384 \\
		0.020 to 0.030 & 1.82 & 2.02 & 0.903 \\
		0.030 to 0.040 & 1.82 & 0.870 & 2.10 \\
		0.040 to 0.050 & 1.25 & 0.347 & 3.61 \\
		0.050 to 0.060 & 1.25 & 0.138 & 9.06 \\
		0.060 to 0.070 & 1.25 & 0.0519 & 24.1 \\
		0.070 to 0.075 & 1.25 & 0.0219 & 57.1 \\
		0.075 to 0.090 & 1.46 & 0.00585 & 249 \\
		0.090 to 0.100 & 1.46 & 0.00283 & 516 \\
		0.100 to 0.121 & 1.74 & 0.00283 & 614 \\
		Above 0.121 & 0.522 & 0.0000731 & 7\,140 \\
		\hline
	\end{tabular}
\end{table}

\subsubsection{Isochronal age likelihoods}
\label{sec:isochronal_age_likelihoods}

\begin{figure}
	\includegraphics[width=\columnwidth]{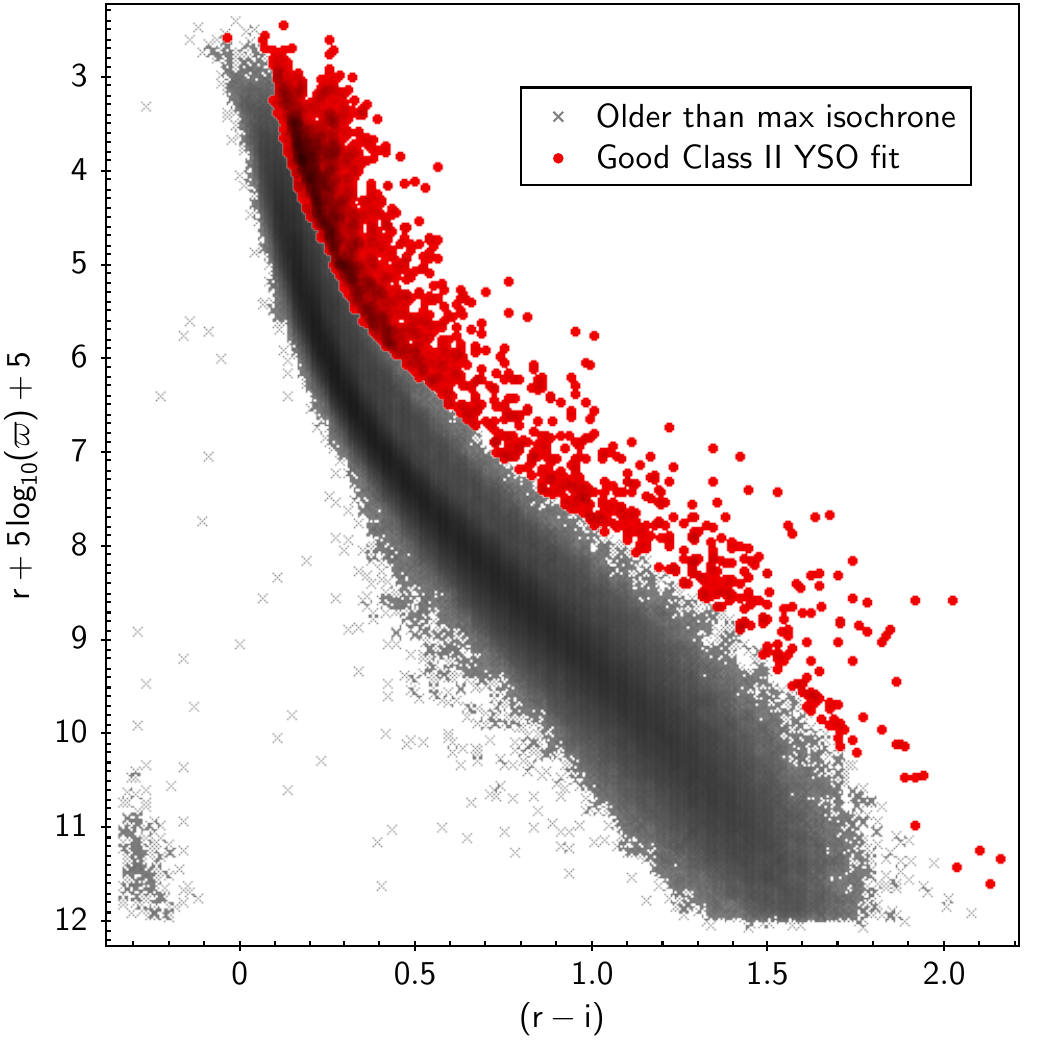}
	\caption{An $(r,r-i)$ CMD of the sources in the NGPn data set used in the Isochronal Age feature. The $r$ magnitude is distance corrected using the \textit{Gaia} EDR3 parallax. The red dots are sources flagged as a \edit{\lq good Class II YSO fit\rq{}} to the young stellar isochrones with an $E(B-V)<0.1$. The grey crosses are sources flagged as \edit{\lq older than the maximum isochronal age\rq{}} with $E(B-V)<0.1$.}
	\label{fig:IsoAge_GoodCII_Older}
\end{figure}

Although the classifier returns an age based on the location of a source relative to young stellar isochrones, this age should not be taken at face value. The feature has inherent problems such as ignoring multiple star systems and the interpolation of the isochrones can give erroneous results. This is where the quality flags come in \edit{(see Table~\ref{tab:isoage_flags})}. The age was only used where the flags indicated a reliable result. This was defined as sources within the bounds of the isochrone colours and magnitudes, the backwards interpolation of colour and magnitude from the derived age and mass yielded values within 0.1 of the input colour and magnitude, and the resulting mass and age were within the limits of the isochrones. The sources with these reliable results were flagged as \lq good Class II YSO fit\rq \edit{. A subset with $E(B-V)<0.1$ are shown in Fig.~\ref{fig:IsoAge_GoodCII_Older}), demonstrating that as expected, their location lies above the MS.}

It should be noted that some of the \textsc{Other} training set fall within the young stellar isochones and are thus assigned ages consistent with YSOs. This is to be expected as some multiple star systems composed of MS stars will lie above the MS, occupying the region where single YSOs are found. Also, stars transition across the YSO region when they evolve off the MS. This is an inherent weakness of our single star isochrone approach, discussed in Sections~\ref{sec:isochronal_age_feature} and \ref{sec:improvements_and_bias}.

Only a small fraction of the total sources in our NGPn data set occupy the young star region of the CMD. We therefore examined the data outside of the young stellar isochrones to see if robust likelihood ratios could be assigned to some of these sources. We found age extrapolation to younger or older values than the isochrone bounds gave robust and useful results even though the exact ages were meaningless. The older and younger flags were only assigned where the original photometry lay within the upper and lower bounds of the isochrone colours and magnitudes. These results were combined into bulk younger and older likelihood bins. However, there were only 2 sources flagged as younger in the \textsc{CII} training data. As we impose a minimum of 10 sources for a likelihood calculation, the younger bin was not used. The grey crosses of Fig.~\ref{fig:IsoAge_GoodCII_Older} are the subset of sources flagged as older, with only sources with $E(B-V)<0.1$ shown to avoid confusion due to reddening. Their location is consistent with an older age as they lie below the red dots of the young stars, and the white dwarfs jump out as a separate group in the bottom left corner.

Once sources with suspect reddening were removed, this left 5\,939\,814 sources with an Isochronal Age likelihood, 95 per cent of those with IGAPS photometry and 74 per cent of our NGPn catalogue. This was split 849\,978 classified as a \lq good Class II YSO fit\rq{} and 5\,089\,836 flagged as \lq older than the maximum isochronal age\rq. The high proportion of sources classified as a \lq good Class II YSO fit\rq\ is not telling us there is a high proportion of young stars. This simply tells us they occupy the region of the CMD where single young stars are found. This same region will include multiple MS stars and stars evolving off the MS. There were 64\,347 sources flagged as \lq younger than the minimum isochronal age\rq\ that we were unable to use.

\begin{figure}
	\includegraphics[width=\columnwidth]{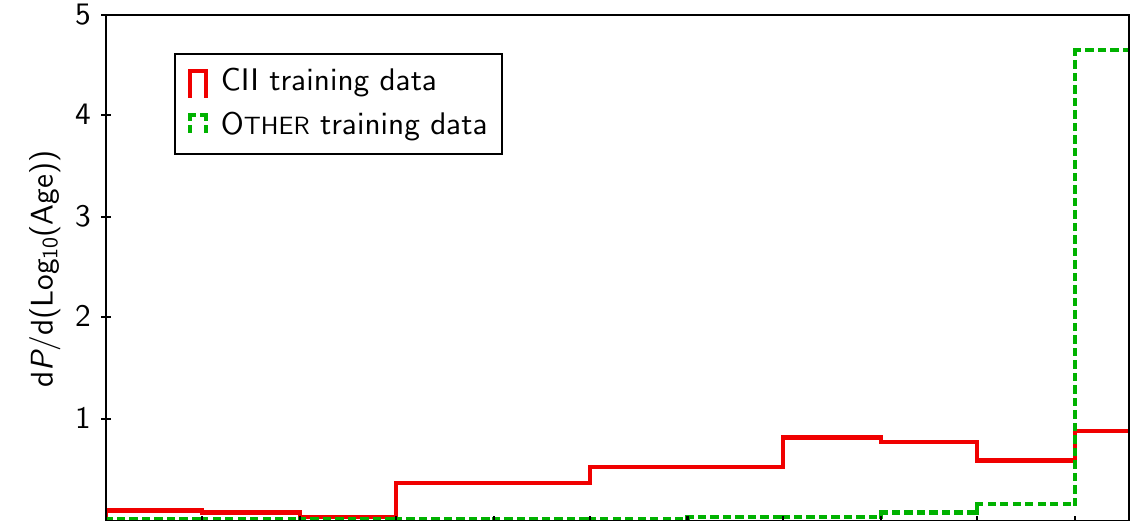}
	\includegraphics[width=\columnwidth]{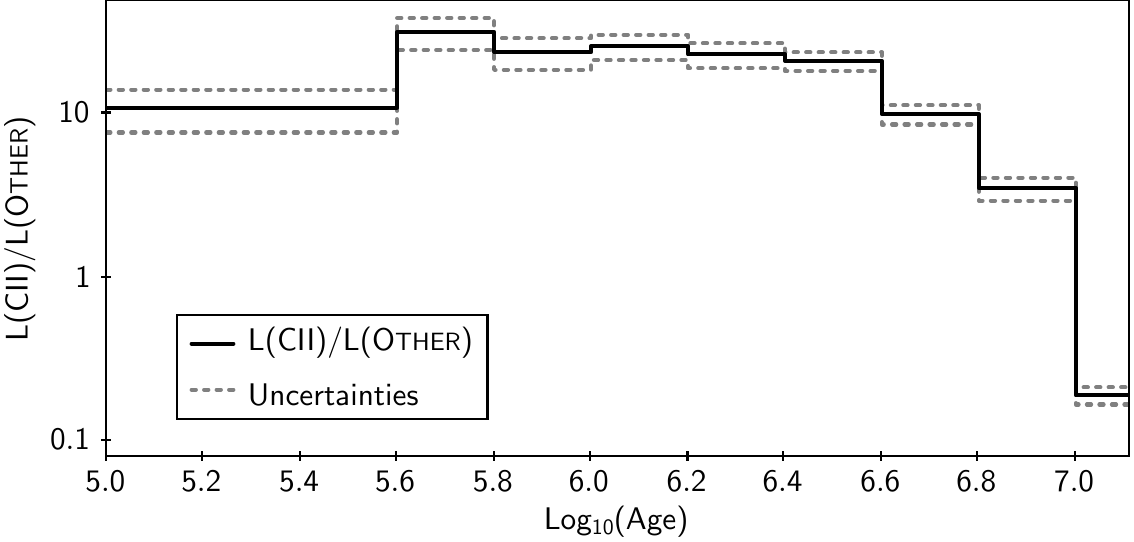}
	\caption{\edit{As Fig.~\ref{fig:WISE_Hist} but for the isochronal age feature. The training data flagged as a \lq good Class II YSO fit\rq{} cover $5.0 \leq log_{10}(Age) \leq 7.0$. The final bin contains all the data flagged as \lq older than the maximum isochronal age\rq{}, the sources with $log_{10}(Age)>7.0$. The bottom panel likelihood ratios are defined in Table~\ref{tab:isoage_likelihoods}.}}
	\label{fig:IsoAge_Training}
\end{figure}

The top \edit{panel} of Fig.~\ref{fig:IsoAge_Training} shows the distribution of the training data flagged as \edit{a \lq good Class II YSO fit\rq{} or \lq older than the maximum isochronal age\rq{}}. The bulk of the \textsc{CII} lie at $5.6 < log_{10}(Age) < 7.0$, with a peak at $6.4 < log_{10}(Age) < 6.6$ \edit{though 18 per cent are flagged as \lq older than the maximum isochronal age\rq}. The \textsc{Other} exhibit increasing numbers with increasing age.

Defining probability densities for $log_{10}(Age)$ would be non-trivial. Instead we imposed identical bins on the two classifications, allowing the likelihoods to be defined as pure probabilities. \edit{The likelihoods and their ratios are given Table~\ref{tab:isoage_likelihoods}, and the bottom panel of Fig.~\ref{fig:IsoAge_Training} shows the ratios with uncertainties.} The likelihoods were calculated as the proportion of the training data flagged as a \edit{\lq good Class II YSO fit\rq{} or \lq older than the maximum isochronal age\rq}.

It can be seen in the bottom \edit{panel} of Fig.~\ref{fig:IsoAge_Training} and Table~\ref{tab:isoage_likelihoods}, that the likelihood ratios favouring \textsc{CII} peak at a $log_{10}(Age)$ of 5.6 to 5.8 with a ratio of 31.8. The sources given a $log_{10}(Age)>7.0$ slightly favour the \textsc{Other} classification. None of the likelihood ratios give a strong result. This feature therefore has a weak affect on the posterior when compared with the more extreme likelihood ratios of the other features.

\begin{table}
	\centering
	\caption{The isochronal age feature likelihoods and their ratio. All the likelihoods are calculated as pure probabilities, using the proportion of sources in the bin, relative to the total number of sources used in training that class.}
	\label{tab:isoage_likelihoods}
	\begin{tabular}{lcccc} % four columns, alignment for each
		\hline
		$log_{10}(Age)$ & \textsc{CII} & \textsc{Other} & Ratio \\
		range & likelihood & likelihood & $L(\textsc{CII})/L(\textsc{Other})$ \\
		\hline
		Younger than 5.0 & Insufficient data & 0.00246 &  \\
		5.0 to 5.6 & 0.0410 & 0.00375 & 11.0 \\
		5.6 to 5.8 & 0.0709 & 0.00223 & 31.8 \\
		5.8 to 6.0 & 0.0709 & 0.00296 & 24.0 \\
		6.0 to 6.2 & 0.104 & 0.00401 & 26.1 \\
		6.2 to 6.4 & 0.104 & 0.00450 & 23.2 \\
		6.4 to 6.6 & 0.164 & 0.00775 & 21.2 \\
		6.6 to 6.8 & 0.153 & 0.0153 & 10.0 \\
		6.8 to 7.0 & 0.116 & 0.0328 & 3.52 \\
		Older than 7.0 & 0.175 & 0.927 & 0.189 \\
		\hline
	\end{tabular}
\end{table}

\subsubsection{Gaia G-band variability likelihoods}
\label{sec:variability_likelihoods}

As the variability feature ($\hat{\sigma}_{\rm O}$) exhibits a complex behaviour, we constructed a model for $\hat{\sigma}_{\rm O}$ to generate the likelihoods. This model captures both the measurement scatter and the variations due to intrinsically variable stars. The two classifications of \textsc{Other} and \textsc{CII} represent separate populations of stars with different proportions and types of variable stars. Hence, the model parameters related to intrinsic variability took separate values for the classes, determined from the training data.

To model the measurement scatter, we consider a theoretical star with no intrinsic variability but an instrumental uncertainty $\sigma_{\rm I}$ (noise). That instrumental uncertainly includes both photon noise and instrumental systematics. For a series of observations of this star we define the root mean square \edit{deviation} of the observations about the mean flux as $\sigma_{\rm O}$. Since the series of observations is a finite sample, $\sigma_{\rm O}$ will only approximate $\sigma_{\rm I}$ (they would become equal for an infinite set of observations). In Appendix~\ref{sec:variability_model} we demonstrate that for multiple sets of observations of this star (or sets of observations of other identical stars) the observed fractional variance ($\hat{\sigma}_{\rm O}^2$) follows a $\chi^2$ distribution. The $\chi^2$ model of $\hat{\sigma}_{\rm O}^2$ has two parameters, the number of observations $N_{\rm{obs}}$ (equal to the number of degrees of freedom) and the fractional uncertainty for a single measurement $\hat{\sigma}_{\rm I}$, we expect $\hat{\sigma}_{\rm I}$ to be a function of $G$ alone.

For our model we transform the fractional variance into the fractional standard deviation space of our variability feature, via the method in Appendix~\ref{sec:variability_model}. In Fig.~\ref{fig:Gvar_FSDF_165G166} we take a slice through Fig.~\ref{fig:Gvar_G_vs_FSDF} at $16.5 \leq G < 16.6$ to examine the ridgeline. We also plot a $\chi^2$ distribution with parameters derived from the \textsc{Other} training data, transformed into the fractional standard deviation space. This gives an excellent fit to the data around the ridgeline at low variability but it is too narrow, not capturing the long tail to higher variability. This is because the $\chi^2$ distribution is purely modelling the instrumental noise, and this long tail is due to variable stars in the data.

\begin{figure}
	\includegraphics[width=\columnwidth]{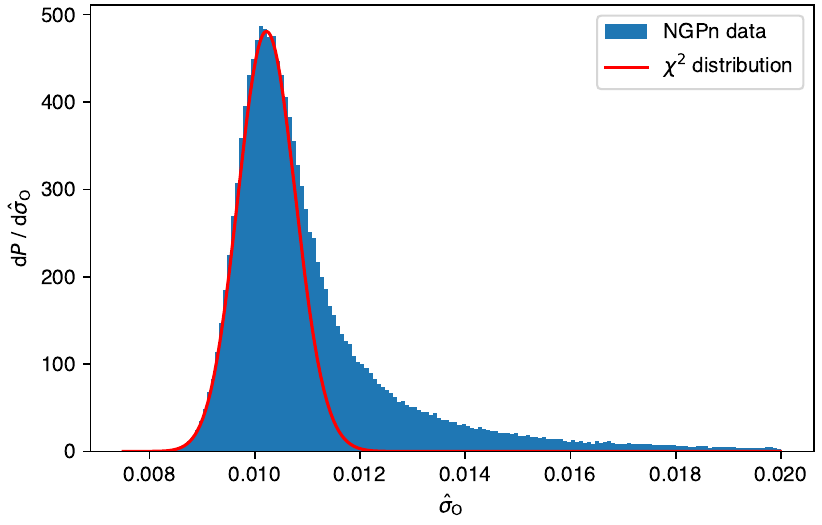}
	\caption{The probability density function of $\hat{\sigma}_{\rm O}$ (fractional standard deviation of the flux) to investigate the behaviour for a narrow slice in $G$-band magnitude. Sources in the NGPn data set in the magnitude range $16.5 \leq G < 16.6$ are in blue. Overlaid in red is a $\chi^2$ distribution that has been fitted to the \textsc{Other} training data in fractional variance of the flux, converted to the fractional standard deviation by the method explained in Appendix~\ref{sec:variability_model}.}
	\label{fig:Gvar_FSDF_165G166}
\end{figure}

\begin{figure}
	\includegraphics[width=\columnwidth]{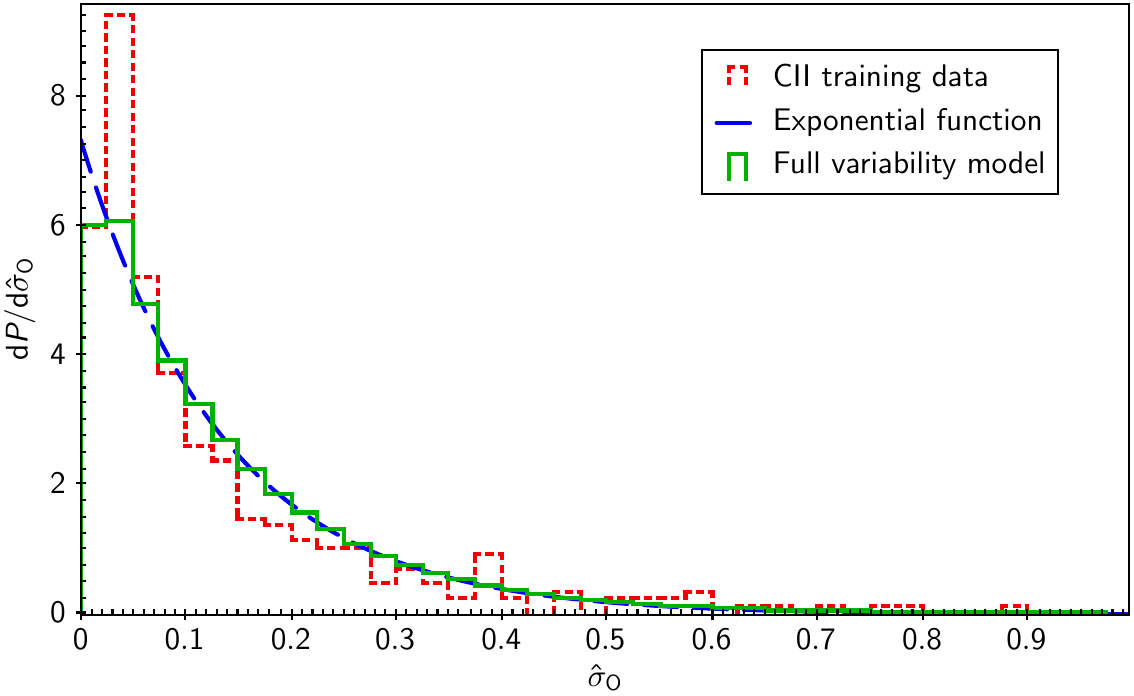}
	\caption{The probability density function of $\hat{\sigma}_{\rm O}$ (fractional standard deviation of the flux) to investigate the Class II YSO distribution. The entire \textsc{CII} training data are the \edit{red} dotted line, with no restriction on $G$-band magnitude or $N_{\rm obs}$. The \edit{blue} dashed line is the exponential function taken from the fit to the \textsc{CII} training data giving $C_{\rm scale}=0.137$. The green solid line is the full variability model with parameters from $17.0 \leq G < 17.1$ with $N_{\rm obs}=324$.}
	\label{fig:Gvar_FSDF_CII_exp_model}
\end{figure}

We conclude the $\chi^2$ distribution is a reasonable model for the measurement scatter in fractional variance space, but a component for variable stars needs to be included in the model. As the majority of YSOs show some form of variability, compared with only a fraction of the general stellar population, we chose an astrophysical variability model based on the behaviour of the \textsc{CII} training data. \cite{ContrerasPena2019} demonstrated the variability amplitude of Class II YSOs follows a decreasing exponential trend with fewer stars exhibiting large amplitude variations. We found the distribution of $\hat{\sigma}_{\rm O}$ for the \textsc{CII} training set was well modelled by such an exponential function (Equation~\ref{eq:var_mag_exp}) \edit{as can be seen by the blue} dashed line of Fig.~\ref{fig:Gvar_FSDF_CII_exp_model}. However, there is an additional component that shifts intrinsically non-variable sources to small non-zero $\hat{\sigma}_{\rm O}$. This is the instrumental noise modelled by the $\chi^2$ distribution. Hence, the probability density function (PDF) for the variable star distribution is calculated by convolving the $\chi^2$ function with the exponential function. The green line of Fig.~\ref{fig:Gvar_FSDF_CII_exp_model} shows the combined \textsc{CII} variability model with components for both the instrumental noise and intrinsic stellar variability. The model parameters had to be chosen for specific values of $G$ and $N_{\rm obs}$, so we used the representative values $17.0 \leq G < 17.1$ and $N_{\rm obs}=324$. Due to the lack of \textsc{CII} training data, no restrictions were placed on the magnitude or number of observations for the \edit{red dotted} line of the data. Hence the green model line is not expected to be a perfect representation of the \edit{dotted red} data line, though it can be seen it mimics the key features of the distribution.

The full model is a combination of the variable star PDF convolved with the $\chi^2$ function, summed with the pure $\chi^2$ function for the non-variable stars. The relative fraction of non-variable stars forms another model parameter, to normalise the summed variable and non-variable star PDFs. A full mathematical derivation of this model is given in Appendix~\ref{sec:variability_model}. 

Taking the model behaviour and data scatter into consideration we decided to limit the applicable range of the model to $13.2 \leq G < 18.0$, retaining 97 per cent of the NGPn data. We set the lower limit to 13.2 rather than 13.0 since data at precisely $G=13.0$ are likely to be a blend of 1D and 2D window classes, and there is a bump in $\hat{\sigma}_{\rm O}$ around this transitional magnitude.

\subsubsection*{Variability model parameter selection}
\label{sec:variability_parameters}

The $\hat{\sigma}_{\rm O}$ model was split into sub-models for the \textsc{CII} and \textsc{Other} classifications. The parameters for these models were the number of $G$-band observations ($N_{\rm{obs}}$), the instrumental noise ($\hat{\sigma}_{\rm I}$), the exponential scale factor ($C_{\rm{scale}}$), and the fraction of non-variable stars ($h$). These parameters were determined by fitting the probability density functions of $\hat{\sigma}_{\rm O}$ using a 1D $\tau^2$ maximum likelihood statistic assuming zero uncertainties \citep{Naylor2006}.

\subsubsection*{Variability model \textsc{Other}  parameter selection}
\label{sec:variability_model_other_parameter_selection}

To calibrate our \textsc{Other} model parameters we took a similar approach to \cite{Vioque2018}, who calibrated their \textit{Gaia} variability indicator to all \textit{Gaia} sources within $\pm 0.1$ magnitudes, whilst we fit our \textsc{Other} parameters in bins of 0.1 magnitude in $G$. We started by examining the fit of our model parameters to the \textsc{Other} training set in both $G$ and $N_{\rm{\rm{obs}}}$. We split the data into three subsets $300 < N_{\rm obs} \leq 400$, $400 < N_{\rm obs} \leq 500$ and $500 < N_{\rm{obs}} \leq 600$, and magnitudes bins of 0.1. To begin with the fit was performed with $\hat{\sigma}_{\rm I}$, $N_{\rm{eff}}$ \edit{(effective number of observations, it will be explained $N_{\rm{eff}} \neq N_{\rm{\rm{obs}}}$)}, $C_{\rm{scale}}$ and $h$ as free parameters.

\begin{figure}
	\includegraphics[width=\columnwidth]{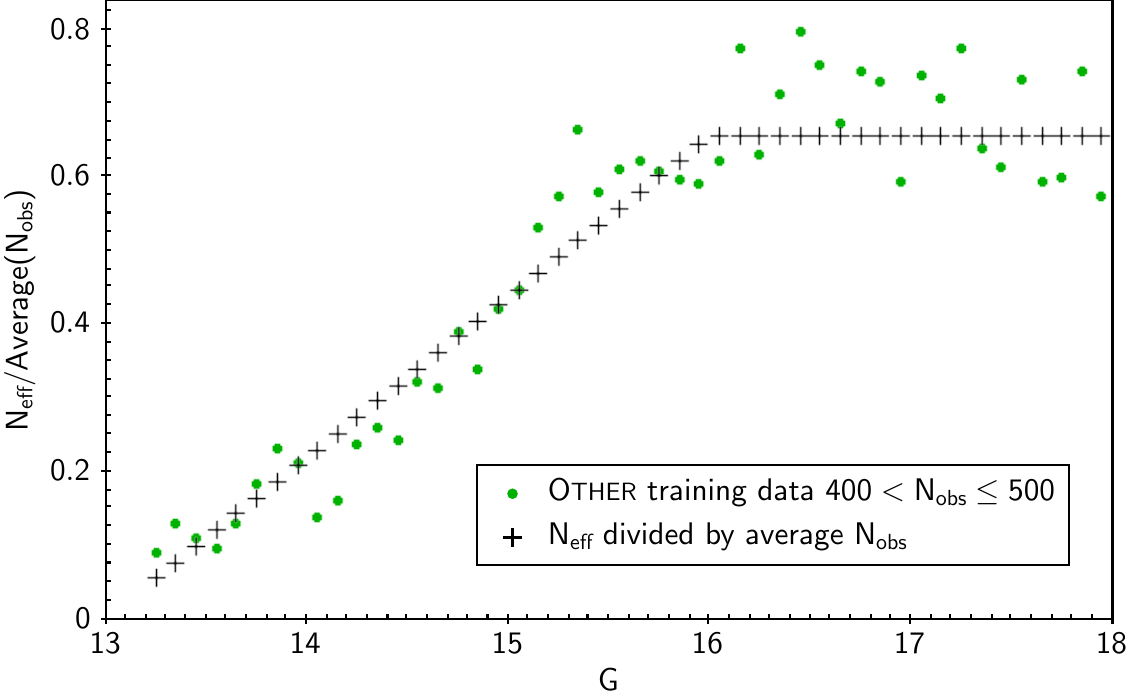}
	\caption{\edit{Illustration of the relationship and scatter between the effective number of observations and the number of observations from the \textit{Gaia} archive, and how this changes with with $G$-band magnitude. The data are the subset of the \textsc{Other} training data with $400 < N_{\rm{obs}} \leq 500$ in bins of 0.1 magnitude. The green dots are the value of $N_{\rm{eff}}$ divided by the average $N_{\rm{obs}}$ in the bin. The black pluses are the value of $N_{\rm{eff}}$ from Equation~(\ref{eq:Neff}) divided by the average $N_{\rm{obs}}$ for the bin.}}
	\label{fig:Gvar_NeffNobs}
\end{figure}

It was immediately apparent the \textit{Gaia} $N_{\rm{obs}}$ did not provide a good fit to the model. The ratio of $N_{\rm{obs}}$ from the \textit{Gaia} archive to $N_{\rm{eff}}$ fitted to the model is plotted in Fig.~\ref{fig:Gvar_NeffNobs} \edit{for the middle subset $400 < N_{\rm{obs}} \leq 500$, with the other two subsets exhibiting similar behaviour}. The ratio has a large scatter but is always significantly less than one, indicating the fitted $N_{\rm{eff}}$ is less than the quoted $N_{\rm{obs}}$. There is also a clear dependence on $G$ to brighter magnitudes. This means the average of $N_{\rm{obs}}$ is not following the normal $\sqrt{n}$ behaviour. The average is being affected by a systematic in the \textit{Gaia} processing pipeline that we are failing to capture. Regardless of the exact cause, it was clear that using the \textit{Gaia} archive $N_{\rm{obs}}$ as an input to our model could lead to poor quality results. Hence, we decided to create a function for $N_{\rm{eff}}$ as the input to the model. At magnitudes brighter than $G=16.0$ the function would be dependent on both $N_{\rm{obs}}$ and $G$, while at fainter magnitudes there is no clear dependence on $G$ so we decided this region would be dependent on $N_{\rm{obs}}$ alone. The change in behaviour at $G=16.0$ may be connected to the change in \textit{Gaia} Window size at this magnitude \citep{Evans2017,Evans2018}.

We decided to adopt a simple function for $N_{\rm{eff}}$, as the scatter in the fits \edit{makes} it difficult to justify a complex function. We started with the function for $16.0 < G < 18.0$ as this would be dependent solely on $N_{\rm{obs}}$. We chose a power law as this gave a consistent fit across the three $N_{\rm{obs}}$ subsets. The power was determined by dividing the logarithm of fitted $N_{\rm{eff}}$ by the logarithm of the average \textit{Gaia} $N_{\rm obs}$ by magnitude bin, and finally averaged across the magnitude range. The values of the power across the three ranges were 0.94 for $300 < N_{\rm{obs}} \leq 400$, 0.94 for $400 < N_{\rm{obs}} \leq 500$, and 0.92 for $500 < N_{\rm{obs}} \leq 600$. While there may be a slight hint the power decreases with increasing $N_{\rm{obs}}$, this is small compared to the scatter in the data and risks over fitting a weak relationship. For the model we used a single valued parameter calculated from the subset $350 < N_{\rm{obs}} \leq 450$ containing 65 per cent of sources in the \textsc{Other} training set, giving a value of 0.93 for the power.

As the brighter observations showed an increasing relationship with $G$ containing a lot of scatter, we took the product of our fit to the fainter region with a linearly increasing function of $G$ for this brighter region. The trend appears to be heading for zero $N_{\rm{eff}}$ at about $G=13.0$, noting the applicable range of our model is limited to $13.2 \leq G < 18.0$. Thus we adopted the straight line from $N_{\rm{eff}}=0.0$ at $G=13.0$ to $N_{\rm{eff}}=N_{\rm{obs}}^{0.93}$ at $G=16.0$. This gave the pair of equations
\begin{equation}
	N_{\rm{eff}} = 
	\begin{cases}
		\frac{1}{3} \left( G - 13 \right)  N_{\rm{obs}}^{0.93} & \quad G < 16.0 , \\
		N_{\rm{obs}}^{0.93} & \quad G \geq 16.0 .
	\end{cases}
	\label{eq:Neff}
\end{equation}
The fit of this function to the \edit{subset $400 < N_{\rm{obs}} \leq 500$} can be seen in Fig.~\ref{fig:Gvar_NeffNobs}.

Having set $N_{\rm{eff}}$ to the output from Equation~\ref{eq:Neff}, we now fitted the PDFs with the 3 remaining free parameters. Our next step was to estimate the instrumental noise $\hat{\sigma}_{\rm I}$ by $G$-band magnitude. The \textsc{Other} training set was suited to this task as it was deliberately located in a bland region, where we would expect non-variable stars to dominate.

\begin{figure}
	\includegraphics[width=\columnwidth]{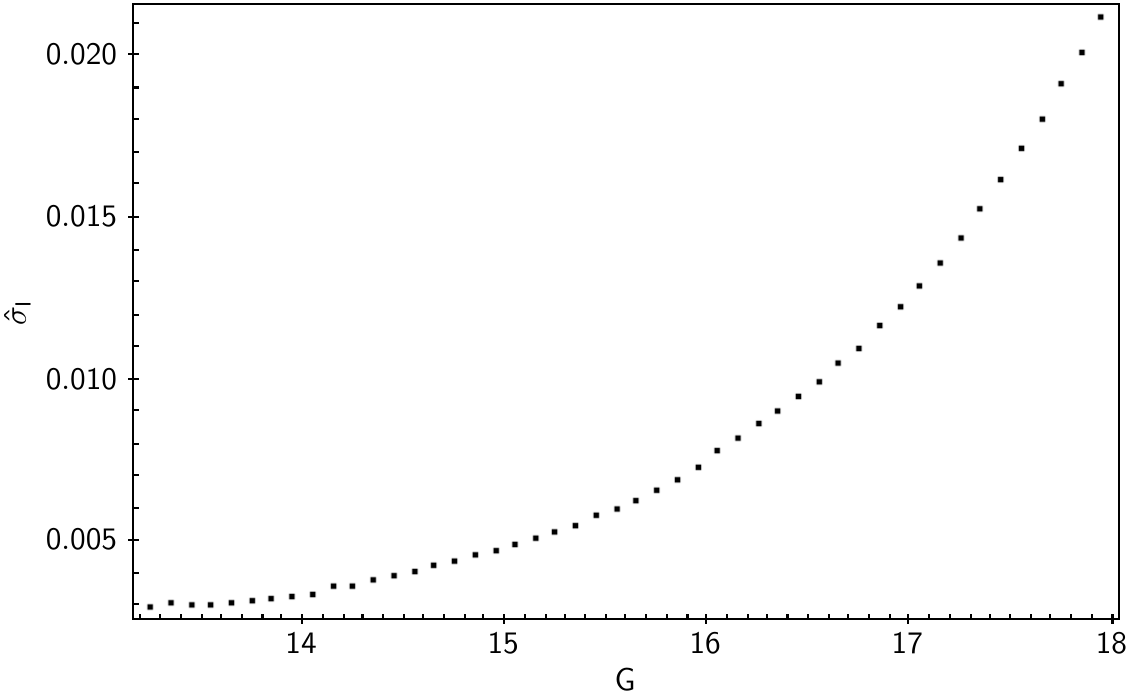}
	\caption{The values for the instrumental noise ($\hat{\sigma}_{\rm I}$) by \textit{Gaia} $G$-band magnitude found by fitting $\hat{\sigma}_{\rm O}$ from the \textsc{Other} training data restricted to $350 < N_{\rm obs} \leq 450$ with a 1D $\tau^2$ maximum likelihood statistic assuming zero uncertainties.}
	\label{fig:Gvar_G_vs_InstNoise}
\end{figure}

To fit the instrumental noise parameter we again used the \textsc{Other} subset $350 < N_{\rm obs} \leq 450$, containing two thirds of the training data. This minimized any small effects from varying $N_{\rm{obs}}$ in the model fitting. The resulting instrumental noise can be seen in Fig.~\ref{fig:Gvar_G_vs_InstNoise}. There is a little scatter at brighter magnitudes where there is more spread in the data, and a slight change at around $G=16.0$ corresponding to the change in the \textit{Gaia} Window Class \citep{Evans2017,Evans2018}. We used these values to assign $\hat{\sigma}_{\rm I}$ at a resolution of 0.1 magnitudes.

To investigate the exponential scale factor and fraction of non-variable star parameters, we performed a fresh 1D $\tau^2$ fit using the \textsc{Other} subset $350 < N_{\rm obs} \leq 450$ in bins of 0.1 magnitudes. We set $N_{\rm eff}$ to the derived function of $N_{\rm obs}$, $\hat{\sigma}_{\rm I}$ as the set of values in Fig.~\ref{fig:Gvar_G_vs_InstNoise}, and left the exponential scale factor and the fraction of non-variable stars as free parameters. 

\begin{figure}
	\includegraphics[width=\columnwidth]{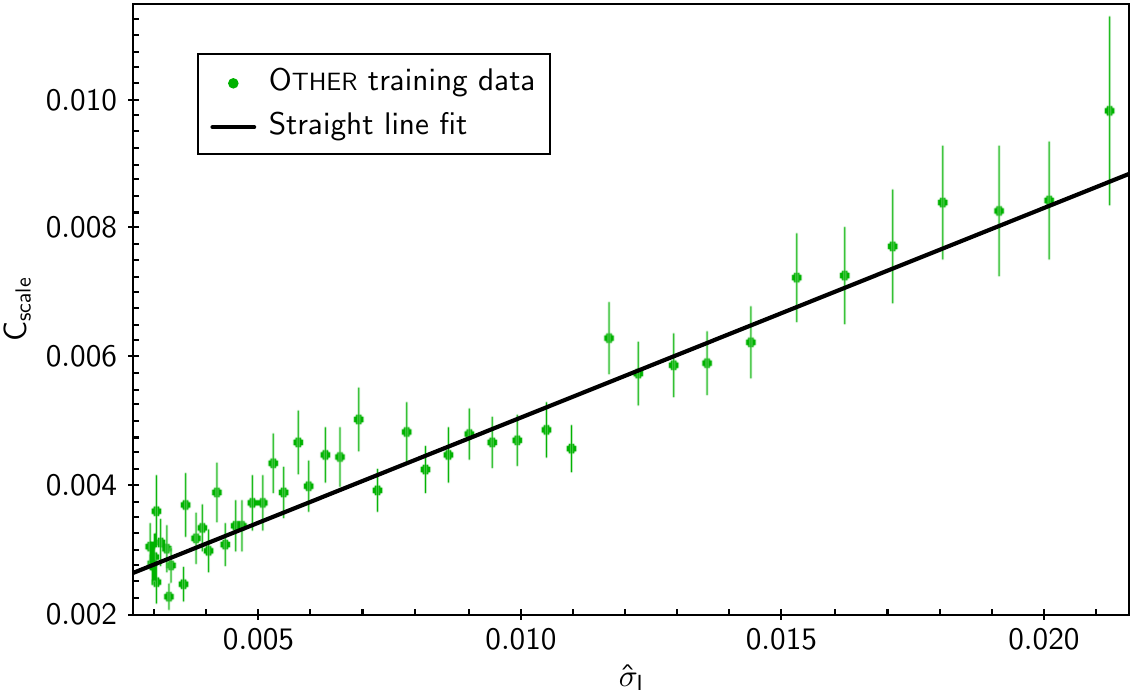}
	\caption{The relation between exponential scale factor $C_{\rm{scale}}$ and the instrumental noise $\hat{\sigma}_{\rm I}$ for the \textsc{Other} model. The \textsc{Other} training data binned by 0.1 magnitudes are green dots with uncertainty in $C_{\rm{scale}}$ from the 1D $\tau^2$ maximum likelihood statistic. The least-squares fit straight line of equation~(\ref{eq:Gvar_Other_Exp}) is in black.}
	\label{fig:Gvar_Other_Exp}
\end{figure}

The exponential scale factor showed an increasing non-linear relationship with $G$, or alternatively an increasing linear relationship with the instrumental noise. As the relationship to the instrumental noise was simplest, we decided to model this linear relationship for the exponential scale factor parameter \edit{(see Fig.~\ref{fig:Gvar_Other_Exp})}. We carried out an unweighted non-linear least-squares fit to derive the straight line
\begin{equation}
	C_{\rm{scale}} = 0.3268 \hat{\sigma}_{\rm I} + 0.0018 .
	\label{eq:Gvar_Other_Exp}
\end{equation}
As it only takes one spurious photometric measurement to increase the variability measure, the tail of the exponential distribution may be contaminated by sources with little or no intrinsic variability but a single bad data point in their light curve. While this is an inherent weakness in our approach, the quality of the \textit{Gaia} data means we should only expect a tiny number of spurious measurements.

\begin{figure}
	\includegraphics[width=\columnwidth]{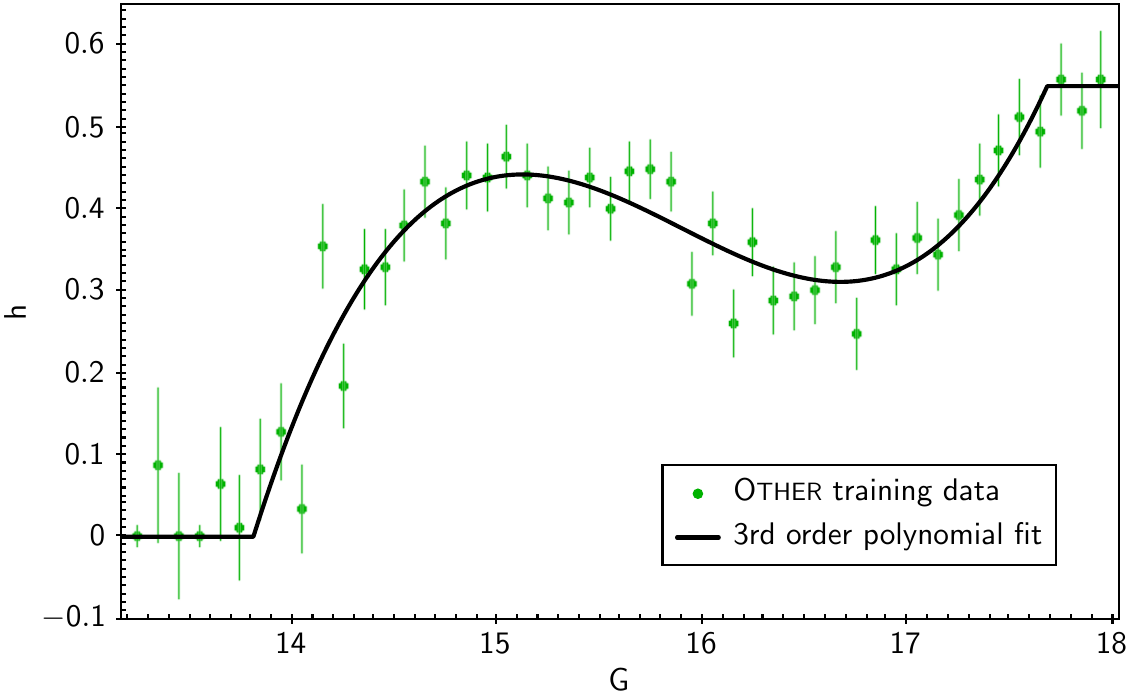}
	\caption{The relation between the fraction of non-variable stars and $G$-band magnitude for the \textsc{Other} model. The \textsc{Other} training data binned in 0.1 magnitudes are green dots, with the uncertainty in the fraction of non-variable stars from the 1D $\tau^2$ maximum likelihood statistic. The black line is the least-squares fit third order polynomial with bounds of equation~(\ref{eq:Gvar_Other_Frac}).}
	\label{fig:Gvar_Other_Frac}
\end{figure}

The final parameter for the \textsc{Other} model was the fraction of non-variable stars, $h$. We again performed a 1D $\tau^2$ fit to the \textsc{Other} subset $350 < N_{\rm obs} \leq 450$ in bins of 0.1 magnitudes. This time the only free parameter was the fraction of non-variable stars, with the other parameters set as per the values and functions we found. The fit exhibited a complex behaviour with $G$ \edit{(see Fig.~\ref{fig:Gvar_Other_Frac})}. At the brightest magnitudes it appears all stars are variable to some extent. This is a reasonable result, since all stars exhibit low level variability and the instrumental noise will be very small at these bright magnitudes. That the relationship is not a monotonic function is a little concerning. It may be due to the model imperfectly matching the systematics in the instrumental noise, though we do not have a good physical explanation. We decided to fit a polynomial limited by upper and lower bounds to avoid it growing beyond the data values at the extremes. An unweighted non-linear least-squares fit to a third order polynomial was performed, giving the function with bounds
\begin{equation}
	h = 
	\begin{cases}
		0.00 & \quad G \leq 13.81 , \\
		\begin{split}
			& -280.055 + 53.1768 G \\
			&\qquad - 3.35316 G^2 + 0.0703122 G^3 
		\end{split} & \quad 13.81 < G < 17.68 \\
		0.55 & \quad G \geq 17.68 .
	\end{cases}
	\label{eq:Gvar_Other_Frac}
\end{equation}

\subsubsection*{Variability model \textsc{CII}  parameter selection}

As the number of observations and instrumental noise parameters should be independent of the type of source, we used the same relations found for the \textsc{Other} model for \textsc{CII}. We would expect the fraction of non-variable stars and the exponential scale factor to take on different values, so we performed a dedicated fit using the \textsc{CII} training data.

As with the \textsc{Other} parameter fitting, we restricted the data to the range $350 < N_{\rm obs} \leq 450$, to limit the changes due to different numbers of observations. We performed a series of 1D $\tau^2$ fits by bins of whole magnitudes except for the brightest magnitude bin $13.2 \leq G < 14.0$. The instrumental noise and effective number of observations were set to the values found for the \textsc{Other} training data at the mean values for the magnitude bins. This consistently gave a tiny number for the fraction of non-variable stars, and we did not find any meaningful relationship for the exponential scale factor within the uncertainties of the fits, with values of $C_{\rm{scale}}$ ranging from 0.10 to 0.20. So we performed a 1D $\tau^2$ fit across the entire magnitude range. This gave the single value of 0.137 for the exponential scale factor that we took for the \textsc{CII} parameter. We adopted a value of zero for the fraction of non-variable stars, consistent with the fit as well our expectation that the majority of young stars will show some form of variability.

\subsubsection*{Variability model summary}

\begin{figure}
	\includegraphics[width=\columnwidth]{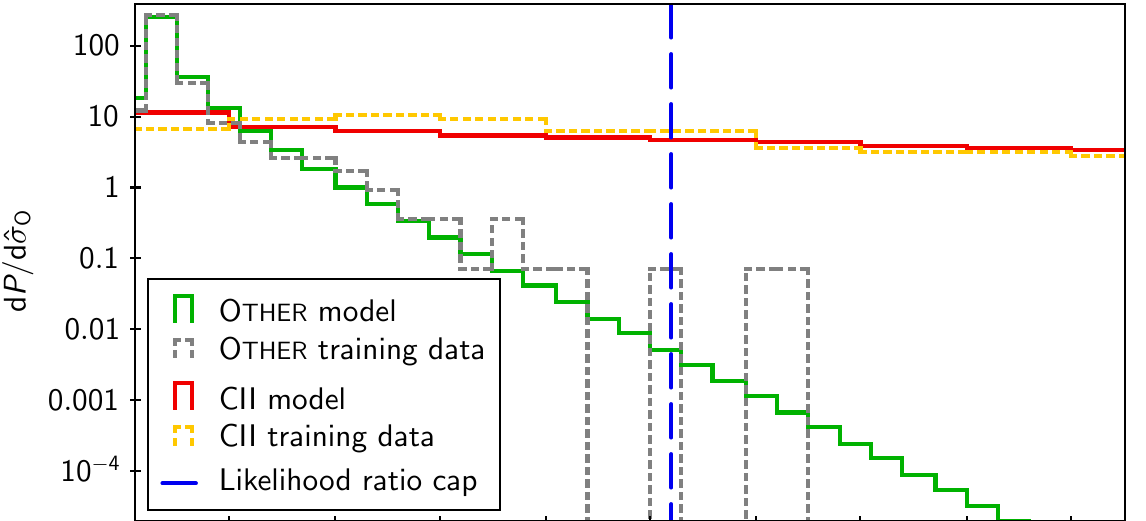}
	\includegraphics[width=\columnwidth]{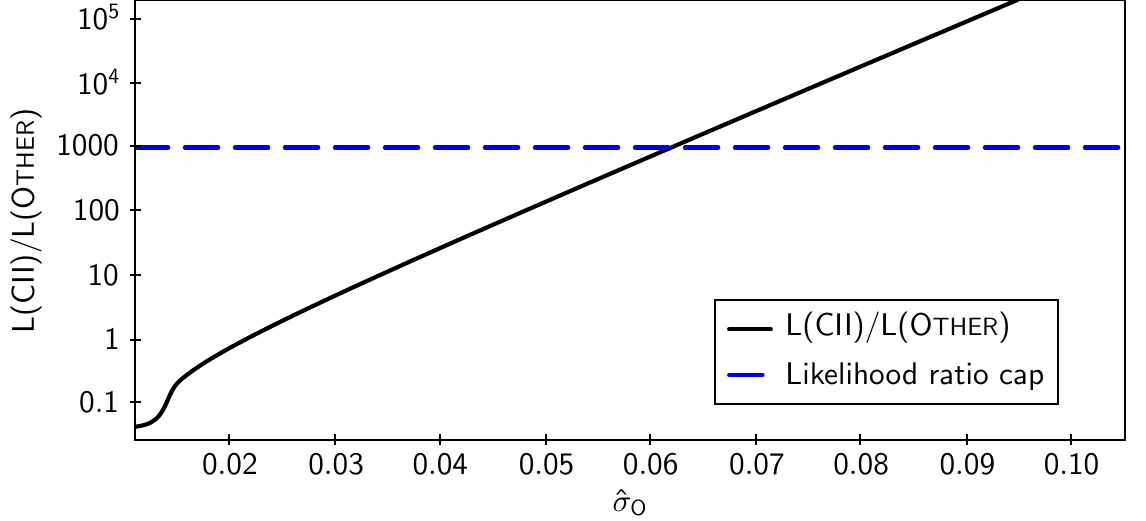}
	\caption{\edit{The variability models for $17.0 \leq G < 17.1$ with $N_{\rm obs}=300$ ($N_{\rm eff}=201$). In the top plot the \textsc{Other} model is the green line, and the \textsc{CII} model is the solid red line. The dotted grey line is the \textsc{Other} training data for $17.0 \leq G <17.1$, and the dotted orange line is the \textsc{CII} training data for $16.0 \leq G < 18.0$, neither with a restriction on $N_{\rm obs}$.  The vertical blue dashed line indicates the $\hat{\sigma}_{\rm O}$ where the ratio of the likelihoods is capped to the ratio of the priors. In the bottom plot the likelihood ratio $L(\textsc{CII}) / L(\textsc{Other})$ is the black line. The horizontal dashed blue line marks the cap on the ratio.}}
	\label{fig:Gvar_Model_Summary}
\end{figure}

Thus we have a semi-analytical model for our $G$-band variability feature. For a given $G$ and $N_{\rm obs}$ it provides the probability of observing a given $\hat{\sigma}_{\rm O}$ (and hence uncertainty in $G$). A comparison of the \textsc{Other} and \textsc{CII} models is shown in Fig.~\ref{fig:Gvar_Model_Summary}. \edit{To retain a sufficient quantity of data for a meaningful histogram there was no restriction placed on $N_{\rm obs}$. It can be seen that the model follows the \textsc{Other} training data reasonably well. Above $\hat{\sigma}_{\rm O}\approx0.5$ there is very little \textsc{Other} data, resulting in gaps in the histogram and jumps to high values caused by single data points. To have sufficient sources for the histogram, it was necessary to accept \textsc{CII} in the range $16.0 \leq G <18.0$ with no restriction on $N_{\rm obs}$, retaining 249 sources. The peak of the \textsc{CII} training data is to slightly higher $\hat{\sigma}_{\rm O}$ than the model. The accurate location of this peak is important for capturing the behaviour of the \textsc{Other} training data but is less critical for the broad low peak of the \textsc{CII}. The important behaviour of the \textsc{CII} is the slow decline of sources with increasing $\hat{\sigma}_{\rm O}$ compared to the \textsc{Other} sources, and this is accurately captured by the model. The ratio of the likelihoods is shown in the bottom panel of Fig.~\ref{fig:Gvar_Model_Summary}. We place a limit on the likelihood ratio, to cap it before the model diverges significantly from the data.}

\section{Results}
\label{sec:results}

The full classifier catalogue contains 8\,080\,045 sources. This includes \editthree{29\,166} sources without any valid feature likelihoods, that simply return the prior for the posterior. We provide a data dictionary for the published catalogue in the \edit{supplementary material}.

\subsection{Classifier performance}

We split our samples of known \textsc{CII} and \textsc{Other} into training and test sets in the ratio $80:20$. The test set was used for an independent check on the classifier performance.

The \textsc{CII} test set contains just 91 sources, compared to 37\,911 in the \textsc{Other} test set. Hence, the size of the \textsc{CII} set limits the precision that can be achieved by the performance statistics. Also, even a small contamination of the \textsc{CII} set would have a significant impact on the statistics, noting undetected contaminants would make the classifier look worse not better.

\subsubsection{Recovery of the training data}

\begin{figure}
	\includegraphics[width=\columnwidth]{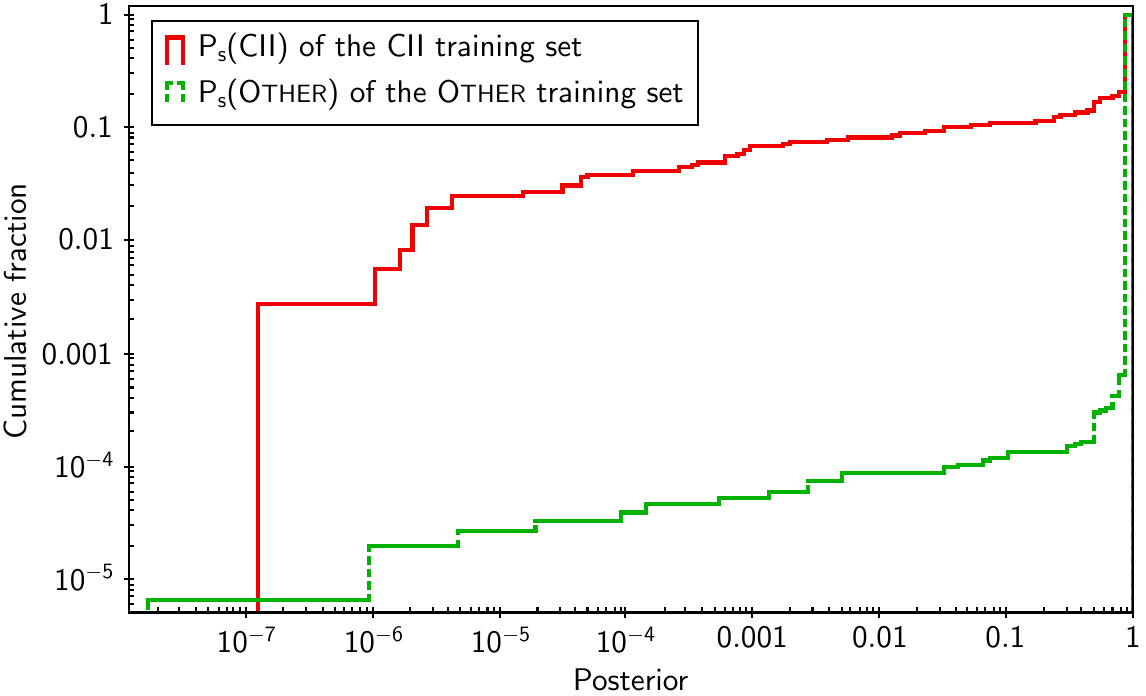}
	\caption{A cumulative histogram of the training set posteriors. The red line is the $P_{\rm s}(\textsc{CII})$ of the \textsc{CII} training set  and the green dotted line is the $P_{\rm s}(\textsc{Other})$ of the \textsc{Other} training set. A \edit{cumulative fraction} is used as the two data sets are of very different size, with log scale for both axes to bring out the detail at low posteriors.}
	\label{fig:Post_HistCum_Train}
\end{figure}

A useful initial step is to review the recovery of the training data by examining the posteriors \edit{with a cumulative histogram (see Fig.~\ref{fig:Post_HistCum_Train})}. While this is not an independent test, it is a useful check on the operation of the classifier and the apparent purity of the training data. The important result is the majority \edit{of} the training data are correctly classified with high posteriors favouring their training class. There are a small number of sources for each training set with a low posterior for their class. This is not surprising and will be due to a mixture of misidentification by the classifier and contamination of the training data. \edit{The \textsc{Other} training set has an extremely low level of misidentification at just over one in six thousand by $P_{\rm s}(\textsc{Other}) \leq 0.5$. In contrast, the \textsc{CII} data has a higher level of misidentification at around one in seven by $P_{\rm s}(\textsc{CII}) \leq 0.5$.} This difference in the sets is partially explained as we are expecting a population of around one \textsc{CII} in a thousand sources. Hence, even by random chance we would expect the \textsc{Other} training set to have an extremely low level of contamination.

\subsubsection{Confusion matrices}
\label{sec:confusion}

A confusion matrix is a tool for analysing the performance of a classifier via a 2-dimensional matrix. The \lq true\rq\ classifications take up one axis and the predicted classifications the other. This makes it simple to assess the success of the classifier at placing objects into their correct classifications. For a perfect classifier, all sources would be placed in elements along the leading diagonal, with no sources off the diagonal. The drawback of a confusion matrix for our naive Bayes classifier is they require binary true/false classifications rather than posteriors with a continuum of values between zero and one. To get around this we specify a threshold \textsc{CII} posterior as the binary cut-off for classifying candidate \textsc{CII} as true/positive and \textsc{Other} as false/negative. As this will only capture the classifier performance at a single threshold posterior, we calculate matrices at two thresholds. This gives a broader understanding of the classifier performance and illustrates two different use cases. For a case where a large number of candidate Class II YSOs are required and a higher level of contamination is acceptable, we select a threshold giving about fifty thousand candidate \textsc{CII}, $P_{\rm s}(\textsc{CII})>0.0109$. This predicts 50\,082 sources as \textsc{CII} and we shall refer to this as our 50k set. To illustrate a set with an expected low contamination rate we chose a threshold of $P_{\rm s}(\textsc{CII})>0.5$, predicting 6\,504 sources as \textsc{CII}. This set has a further quality advantage as it is only possible to have a $P_{\rm s}(\textsc{CII})>0.5$ where two or more features contribute to the posterior and at least two favour a \textsc{CII} classification. These confusion matrices are displayed in Table~\ref{tab:confusion_matrix_table_count}.

\begin{table}
	\caption{Confusion matrix for the classifier with two illustrative threshold posteriors, \textsc{CII} is considered positive and \textsc{Other} as negative. The top table has a posterior Class II YSO threshold of 0.0109 corresponding to 50\,082 objects, and the middle table has a threshold posterior of 0.5. The bottom table gives the confusion matrix element labels.}
	\label{tab:confusion_matrix_table_count}
	\begin{tabular}{ c c c c }
		$P_{\rm s}(\textsc{CII})>0.0109$ & Predicted \textsc{CII} & Predicted \textsc{Other} & \edit{Actual} Total \\
		Actual \textsc{CII} & 89 & 2 & 91 \\
		Actual \textsc{Other} & 78 & 37\,833 & 37\,911 \\
		\edit{Predicted} Total & 167 & 37\,835 & 38\,002 \\
		\hline
		$P_{\rm s}(\textsc{CII})>0.5$ & Predicted \textsc{CII} & Predicted \textsc{Other} & \edit{Actual} Total \\
		Actual \textsc{CII} & 79 & 12 & 91 \\
		Actual \textsc{Other} & 7 & 37\,904 & 37\,911 \\
		\edit{Predicted} Total & 86 & 37\,916 & 38\,002 \\
		\hline
		& Predicted \textsc{CII} & Predicted \textsc{Other} & \edit{Actual Total} \\
		Actual \textsc{CII} & TP & FN & P \\
		Actual \textsc{Other} & FP & TN & N \\
		\edit{Predicted Total} & \edit{PP} & \edit{PN} & \edit{T} \\
	\end{tabular}
\end{table}

\begin{table}
	\caption{Confusion matrix results as proportions for the classifier. \edit{The notes and rules from Table~\ref{tab:confusion_matrix_table_count} apply.} The bottom table gives the confusion matrix element labels and formulae. The small size of the \textsc{CII} test set means small number statistics are a problem, limiting the precision of the TPR and FNR. Brackets give the range in values from ten random samples of half the sources from the test data.}
	\label{tab:confusion_matrix_table_proportion}
	\begin{tabular}{ c c c }
		$P_{\rm s}(\textsc{CII})>0.0109$ & Predicted \textsc{CII} & Predicted \textsc{Other} \\
		Actual \textsc{CII} & 0.98 (0.96--1.00) & 0.02 (0.00--0.04) \\
		Actual \textsc{Other} & 0.0021 (0.0017--0.0026) & 0.9979 (0.9974--0.9983) \\
		\hline
		$P_{\rm s}(\textsc{CII})>0.5$ & Predicted \textsc{CII} & Predicted \textsc{Other} \\
		Actual \textsc{CII} & 0.87 (0.82--0.93) & 0.13 (0.07--0.18) \\
		Actual \textsc{Other} & 0.0002 (0.0001--0.0003) & 0.9998 (0.9997--0.9999) \\
		\hline
		& Predicted \textsc{CII} & Predicted \textsc{Other} \\
		Actual \textsc{CII} & $\rm{TPR=TP/P}$ & $\rm{FNR=FN/P}$ \\
		Actual \textsc{Other} & $\rm{FPR=FP/N}$ & $\rm{TNR=TN/N}$ \\
	\end{tabular}
\end{table}

The binary positive/negative results used to create the confusion matrices can be used to calculate a variety of standard statistics. In the confusion matrices of Table~\ref{tab:confusion_matrix_table_count} we have the true positives (TP), the false positives (FP), the true negatives (TN), and the false negatives (FN). Combining these allows us to generate proportional statistics, some of which are included in the proportional confusion matrices of Table~\ref{tab:confusion_matrix_table_proportion}.

The precision of these statistics is limited by the finite size of the test sets. The small size of the \text{CII} test set means small number statistics limit the precision. By taking ten random samples of half the sources from the test sets, we were able to analyse how Poisson statistics affect the confusion matrix and other statistics. These subsamples gave us a range of values we quote in brackets next to the statistics.

The true positive rate (TPR) or recall rate, tells us the correctly classified fraction of the test set. Our 50k set has a very high TPR of 0.98 (0.96--1.00) while our more conservative $P_{\rm s}(\textsc{CII})>0.5$ set has a rate of 0.87 (0.82--0.93). So both find the majority of the Class II YSOs from our test set, with the 50k set finding nearly all of them.

The false negative rate (FNR) tells us the fraction of \textsc{CII} the classifier will miss as they are being misclassified as \textsc{Other}. It has a very low value of 0.02 (0.00--0.004) for the 50k set. While the $P_{\rm s}(\textsc{CII})>0.5$ set has a higher but none the less small value of 0.13 (0.07--0.18).

The false positive rate (FPR) tells us the fraction of \textsc{Other} sources that are being misidentified as \textsc{CII} by the classifier. The 50k set has an FPR ten times larger at 0.0021 (0.0017--0.0026) than the $P_{\rm s}(\textsc{CII})>0.5$ set at 0.0002 (0.0001--0.0003). This indicates the 50k set has a much higher contamination rate.

As our test and training sets are non-overlapping selections from master labelled data sets, \edit{these confusion matrices are telling us how well the classifier represents the training data}. The low values for the FNR and FPR, with high values for the TPR and TNR imply our classifier results for the test set are accurately following the training data. The TPR \edit{and FPR} are good at illustrating the key differences between our two example \edit{selections}.

These figures should be taken as indicative, importantly they include systematics from the way we compiled our labelled data. For example, any impurities in our labelled data will affect these statistics. While we can have confidence the majority of our \textsc{CII} are young stars, it is plausible \edit{that} young stars at other stages of their evolution such as Class III YSOs may be contaminating the labelled set. It is also possible our \textsc{Other} set contains a low level contamination of young stars including Class II YSOs. There may be systematics that vary with Galactic location, such as effects on the features from interstellar dust, or our training data may not accurately reflect the populations across the entire survey region. While these are difficult to quantify, our comparison to other classifiers in Section~\ref{sec:comparisons_to_other_catalogues} provides an independent assessment of our classifier.

\subsubsection{ROC curve}
\label{sec:roc}

\begin{figure}
	\includegraphics[width=\columnwidth]{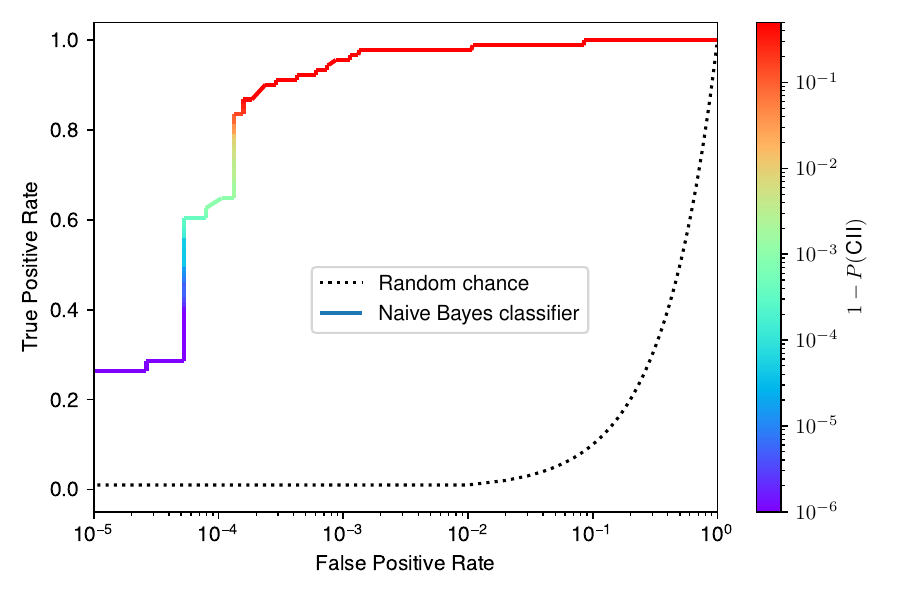}
	\caption{\edit{ROC curve with a solid coloured line for the classifier and a dotted black line for random chance. The false positive rate is plotted on a logarithmic scale. The classifier line is coloured by $1-P_{\rm s}(\textsc{CII})$ in log space. This brings out the change in posterior at the critical region, showing the structure at high $P_{\rm s}(\textsc{CII})$.}}
	\label{fig:ROC}
\end{figure}

To analyse the behaviour of the classifier across the full range of posteriors we use a ROC (Receive Operating Characteristic) curve \edit{(see Fig.~\ref{fig:ROC})}. This shows how the TPR varies with the FPR. The solid coloured line is the result for our classifier and the dotted black line is random chance. Our classifier performs well as it is always far above the random chance line and has a steep rise to a TPR of nearly one for small FPR. A perfect classifier would rise vertically to a TPR of one at an FPR of zero.

\edit{The log axis for the FPR brings out the detail at low FPR.} This is important for our work, since even a low level FPR can overwhelm a population of correctly classified \textsc{CII} with incorrectly classified \textsc{Other}. If our prior is accurate, then we should expect around 1 \textsc{CII} for 999 \textsc{Other}, so the critical region in assessing our classifier is an FPR of around $10^{-3}$. At an FPR below $5 \times 10^{-5}$ the classifier misses about two thirds of the \textsc{CII} in the test set. The success of the classifier at identifying \textsc{CII} increases in a couple of steps until it identifies most \textsc{CII} in the test set by $10^{-3}$. The classifier results are coloured in a log scale of $1-P_{\rm s}(\textsc{CII})$. This brings out the change in posterior $\textsc{CII}$ at high values of the posterior. At $P_{\rm s}(\textsc{CII})=0.9$ the TPR is 0.84, and by $P_{\rm s}(\textsc{CII})=0.999999$ the TPR drops to 0.35.

The area under the ROC curve (AUC) is another measure of the classifier performance. An area above 0.5 indicates better than random chance while 1.0 would be a perfect classifier. The AUC for our classifier is at least 0.99, with an indicative range of 0.998--1.000 calculated by taking ten random half size samples from the test data. The precise value of the AUC is uncertain due to the unknown level of contamination in the labelled sets. The size of the \textsc{CII} set is more important in this regard, as small number statistics are a factor.

\subsection{Classifier catalogue}

The classifier results take the form of posterior probabilities of \textsc{CII} and \textsc{Other}. To create a list of candidate \textsc{CII} a threshold posterior must be chosen to suit the needs of the task at hand. When doing this, it is important to understand the posteriors cannot be taken at face value as genuine probabilities, since they are heavily influenced by the choice of priors \edit{(see Section~\ref{sec:naive_bayes_classifier})}. It is also useful to review how many and which features have been used to classify sources. A source where multiple features have been used in the classification might be considered more robust than a source classified by a single feature. Due to the bounds we placed on the likelihood ratios, sources classified by a single feature are limited to $P_{\rm s}(\textsc{CII}) \leq 0.5$. There are 503\,320 posteriors based on a single feature.

\begin{figure}
	\includegraphics[width=\columnwidth]{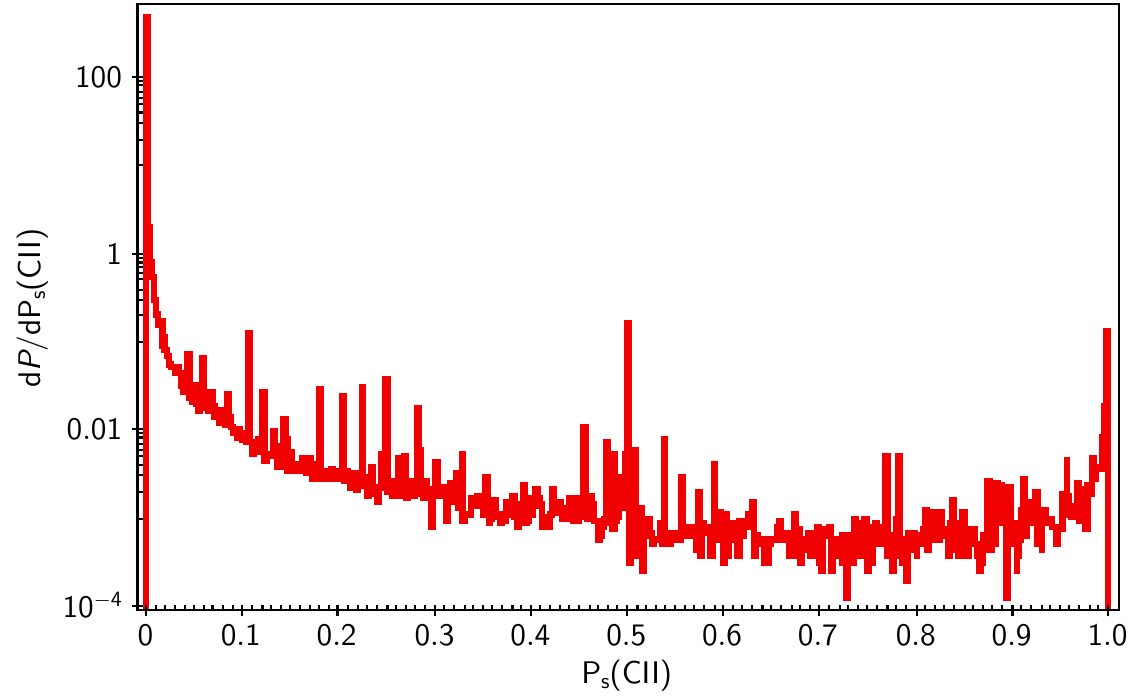}
	\caption{Histogram of the posterior \textsc{CII} for the full NGPn catalogue. A logarithmic scale has been used for \edit{vertical axis to bring out the detail.}}
	\label{fig:PCII_hist}
\end{figure}

A histogram of posterior \textsc{CII} with the normalised count on a logarithmic scale is presented in Fig.~\ref{fig:PCII_hist}, showing a bulk shape similar to a \emph{U}. This \emph{U} shape demonstrates the classifier is working well. Objects are being swept to a classification of either \textsc{Other} or \textsc{CII}, with fewer left with an ambiguous classification in the middle. Due to the larger number of \textsc{Other} sources compared to \textsc{CII}, corresponding to $P_{\rm s}(\textsc{CII}) \approx 0$, the relative count is around a thousand times larger than at $P_{\rm s}(\textsc{CII}) \approx 1$. The spikes are due to the discrete nature of the four feature likelihoods based on the distributions of the training data, and the bounds placed \edit{on} the maximum likelihood ratios. \edit{The largest spike at $P_{\rm s}(\textsc{CII}) = 0.5$ is due to sources classified by a single feature, where the likelihood ratio is capped by the upper bound favouring \textsc{CII}.}

The number of features used to classify sources is summarised \edit{in Table~\ref{tab:FeatureProportions}. It can be seen that 80 per cent of our NGPn catalogue have robust posteriors calculated from a combination of at least three features.} Less than 1 per cent of sources are not classified by any feature. Their results should be ignored since the posterior will simply return the prior.

\begin{table}
	\caption{\edit{The proportion of sources by the number of features used to classify them in our NGPn catalogue.}}
	\label{tab:FeatureProportions}
	\begin{tabular}{lrc}
		\hline
		Features & Sources & Percentage of catalogue \\
		\hline
		0 & 29\,166 & 0\% \\
		1 & 503\,320 & 6\% \\
		2 & 1\,123\,490 & 14\% \\
		3 & 1\,122\,703 & 14\% \\
		4 & 3\,382\,596 & 42\% \\
		5 & 1\,918\,770 & 24\% \\
		\hline
	\end{tabular}
\end{table}

\subsubsection{Identified star forming regions}
\label{sec:identified_star_forming_regions}

\begin{figure*}
	\includegraphics[width=1\textwidth]{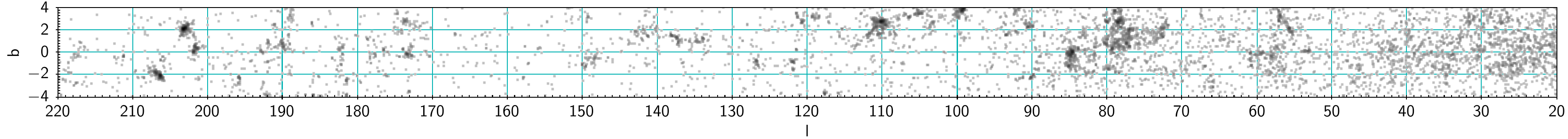}
	\includegraphics[width=1\textwidth]{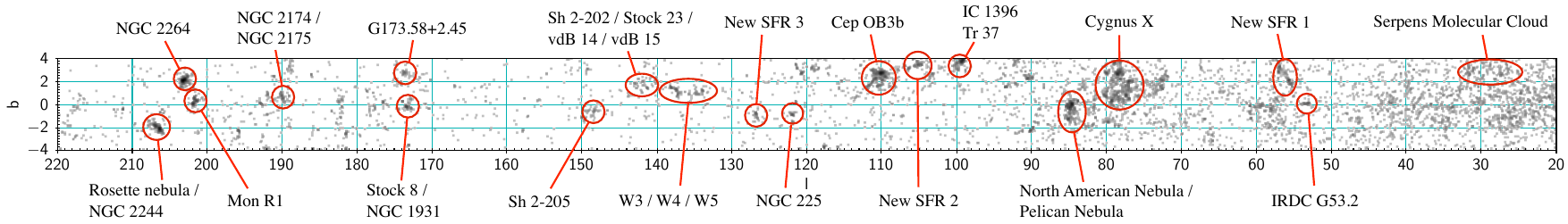}
	\includegraphics[width=1\textwidth]{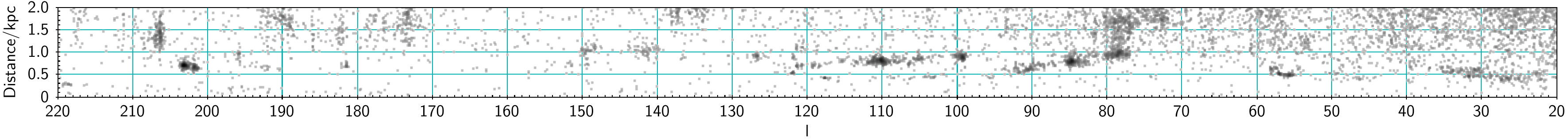}
	\caption{The data set defined $P_{\rm s}(\textsc{CII})>0.5$ containing 6\,504 sources. The top plot is Galactic longitude and latitude. The middle plot is identical except several well known star forming regions and young clusters are identified. The bottom plot is Galactic longitude and distance in kpc from the reciprocal of the \textit{Gaia} EDR3 corrected parallax.}
	\label{fig:bayes_high_PCII_gal_coords}
\end{figure*}

Plots of high posterior \textsc{CII} in Galactic coordinates, and Galactic longitude with distance, are useful for reviewing the spatial distribution of candidate Class II YSOs. We chose a threshold of $P_{\rm s}(\textsc{CII})>0.5$ giving 6\,504 candidate Class II YSOs. This threshold shows many concentrations of candidates while keeping a background of scattered sources to a low level. If the classifier is working well then we would expect known star forming regions to be well represented, and this is precisely what we see in Fig.~\ref{fig:bayes_high_PCII_gal_coords}. In Galactic coordinates there are several very obvious knots of candidate \textsc{CII}, many of which coincide with well known star forming regions, and a background of scattered candidates that increases towards direction of the Galactic centre. The bottom plot of Fig.~\ref{fig:bayes_high_PCII_gal_coords} shows Galactic longitude combined with distance by inverting the \textit{Gaia} EDR3 parallax. It is a testament to the quality of the \textit{Gaia} parallaxes that many of the same knots in Galactic coordinates also show as concentrations with distance. At around 1\,kpc there is a transition with closer sources clearly concentrated into knots, while more distance groupings appear smeared in distance. This indicates the uncertainty in the \textit{Gaia} parallaxes start to dominate the measurements by around 1 mas.

A similar useful diagnostic is the \textsc{TOPCAT} \citep{Taylor2005} weighted mean of all posterior \textsc{CII} in our NGPn catalogue \edit{(the top plot of Fig.~\ref{fig:bayes_sky_plots})}. These have many features in common, though not all. This plot and the individual feature plots are discussed in more detail in Section ~\ref{sec:FeatureAnalysis}.

The middle plot of Fig.~\ref{fig:bayes_high_PCII_gal_coords} identifies many of the star forming regions visible in our $P_{\rm s}(\textsc{CII})>0.5$ data set, noting \edit{the identifications are} not exhaustive. We briefly describe these regions.

\begin{description}
	
	\item The Serpens molecular cloud including the HII region W40 is located at $l\approx30$ and $b\approx3.5$ at a distance of 380--480\,pc \citep{Herczeg2019}. This does not show up in the $P_{\rm s}(\textsc{CII})>0.5$ Galactic coordinate plot but does appear as a slight over density at 0.5\,kpc in the Galactic longitude with distance plot. Interestingly it is visible in the weighted mean posterior \textsc{CII} of Fig.~\ref{fig:bayes_sky_plots}. \edit{This region has a paucity of sources in the master catalogue that is likely caused by dust obscuration}. Hence, the weighted mean posterior plot is capturing a larger than average number of high probability \textsc{CII} while the number of sources are insufficient to show up as an over density in the $P_{\rm s}(\textsc{CII})>0.5$ set in Galactic coordinates.
	
	\item There is a linear grouping of candidate \textsc{CII} at $l\approx53.5$ and $b\approx0.0$. This location corresponds to the infrared dark cloud (IRDC) G53.2, where \cite{Kim2015} found 302 young stars including 129 \textsc{CII}.
	
	\item There is weak diagonal line of candidate \textsc{CII} in Vulpecula around $55 < l < 57$ and $1.5 < b < 4.0$, showing up well in the distance plot at around 0.5\,kpc. Our $P_{\rm s}(\textsc{CII})>0.5$ set has 83 candidate \textsc{CII} in this region at this distance, with 14 found in the \cite{Zari2018} 3D mapping of the solar neighbourhood. An over density of PMS sources at this location in Vulpecula is identified but not discussed in \citet[figure 8]{Zari2018}. \edit{Around this location is the cluster NGC 6793 at $l=56.2$ and $b=3.3$ with a distance of about 0.6\,kpc \citep{2018A&A...618A..93C}. However, \cite{2018A&A...616A..10G} find an age of about 600\,Myr and \cite{2019Ap&SS.364..152Y} over 400\,Myr, indicating the cluster is too old to contain YSOs. Hence, this may be a new association which we label as New SFR 1, first picked out by \cite{Zari2018}.}
	
	\item Cygnus X \citep{Piddington1952} is composed of a large molecular cloud complex with active star formation including several OB associations, covering $77 \loa l \loa 82$ around the Galactic midplane  at a distance of approximately 1.7\,kpc \citep{Schneider2006}. There are clearly many high posterior \textsc{CII} in this region at around this distance.
	
	\item The North American Nebula (NGC 7000) and Pelican Nebula (IC 5070) at $l\approx85$ and $b\approx-1$ are a star forming complex referred to as NANeb by \cite{Guieu2009}. In the distance plot we can see this region is around 0.7--0.8\,kpc from the Sun, consistent with the older distance estimate of 0.6\,kpc from \cite{Laugalys2002}. It appears this may be part of a larger structure starting from 0.6--0.7\,kpc at $l\approx90$ and reaching 0.8--1.0\,kpc by $l\approx80$.
	
	\item There is an obvious grouping of candidate \textsc{CII} at the location of the HII region IC 1396 and associated young cluster Trumpler 37 at $l\approx99$ and $b\approx3.7$ and a distance $\approx0.9$\,kpc \citep{Contreras2002}. These form part of Cep OB2 \citep{Kun2008a,Barentsen2011} and are at the edge of the Cepheus Bubble \citep{Kun2008a}, most of which is above our $b=4$ cutoff. At the other side of the bubble is a grouping of 39 candidates at $l\approx105$ and $b\approx3.5$ with a distance $\approx 0.9$\,kpc. This is close to the dark cloud LDN 1188 located at $l=105.7$ and $b=4.2$ where \cite{2019MNRAS.484.1800S} identify 4 of our candidates as young stars. The open cluster \edit{MWSC 3607 \citep{2012A&A...543A.156K}} at $l=105.1$ and $b=3.4$ is very close in Galactic coordinates, though \cite{Kharchenko2013} estimate the distance to be around 4.4\,kpc with an age of about a Gyr. \edit{\cite{2020A&A...640A...1C} identify three clusters in this general area. Alessi-Teutsch 5 at $l=104.5$ and $b=4.2$ with a distance of 0.9\,kpc and an age around 19\,Myr, UPK 178 \citep{2019JKAS...52..145S} at $l=104.9$ and $b=2.7$ with a distance of 1\,kpc and an age around 200\,Myr, and cluster 30 from \cite{2003A&A...404..223B} at $l=105.3$ and $b=4.1$ with a distance of 0.9\,kpc \citep{2018A&A...618A..93C} and an age around 3.5\,Myr from \cite{Kharchenko2013}. Cluster 30 from \cite{2003A&A...404..223B} is the only cluster young enough to contain Class II YSOs, though with a radius containing half its members of 1.4\arcmin \citep{2020A&A...640A...1C} and a location about half a degree away from the grouping's centre it seems unlikely to be a match. Alessi-Teutsch 5 is a little old if the age of 19\,Myr is correct, though it has a larger half radius of 0.43\degr \citep{2020A&A...640A...1C} and there are three matches to this cluster from our 39 candidate \textsc{CII}. Hence, this may be a newly discovered young cluster or star formation associated with Alessi-Teutsch 5 and cluster 30 from \cite{2003A&A...404..223B}, which we label as New SFR 2.} Following a roughly linear feature of star formation leads us to the young cluster Cep OB3b at $l\approx110.0$ and $b\approx2.5$ at a distance $\approx0.7$\,kpc \citep{Moscadelli2009,Allen2012}. This cluster stands out clearly with halo of candidate \textsc{CII}.
	
	\item NGC 225 is just detectable at $l\approx127$ and $b\approx-1$ with a distance $\approx850$\,pc from \cite{Kharchenko2013}.
	
	\item There is a group of 23 candidate \textsc{CII} at $l\approx127$, $b\approx-1$ and a distance $\approx0.9$\,kpc. This location is very close to \edit{\cite{2007MNRAS.374..399F} cluster 541} located at $l=127$ and $b=-0.8$. However, \cite{2013MNRAS.436.1465B} estimate its distance at 12.9\,kpc, well beyond our cutoff. \edit{The cluster COIN-Gaia 31 \citep{2019A&A...624A.126C} is nearby in Galactic coordinates at $l=127.0$ and $b=-1.5$, though \cite{2019A&A...624A.126C} find a distance of 2.4\,kpc, too far away to be a match.} There is also the dark cloud LDN 1317 at $l=126.7$ and $b=-0.8$. We were unable to find further reference to this location in the literature, so this may be a newly discovered young cluster\edit{, which we label as New SFR 3}.
	
	\item The HII regions W3 / W4 / W5 form a chain in the Perseus arm \citep{Lada1978} covering approximately $132 < l < 138$ centred \edit{around} $b \approx 1$ \citep{Heyer1998} with active star formation in and around these clouds \citep{Carpenter2000}. The distance to these clouds is at the limit of our survey $\approx 2$\,kpc \citep{Carpenter2000}, though we nonetheless see over densities of young stars in this area.
	
	\item There are a few groupings of candidate \textsc{CII} in the vicinity of Cam OB1 at a distance $\approx 1$\,kpc. First, a grouping around $l \approx 142$ and $b \approx 2$, where \cite{Straizys2008} identify the HII region Sh 2-202 (l=140.6 b=1.9) at a distance of 0.8\,kpc \citep{Fich1984} illuminated by HD 19820\edit{. Then the open cluster Stock 23 (l=140.1 b=2.1), though \cite{Straizys2008}} doubt this is a genuine cluster\edit{. Finally,} the reflection nebulae vdB 14 and vdB 15 (l=142, b=1.8--3.0) illuminated by HD 21291 and HD 21389 who \cite{Straizys2008} consider to be members of Cam OB1.
	
	\item At $l \approx 148$ and $b \approx -0.4$ is another little grouping of candidates. In this area lies the emission nebula Sh 2-205 \citep{Straizys2008} at a distance of around 0.9\,kpc \citep{Fich1984}. The distance from the inverted \textit{Gaia} EDR3 parallaxes indicates all these candidates lie at around 1\,kpc, consistent with these older estimates.
	
	\item There are some groupings of \textsc{CII} candidates in Auriga at around $l \approx$ 173--174 and $b \approx$ -1--3. They are smeared out in distance from 1--2\,kpc, concentrated to the upper end of this range. At $l\approx173.4$ and $b\approx-0.2$ is the young cluster Stock 8 with median age $\approx3$\,Myr \citep{Jose2017} and a distance $2.05 \pm0.10$\,kpc \citep{Jose2008}. At $l\approx173.9$ and $b\approx+0.3$ is the young cluster NGC 1931. \cite{Pandey2013} found a mean YSO age of $2\pm1$\,Myr with a distance of $2.3\pm0.3$\,kpc for this cluster. G173.58+2.45 is a young cluster located at $l\approx173.6$ and $b\approx+2.4$. It appears to be associated with the surrounding HII regions Sh 2-231, Sh 2-232, Sh 2-233 and Sh 2-235 whose distances vary from 1.0 to 2.3\,kpc \citep{Shepherd2002}. All of these clusters are around the distance limit of our catalogue. This is consistent with the spread of distances we see for our \textsc{CII} candidates, since the individual \textit{Gaia} EDR3 parallaxes are becoming dominated by their uncertainties.
	
	\item At $l\approx190.0$ and $b\approx0.5$ is the open cluster NGC 2175 and the HII region NGC 2174. \cite{Bonatto2011} found this cluster and HII region to be part of a star forming complex with multiple young clusters at a distance of $1.4 \pm 0.4$\,kpc, consistent with our distance plot.
	
	\item \cite{Montillaud2019a,Bhadari2020} consider the young cluster NGC 2264 (l=203, b=2), Mon OB1 and the reflection nebula filament Mon R1 ($201 < l <202$, $0 < b < 1$) containing IC 446, IC 447, NGC 2245, NGC 2247 to be a large related star forming complex. \cite{Sung1997} found the distance to NGC 2264 to be 760\,pc using the distance moduli of 13 B-type stars in the cluster. This figure has been adopted by \cite{Montillaud2015,Montillaud2019b,Bhadari2020} for Mon OB1 and Mon R1. This distance is at the upper end of what we find for our candidate \textsc{CII} in this region, but nonetheless supports the notion this is part of a single star forming region.
	
	\item A group of candidate \textsc{CII} are clearly visible in the vicinity of the Rosette nebula and the young cluster NGC 2244 ($l\approx206.3, b\approx-2.1$). \cite{Lim2021} estimate the age of NGC 2244 at 2\,Myr and found other groups of young stars around the edges of the Rosette. They obtained a distance to the cluster of $1.4 \pm 0.1$\,kpc from the \textit{Gaia} EDR3 parallaxes. Unsurprisingly this is a good match to our distance with Galactic longitude plot.
	
\end{description}

\subsubsection{Feature analysis}
\label{sec:FeatureAnalysis}

\begin{figure*}
	\includegraphics[width=1\textwidth]{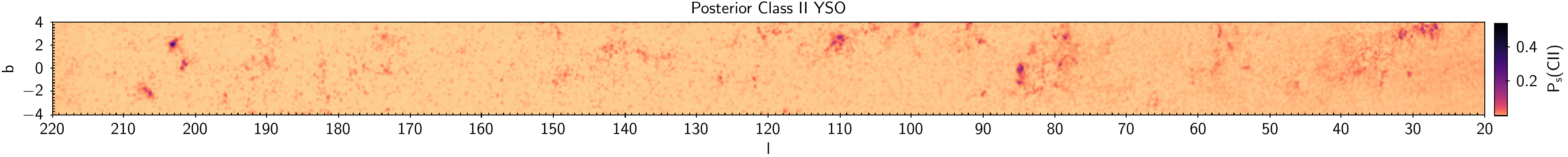}
	\includegraphics[width=1\textwidth]{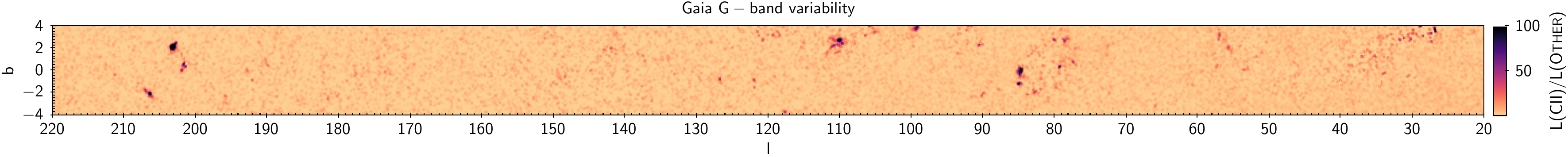}
	\includegraphics[width=1\textwidth]{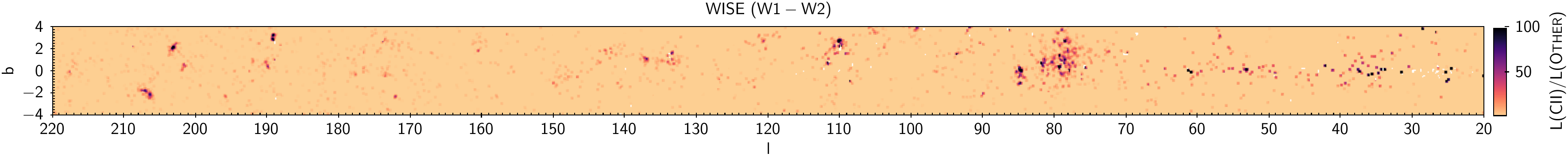}
	\includegraphics[width=1\textwidth]{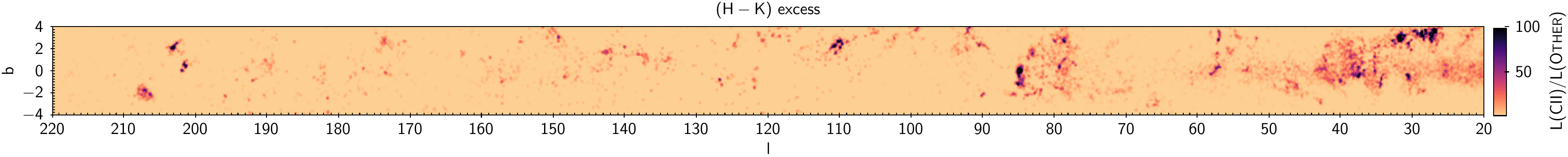}
	\includegraphics[width=1\textwidth]{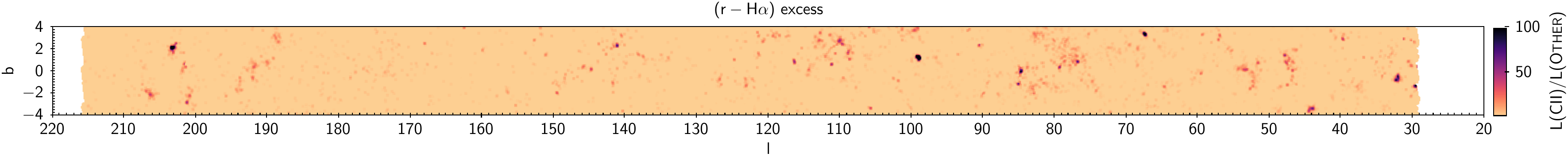}
	\includegraphics[width=1\textwidth]{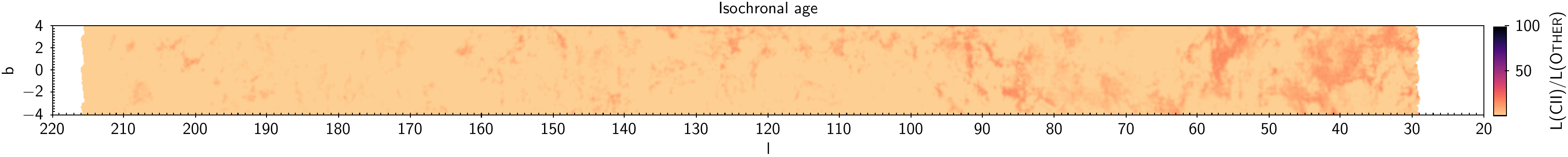}
	\caption{A series of plots in Galactic coordinates of the results set and the five features. The top plot is coloured by the posterior \textsc{CII} with a square root colour range to bring out the star forming regions. The subsequent plots are coloured by the feature likelihoods ratios $L(\textsc{CII})/L(\textsc{Other})$ on a fixed linear colour axis range from 1 to 100 to allow comparison between features. Only sources with a likelihood calculation for a feature are included in the mean plot of that feature. The plots were produced in \textsc{TOPCAT} using a weighted mean to remove the effect of varying source density with Galactic coordinates.}
	\label{fig:bayes_sky_plots}
\end{figure*}

To analyse how different features influence the posterior results, we created a series of Galactic coordinate plots in \textsc{TOPCAT} \edit{(see Fig.~\ref{fig:bayes_sky_plots})}. The top plot is the posterior \textsc{CII} followed by plots of the likelihood ratio $L(\textsc{CII})/L(\textsc{Other})$ for the five features. The weighted mean is used to colour the plots as it removes the effect of changing source density with coordinates, allowing regions of higher posterior and likelihood ratios to be picked out. The posterior \textsc{CII} plot uses a square root colour range to bring out star forming regions. The feature plots use a fixed linear range to allow easy comparison of the contributions by different features.

Many of the known star forming regions identified using the $P_{\rm s}(\textsc{CII})>0.5$ data set in Fig.~\ref{fig:bayes_high_PCII_gal_coords} are also visible in the weighted mean $P_{\rm s}(\textsc{CII})$ top plot of Fig.~\ref{fig:bayes_sky_plots}. Such as the Cygnus star forming complex at $l \approx 80$, Cep OB3b is obvious at $l \approx 110$, and the region $200 < l < 210$ of star forming complexes around NGC 2264 and the Rosette nebula.

The variability feature $\hat{\sigma}_{\rm O}$ \edit{(second from top in Fig.~\ref{fig:bayes_sky_plots})}, shows a strong set of results matching known star forming regions. There is also a general background of higher likelihood ratio sources. The variability feature would be expected to pick out variable stars of many types, not just Class II YSOs and other classes of young stars. Hence it is not surprising that a background population of variable stars are visible.

The \textit{WISE} $(W1-W2)$ plot \edit{(third from top in Fig.~\ref{fig:bayes_sky_plots})}, also indicates good results by picking out some known star forming regions. There are fewer sources with the \textit{WISE} feature towards the Galactic centre as the increased source confusion reduces the number of sources with a reliable match. This allows a few individual sources with a strong \textit{WISE} result to show up in the weighted mean close to the Galactic plane where $l < 65$. There is an interesting over density of higher likelihood ratios favouring \textsc{CII} at $l \approx 189$ and $b \approx 3$ that does not correspond to strong results in the other features. It is caused by 78 sources with a \textit{WISE} likelihood ratio above 50. Only 4 of these sources have $P(\textsc{CII})>0.5$, though 31 have $P(\textsc{CII})>0.0109$ (threshold of our 50k set), confirming the other features do not strongly reinforce the \textit{WISE} results. This location corresponds to the supernova remnant IC 443, where \cite{Su2014} identified 98 candidate Class I and II YSOs using \textit{WISE} and 2MASS photometry. Our NGPn catalogue matches to 19 of the \cite{Su2014} candidates. Only 4 of these have $P(\textsc{CII})>0.5$ and 10 have $P(\textsc{CII})>0.0109$ from our catalogue. This is an intriguing result. As \textit{WISE} is our longest wavelength feature, with weaker results from our other features, this might be an indication that many of the YSOs in this region are younger than our Class II targets.

The $(H-K)$ excess feature plot \edit{(third from the bottom in Fig.~\ref{fig:bayes_sky_plots})}, shows a wealth of structure where the ratio of the likelihoods favours \textsc{CII}. Many of these correlate with structures found in the $P_{\rm s}(\textsc{CII})$ plot. The section for $l<60$ displays more structure and a stronger \textsc{CII} signal than in any of the other plots. \edit{There is a loose correlation between this structure and regions with higher reddening, indicating these structures may have some relation to interstellar dust.} By itself this does not mean all the structures relate to star formation, though the pronounced structure at $l\approx30$ and $b\approx3.5$ corresponds to the Serpens molecular cloud. The $(H-K)$ feature is a mix of UKIDSS and 2MASS, with UKIDSS of higher quality but missing data for $107 < l < 141$ as UKIRT is unable to observe in that direction. It is encouraging that this lack of UKIDSS data does not show up in this plot, though we were able to see it by stretching the colour axis.

The $(r-H \alpha)$ excess feature is second from bottom of Fig.~\ref{fig:bayes_sky_plots}. Due to the IGAPS survey limits, the results only extend across $30 < l < 215$. The feature shows a weaker signal than some other features, though there are clear structures favouring \textsc{CII} that coincide with known star forming regions. These include the star forming complexes in Cygnus, Cep OB3b, NGC 2264 and the Rosette nebula. There are others that are unique to the $H\alpha$ feature, such as groupings favouring \textsc{CII} at $l\approx67$ and $b\approx3$, and $l\approx99$ and $b\approx1$, as well as an arc centred on $l\approx53$ and $b\approx-1$. The grouping at $l\approx67$ and $b\approx3$ has a rectangular shape with edges aligned with the IGAPS images, indicating it must be an artefact, probably where the IGAPS photometric calibrations are lower quality. The grouping at $l\approx99$ and $b\approx1$ has a less regular shape but with straight lines and right angles that match to IGAPS image tilings, so this must also be an IGAPS calibration artefact. The arc centred on $l\approx53$ and $b\approx-1$ is located in Sagitta. It is more natural looking with an extent of 2\degr in Galactic latitude. The top of this arc along the Galactic equator corresponds to the infrared dark cloud G53.2, where \cite{Kim2015} found 302 young stars including 129 \textsc{CII}. This top portion of the arc shows up in the $P_{\rm s}(\textsc{CII})>0.5$ plot of Fig.~\ref{fig:bayes_high_PCII_gal_coords}.

The bottom plot of Fig.~\ref{fig:bayes_sky_plots} shows the weighted mean likelihood ratios for the isochronal age feature. As with the $H\alpha$ feature this uses IGAPS, so results only extend across $30 < l < 215$. Since the colour scaling between the features is the same, it is clear that the isochronal age is the weakest. There are still a good number of the known star forming regions visible, such as Cygnus, Cep OB3b and the star formation in the area $200 < l < 210$. There is also structure at a variety of other locations, most notably $30 < l < 60$. \edit{This region has a higher level of reddening, implying there is a large quantity of dust. This could indicate the presence of molecular clouds that are associated with star formation.} We should treat the results at this location with caution, as it may simply be a problem with the approximate reddening we apply to individual sources, causing the isochronal age to give a false result.

Our overview of the features gives us confidence in the operation of the features and the classifier as a whole. As expected, the strong posterior results favouring \textsc{CII} come about where different features reinforce the same result. Where features differ or do not give a strong result, then we tend not to see it in the posterior. This prevents any single feature from dominating the results, giving strength to the outcomes, and preventing survey artefacts from creating high posterior results.

\subsection{Comparison to other catalogues}
\label{sec:comparisons_to_other_catalogues}

We compared our candidate Class II YSOs to the young stars found by other classifiers. A summary of these comparisons is given in Table~\ref{tab:comparisons}.

To analyse each literature catalogue, the first step was to cross-match it to our NGPn data set, creating a combined data set of sources present in both catalogues, labelled $M$ in Table~\ref{tab:comparisons}. We chose to compare good quality YSO samples as these are relatively straightforward to interpret. From our full catalogue we chose the high purity set $P_{\rm s}(\textsc{CII})>0.5$, returning 6\,504 candidates from our catalogue. \footnote{This is relatively insensitive to changes in posterior, for example $P_{\rm s}(\textsc{CII})>0.75$ yields 4\,738 candidates and $P_{\rm s}(\textsc{CII})>0.25$ 12\,832.} The number of $P_{\rm s}(\textsc{CII})>0.5$ in the combined data set is $A$ in Table~\ref{tab:comparisons}. To make a meaningful comparison we need to choose an equivalent data set of candidate YSOs from the other classifier. Where catalogues simply provide lists of candidates, we use the members of that list which appear in $M$. Where the catalogue provides a young star probability, we chose a threshold to provide a comparable number of young stars to set $A$. These other classifier YSO samples we refer to as data set $B$. We did not use a probability threshold of 0.5 to create the $B$ data sets from the literature catalogues, as our posterior does not translate into a simple probability \edit{(see Section~\ref{sec:naive_bayes_classifier})}. This approach inevitably gives a narrow view of the comparison, since different choices of threshold will give different results. Hence, this approach merely investigates \edit{the} performance of good quality sets between the catalogues. It must also be emphasized that we are not comparing entirely equivalent classifications between catalogues. We are specifically trying to identify Class II YSOs, while many of the other catalogues are designed to search for a broader age range of PMS stars.

The middle part of the table shows the comparison statistics. The first row is the number of sources in common between $A$ and $B$, i.e. both catalogues consider the source a probable \textrm{CII} or young star by the quoted measures. The next statistic $A \notin B$ tells us how many of our candidate \textrm{CII} are not labelled as probable young stars by the other catalogue measure. We also provide this as a percentage of our candidate \textrm{CII}. The final statistic $B \notin A$ gives how many of the candidate young stars from the other catalogue are not considered probable \textrm{CII} by our classifier, also given as a percentage of the matched candidate young stars from the other catalogue. Since we aim to find roughly equal matched candidates from both catalogues, these two statistics are naturally very similar.

The bottom part of the table gives background information. Starting with the method used to match the catalogues, all performed in \textsc{TOPCAT}. Then a list of the source surveys utilised by the other classifiers.

\begin{landscape}
	\begin{table}
		\vspace{4.0cm}
		\caption{A summary of the comparisons to other catalogues, baselined to our $P_{\rm s}(\textsc{CII})>0.5$ classifier results. The label $M$ is used for the combined data set from cross-matching the Bayes NGPn catalogue to the literature catalogue. The label $A$ is used for the candidate \textrm{CII} from our catalogue. The label $B$ is used for the candidate young star measure from the other catalogue. * indicates surveys used by our classifier.}
		\label{tab:comparisons}
		\begin{tabular}{lcccccccc}
			\hline
			& \protect\cite{Robitaille2008} & \protect\cite{Marton2019} & \protect\cite{Vioque2020} & \protect\cite{Chiu2021} & \protect\cite{Kuhn2021} & \protect\cite{McBride2021} & \protect\cite{2021AA...647A.116C} & \protect\cite{2022arXiv220606416E} \\
			& & & & Model IV & & $P(\rm PMS) \geq 0.7$ & & \protect\cite{2022gdr3.reptE..10R} \\
			\hline
			Method & Non-ML & Random forest & Neural net & Neural net & Random forest & Neural net & Neural net & Random Forest and\\
			& & & & & & & & Extreme Gradient Boost \\
			Target & YSOs / AGB stars & YSOs & Herbig Ae/Be & YSOs & YSOs & PMS & YSOs & YSOs \\
			& & & \edit{Classical Be / PMS} & & & & & \\
			NGPn footprint & 4\,757 & 20\,270\,985 & 2\,321\,644 & 172\,680 & 32\,891 & 104\,627 & \edit{7\,789} & \edit{22\,162}\\
			Matched ($M$) & 77 & 3\,559\,300 & 409\,991 & 16\,101 & 599 & 32\,617 & \edit{2\,668} & \edit{9\,772} \\
			Other measure & Present & $P(\rm YSO)>0.973$ & $p(\rm{PMS})>0.3$ & $P(\rm{YSO})>0.5$ & Class II & $P(\rm PMS)>0.861$ & \edit{$P(\rm CII)>0.0079$} & \edit{Best class = YSO} \\
			& & & & & & $6.3<log_{10}(Age)$ & & \edit{Best class score $>0.8045$} \\
			& & & & & & $log_{10}(Age)<6.7$ & & \\
			Other count ($B$) & 77 & 3\,423 & 2\,477 & 697 & 389 & 2\,227 & \edit{182} & \edit{1\,345} \\
			$P_{\rm s}(\textsc{CII})>0.5$ ($A$) & 46 & 3\,337 & 2\,481 & 706 & 371 & 2\,216 & \edit{182} & \edit{1\,345} \\
			\hline
			$A$ and $B$ & 46 & 1\,224 & 1\,412 & 459 & 254 & 672 & \edit{137} & \edit{351} \\
			$A \notin B$ ($/A$\%) & 0 (0\%) & 2\,113 (63\%) & 1\,069 (43\%) & 247 (35\%) & 117 (32\%) & 1\,544 (70\%) & \edit{45 (25\%)} & \edit{994 (74\%)} \\
			$B \notin A$ ($/B$\%) & 31 (40\%) & 2\,199 (64\%) & 1\,065 (43\%) & 238 (34\%) & 135 (35\%) & 1\,555 (70\%) & \edit{45 (25\%)} & \edit{994 (74\%)} \\
			\hline
			Match method & 1\arcsec & \textit{Gaia} DR2 Id & \textit{Gaia} DR2 Id & 1\arcsec & 1\arcsec & \textit{Gaia} EDR3 Id & \edit{1\arcsec} & \edit{\textit{Gaia} EDR3 Id} \\
			\hline
			\textit{Gaia} parallax* &  &  &  &  &  & DR2/EDR3 & & \\
			\textit{Gaia} photometry* &  & DR2 & DR2 &  &  & DR2/EDR3 & & \edit{DR3} \\
			\textit{WISE}* &  & Y & Y &  &  & & & \\
			\textit{Spitzer} & Y &  &  & Y & Y & & \edit{Y} & \\
			UKIDSS* &  &  &  & Y & Y & & & \\
			2MASS* &  & Y & Y & Y & Y & Y & & \\
			IGAPS/IPHAS* &  &  & Y &  &  & & & \\
			VPHAS &  &  & Y &  &  & & & \\
			VVV &  &  &  &  & Y & & & \\
			Chandra &  &  &  &  & Y (MYStIX) & & & \\
			\textit{Planck} dust map &  & Y &  &  &  & & & \\
			\hline
		\end{tabular}
	\end{table}
\end{landscape}

There are many high probability young stars from the other catalogues that do not match to any counterpart in our catalogue. This is an inevitable outcome of our decision to only select high quality data from the source surveys for our classifier. This means we miss many young stars. We accept this as it allows us to judge our classifier performance based on the approach, without needing to untangle the effects of poor quality input data.

\subsubsection{Comparison with \protect\cite{Robitaille2008}}

\cite{Robitaille2008} compiled a catalogue of red mid-infrared sources in the Galactic midplane. The fundamental definition of their red sources is \textit{Spitzer} $[4.5]-[8.0] \geq 1$. We only achieve a match rate to their catalogue of just under 2 per cent ($M$ in Table~\ref{tab:comparisons}). There are two reasons for this low match rate. First, we expect these red sources to emit predominantly in the infrared of the \textit{Spitzer} survey, and many may not be sufficiently luminous in the optical to be detected by \textit{Gaia} or to pass our $G_{\rm{RP}}<18.0$ cut used for our base set. A crude 1$\arcsec$ match of \cite{Robitaille2008} red sources to \textit{Gaia} EDR3 returns 5\,535 matches, only 29 per cent of their catalogue, thus supporting this notion. Second, our quality cuts eliminate 96 per cent of sources from the \textit{Gaia} catalogue. The combination of these two factors would give around a 1 per cent match rate, close to what we find.

As the \cite{Robitaille2008} catalogue does not provide a young star probability the $A \notin B$ has no meaning, since all matches count as candidates in \cite{Robitaille2008}. The $B \notin A$ has relevance, showing we consider 40 per cent of the matches to be unlikely or low probability candidate \textrm{CII}. \cite{Robitaille2008} estimate 50--70 per cent of their catalogue are YSOs, with the rest AGB stars along with a few galaxies and planetary nebulae. As our \textrm{CII} should be a subset of their YSOs, our results are roughly inline with their expectations. That our candidate \textsc{CII} constitute over half the \cite{Robitaille2008} candidate YSOs could indicate we are finding YSOs at other stages in their evolution.

\subsubsection{Comparison with \protect\cite{Marton2019}}
\label{sec:marton2019_comparison}

A classifier for identifying candidate YSOs was created by \cite{Marton2019}. We compare to their published results based on a random forest of 500 trees. Their input data is \textit{Gaia} DR2 combined with AllWISE and 2MASS. When comparing to our work, it is worth noting they use the \textit{Gaia} $G$-band photometry but not the parallax measurements. They supplement their data with the \textit{Planck} dust opacity map as a proxy for the interstellar extinction to individual sources. Their input data is limited to regions where they expect YSOs by selecting areas with a dust opacity above $1.3 \times 10^{-5}$.

They classify their results into YSOs, extragalactic objects, main-sequence stars and evolved stars. Their YSO training sets were compiled from VizieR YSO photometric and spectroscopic catalogues, \textit{Spitzer} YSO related publications and the Molecular Cores to Planet-forming discs (c2d) catalogue \citep{Evans2003,Evans2009}.

When selecting data for our comparison we follow the approach of \cite{Marton2019} by selecting between their runs with and without $W3$ and $W4$ based on the probability this longer wavelength photometry are real detections. After matching their catalogue to our NGPn data set ($M$ in Table~\ref{tab:comparisons}), we found a 97.3 per cent probability threshold for their YSO probability gave a comparable number of candidates to our $P_{\rm s}(\textsc{CII})>0.5$ set.

\begin{figure*}
	\includegraphics[width=1\textwidth]{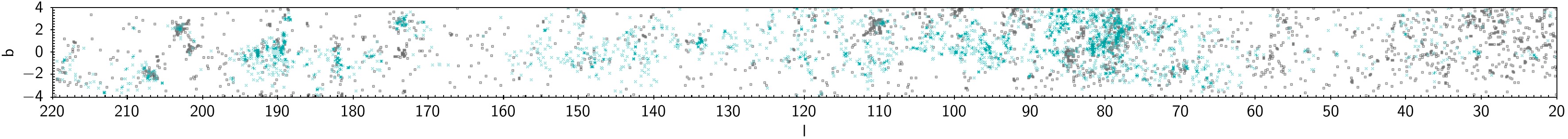}
	\includegraphics[width=1\textwidth]{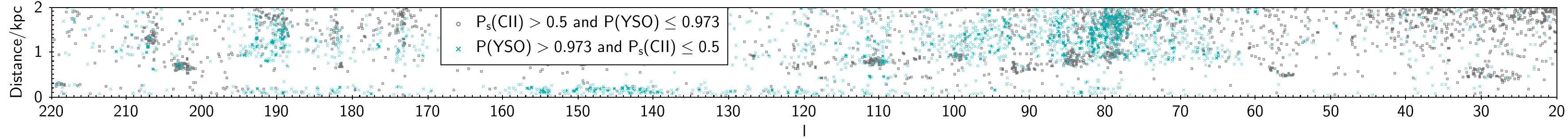}
	\caption{\edit{Matches between our classifier and \protect\cite{Marton2019}. Grey circles are our high confidence \textsc{CII} candidates not in the equivalent \protect\cite{Marton2019} candidate list. Cyan crosses are the \protect\cite{Marton2019} candidates not in our candidate list.} The top plot is Galactic longitude and latitude and the bottom plot is Galactic longitude and distance in kpc by inverting the \textit{Gaia} parallax.}
	\label{fig:Marton2019_l}
\end{figure*}

As the \cite{Marton2019} candidates cover the full range of evolutionary stages, we could reasonably expect them to find more candidates than we do. Their candidates we consider low probability \textrm{CII} could thus be YSOs at other stages of evolution. The additional candidates we find are harder to explain unless \cite{Marton2019} are missing YSOs or our catalogue is contaminated. To give insight into these data sets we examine their distribution in the Galactic plane \edit{(see Fig.~\ref{fig:Marton2019_l})}. The grey circles of our \edit{high confidence \textrm{CII} not in the equivalent \cite{Marton2019} set} generally align with known star forming regions in Cygnus, Cepheus and around the Rosette Nebula and NGC 2264, along with a scattering of sources that increase towards the Galactic centre. The clustering of these sources indicates a significant proportion are genuine young stars. \edit{The cyan crosses of the high confidence \cite{Marton2019} candidates not in our candidate list} also follow star forming regions in Cygnus with concentrations at the Rosette Nebula and NGC 2264, though in general their distribution is less concentrated than ours. They also pick out some very nearby candidates within 200\,pc, most notable at $120<l<155$ and $180<l<195$, that we do not consider probable \textrm{CII}. This implies our lists are finding different types of source, and we may be better at finding concentrations of Class II YSOs.

\subsubsection{Comparison with \protect\cite{Vioque2020}}
\label{sec:vioque2020_comparison}

A catalogue of candidate Herbig Ae/Be was created by \cite{Vioque2020}. They perform a 3-way classification, giving probabilities that each source is a PMS star, Classical Be star and Other. Their PMS classification is trained on a combination of Herbig Ae/Be stars and T-Tauri stars. Although our classifier will be less likely to pick out the higher mass Herbig Ae/Be stars, the T-Tauri stars are a good match to our \textsc{CII}. Hence, their PMS probability should provide a useful comparison to our posterior \textsc{CII}. Their results are of particular interest as they use nearly identical input surveys to our work (\textit{Gaia} DR2, 2MASS, \textit{WISE}, IPHAS and VPHAS+), using an artificial neural network for their classifier.

The artificial neural network cannot cope with missing data, so they require all of their features to be present for all sources, limiting the size of their catalogue to the common sources across the input surveys. They do not impose quality cuts on the photometry or catalogue cross-matching, instead relying on the neural network to deal with poor data and false matches. This contrasts with our naive Bayes classifier which only requires a single feature to give a meaningful result, but is limited in size by the data quality restrictions we impose. These differing approaches result in the combined data set ($M$ in Table~\ref{tab:comparisons}) composed from 18 per cent of \cite{Vioque2020} and 5 per cent of our NGPn catalogue.

A further difference is their neural network learns to distinguish between the classes using 48 features without preconceptions, while we carefully chose a set of 5 features known for their utility in identifying Class II YSOs. We have the common feature of \textit{Gaia} variability, though the implementation of this feature is different between the two classifiers.

We found a $P(\rm PMS)>0.3$ provided a close number of matched candidates to our $P_{\rm s}(\textsc{CII})>0.5$ set. A little over half are mutual candidates between the two sets, with 43 per cent of our \textsc{CII} candidates not considered PMS by \cite{Vioque2020}, and the same percentage of \cite{Vioque2020} candidates we do not consider \textsc{CII}.

\begin{figure*}
	\includegraphics[width=1\textwidth]{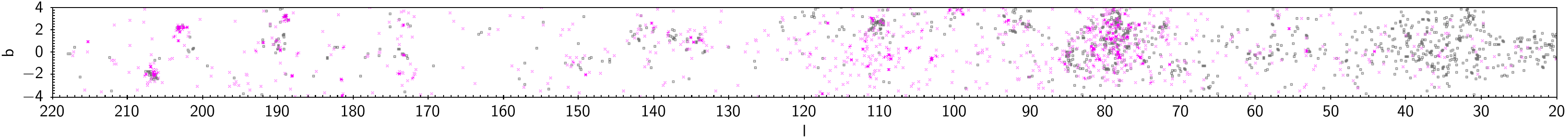}
	\includegraphics[width=1\textwidth]{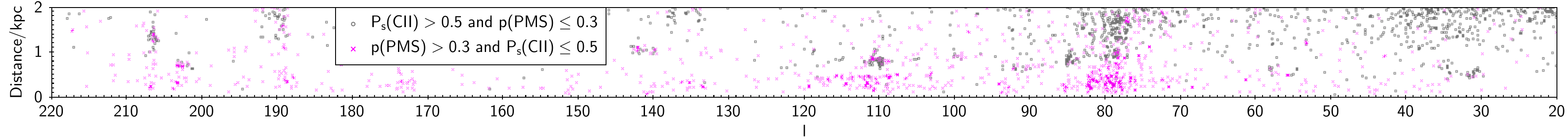}
	\caption{\edit{As Fig.~\ref{fig:Marton2019_l} but for the matches between our classifier and \protect\cite{Vioque2020}. Grey circles are our high confidence \textsc{CII} candidates not in the equivalent \protect\cite{Vioque2020} candidate list. Magenta crosses are the \protect\cite{Vioque2020} candidates not in our candidate list.}}
	\label{fig:Vioque2020_l}
\end{figure*}

In Fig.~\ref{fig:Vioque2020_l} we review the spatial distributions of the two subsets \edit{ of distinct sources}. There are some interesting similarities and surprising differences. A good proportion of both \edit{subsets} line up with known star forming regions such as Cygnus and the Rosette Nebula. However, \cite{Vioque2020} find additional nearby sources within 500\,pc, predominantly in Cygnus and a diffuse grouping around longitude of $110\degr$ close to Cepheus. On the other hand, the additional sources from our classifier which are not in \cite{Vioque2020} form a concentration in Galactic coordinates at or very close to Cep OB3b, and our additional Cygnus sources are more distant. As our extra Cygnus candidates are at a greater distance they could be due to incorrect reddening in our data caused by a wall of dust at this location, or we could be finding sources in the more distant Cygnus X star forming complex missed by \cite{Vioque2020}. The additional \cite{Vioque2020} candidates in Cygnus are about half the distance of the star formation we identify in Fig.~\ref{fig:bayes_high_PCII_gal_coords}. It is possible both sets of candidates in Cygnus are spurious, though we have insufficient information to draw firm conclusions. That both data sets identify sources concentrated around known star forming regions lends to the conclusion that at least some are genuine young stars that are missed by one or other of the classifiers. The scattered sources in both data sets are more likely spurious candidates.

\subsubsection{Comparison with \protect\cite{Chiu2021}}
\label{sec:chiu2021_comparison}

The fully connected neural network SCAO (Spectrum Classifier of Astronomical Objects) built by \cite{Chiu2021} identifies candidate YSOs amongst galaxies and other stars. Their classifier uses SEDs constructed from \textit{Spitzer}, 2MASS and UKIDSS photometry. They trained on the Molecular Cores to Planet-forming discs (c2d) catalogue \citep{Evans2003,Evans2009} to provide good quality labelled data. This means it is a classifier for YSOs at all evolutionary stages rather than the Class II YSOs of our classifier. We compare to the \cite{Chiu2021} SCAO run on the SEIP (\textit{Spitzer} Enhanced Imaging Products) catalogue.

In constructing their classifier they created four models using different input data. We compare to Model IV as this was designed to give the largest number of predictions. It performs the classification using partial \edit{SEDs} from the IRAC 3, IRAC 4 and MIPS 1 photometry bands.

The SEIP catalogue on which SCAO is based is not a full sky catalogue. It covers patches of our NGPn footprint with many of these patches coinciding with star forming regions such as Cygnus and Cepheus, as well as many smaller targets such as NGC 2264 and the Rosette Nebula. This means a spatial examination of the matches would be biased and uninformative.

We found a $P(\rm YSO)>0.5$ gave a similar number of candidates to our $P_{\rm s}(\textsc{CII})>0.5$ in the matched data set ($M$ in Table~\ref{tab:comparisons}). About two thirds of the candidates are common between these sets, with each having just over a third uniquely classified as candidate young stars. This is a good level of agreement between our catalogues and encouraging as they used high quality training data. However, as they expect to find YSOs at all evolutionary stages, this could be another indication that our classifier is finding a broader age range of young stars than the Class IIs we are targetting.

\subsubsection{Comparison with \protect\cite{Kuhn2021}}
\label{sec:kuhn2021_comparison}

A catalogue of around 120\,000 candidate YSOs was presented by \cite{Kuhn2021}. They start by selecting sources with significant MYStIX \citep{Feigelson2013} IR-excess along with a quality constraint based on \textit{Spitzer}/IRAC photometry. Next they fit SEDs to JHK (from 2MASS, UKIDSS and VVV) and IRAC photometry. They only retain sources where the SED does not fit a reddened stellar photosphere. Finally, they run a random forest classifier to identify YSOs. As their classifier requires photometry in all bands, they impute missing values from the available photometry. The YSO Class is determined from a spectral index calculated on \textit{Spitzer} and \textit{WISE} photometry.

We have a relatively small number of matches to \cite{Kuhn2021} ($M$ in Table~\ref{tab:comparisons}), primarily due to our quality cuts on the \textit{Gaia} EDR3 catalogue. A detailed breakdown of the \cite{Kuhn2021} classifications is given in Table~\ref{tab:Kuhn2021}. It can be seen that we match more and a greater percentage of Class II YSOs than the other evolutionary stages. Looking at the percentage of the matched sources with $P_{\rm s}(\textsc{CII})>0.5$, about two thirds of the Flat SED, Class II and Class III are labelled as candidate \textrm{CII} by our classifier. The fraction of Class I is noticeably smaller at around a third. These results make sense in evolutionary terms as the Flat SED\edit{, Class II} and Class III YSOs are closest to our Class II classification, sharing some of their features. So overall there is reasonable agreement between our high posterior \textsc{CII} and the YSO classification by \cite{Kuhn2021}.

\begin{table}
	\caption{The YSO classifications from the \protect\cite{Kuhn2021} catalogue. The figures are the number from their catalogue within our footprint, the number matched to our catalogue, and the number with a posterior \textsc{CII} above 0.5 in our catalogue. The number of matched sources is also given as a percentage of sources in that class. The number of our candidate \textsc{CII} is also give as the percentage of matched sources in that class.}
	\label{tab:Kuhn2021}
	\begin{tabular}{lrrr}
		\hline
		Kuhn Type & NGPn footprint & Matched & $P_{\rm s}(\textsc{CII})>0.5$ \\
		\hline
		Class I & 4\,354 & 45 (1.0\%) & 16 (36\%) \\
		Flat SED & 7\,063 & 117 (1.7\%) & 73 (62\%) \\
		Class II & 17\,261 & 389 (2.3\%) & 254 (65\%) \\
		Class III & 1\,751 & 37 (2.1\%) & 25 (68\%) \\
		Uncertain & 2\,462 & 11 (0.4\%) & 2 (15\%) \\
		\hline
		Total & 32\,891 & 599 & 371 \\
	\end{tabular}
\end{table}

\subsubsection{Comparison with \protect\cite{McBride2021}}
\label{sec:mcbride2021_comparison}

A classifier named Sagitta was created by \cite{McBride2021} for identifying PMS stars and assigning them an age. It is composed of three convolution neural networks, one to create an extinction map, another to assign PMS probabilities, and the third to estimate age. Their input data consists of parallax measurements from \textit{Gaia} with photometry from \textit{Gaia} and 2MASS.

The majority of their training data set is from \cite{Kounkel2020}. They use a hierarchical clustering analysis of \textit{Gaia} DR2 to identify clusters and comoving groups, along with a neural network named Auriga to estimate ages, extinction and distances. Probable PMS stars are identified by \cite{Kounkel2020} from extinction corrected \textit{Gaia} photometry and age estimates.

As neural networks cannot handle missing data, \cite{McBride2021} replace missing data with values slightly below the detection limit of the feature. This is based on the assumption that most non-detections are due to the source being too faint.

For our comparison we use the data from \citet[table 5]{McBride2021}. This is the output from a run of Sagitta on \textit{Gaia} EDR3, hence more comparable to our catalogue. It only contains the subset of results where $P(\rm{PMS})\geq 0.7$, but this is sufficient for a meaningful comparison. There is a high match rate to our NGPn catalogue at 31 per cent of sources within the NGPn footprint ($M$ in Table~\ref{tab:comparisons}), especially notable as this is only a subset of their full catalogue where $P(\rm{PMS})\geq 0.7$. This high match rate is due to \cite{McBride2021} imposing similar criteria on the \textit{Gaia} archive to our base data set. We found a $P(\rm PMS)>0.861$ combined with $6.3<log_{10}(Age)<6.7$ gave a similar number of candidate young stars to our $P_{\rm s}(\textsc{CII})>0.5$ set. We introduced the age criteria on the \cite{McBride2021} data, as this allows us to select their candidates with an age roughly consistent with Class II YSOs, about 2 to 5\,Myr. There is a relatively small number of young stellar candidates in common between the two sets, they are about 70 per cent distinct.

\begin{figure*}
	\includegraphics[width=1\textwidth]{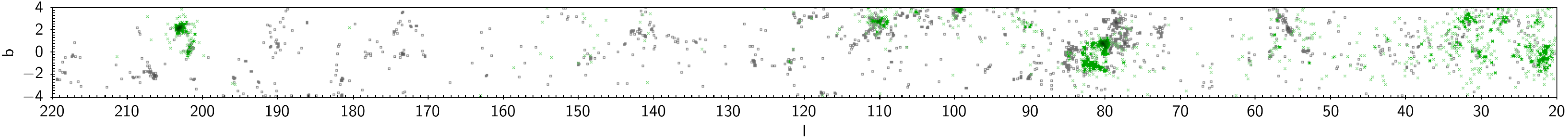}
	\includegraphics[width=1\textwidth]{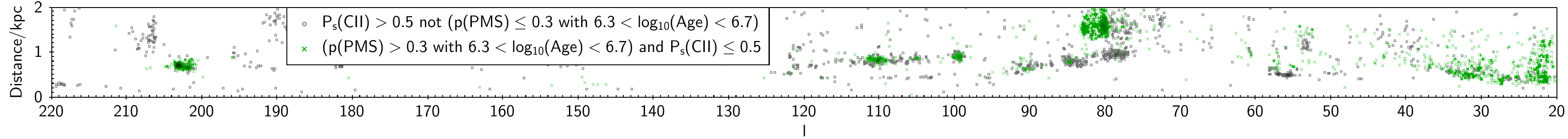}
	\caption{\edit{As Fig.~\ref{fig:Marton2019_l} but for the matches between our classifier and \protect\cite{McBride2021}. Grey circles are our high confidence \textsc{CII} candidates not in the equivalent \protect\cite{McBride2021} candidate list. Green crosses are \protect\cite{McBride2021} candidates not in our candidate list.}}
	\label{fig:McBride2021_l}
\end{figure*}

In Fig.~\ref{fig:McBride2021_l} we present the spatial distribution of the sources \edit{from the distinct subsets}. Both of these sets show strong correlation with star forming regions but not always the same regions. \cite{McBride2021} clearly pick out additional candidates in Cygnus at around $l\approx80$. It is plausible we are  missing these candidates due to a wall of dust at this location causing bad extinction estimates in our classifier, though this is far from certain. \cite{McBride2021} find additional sources towards the Galactic centre around $l<25$, where we lack IGAPS data so our classifier works with at most three features. Neighbouring this region they also find additional candidates in the Serpens region around $l\approx30$. Reviewing our features for the additional \cite{McBride2021} candidates there is no clear reason why we do not consider them to be candidate \textsc{CII}. In our classifier they tend to have weaker results across all features. Our classifier picks out additional candidate \textsc{CII} in Vulpecula at $l \approx 56$ and the Rosette nebula at $l \approx 207$. We both identify unique groupings of candidate young stars around Cep OB3b at $l \approx 110$ and NGC2264 at $l \approx 203$. That both sets of unique candidates tend to be grouped around known star forming regions lends to the notion that both classifiers are finding genuine young stars missed by the other.

The lack of agreement between the classifiers is nonetheless unexpected. To investigate whether this might be due to the age criteria we imposed on the \cite{McBride2021} data set, we repeated the comparison without the age criteria and a revised threshold of $P(\rm PMS)>0.926$ to give a comparable number of candidates to our $P_{\rm s}(\textsc{CII})>0.5$ set. This again gave two data sets with 70 per cent distinct sources. There was a similar spatial pattern for the common sources, and for our \textsc{CII} candidates not considered candidates by \cite{McBride2021}. However, the set of \cite{McBride2021} candidates not in our \textsc{CII} candidate list showed a greater scatter in Galactic coordinates and contained a set of additional candidates within 500\,pc. We conclude imposing the age criteria on the \cite{McBride2021} data set makes it more comparable to our \textsc{CII} candidate list, and has nothing to do with the lack of agreement.

\subsubsection{\edit{Comparison with \protect\cite{2021AA...647A.116C}}}
\label{sec:cornu_and_montillaud_2021_comparison}

\edit{A multilayer perceptron neural network for identifying YSOs was constructed by \cite{2021AA...647A.116C}. They created labelled data using a simplification of the \cite{Gutermuth2009a} method, using only Spitzer photometry without 2MASS. This gave a homogeneous data set while introducing the limitation that only Class I and Class II YSOs could be identified.}

\edit{Their labelled data were compiled from three data sets in Orion \citep{Megeath2012}, NGC 2264 \citep{Rapson2014} and a sample of clouds within 1\,kpc \citep{Gutermuth2009a}. These data sets were split $80:20$ for training and testing. They have not applied the classifier to other regions with Spitzer data as they expect a large number of false positives at distances beyond nearby star forming regions. Their published catalogue is comprised of the Orion and NGC 2264 regions, including both their classifier result and the labels from their modified \cite{Gutermuth2009a} method. As with our catalogue, their classification probabilities do not translate to genuine probabilities, i.e. a $P(\rm CII)=0.9$ does not mean the source has a 90 per cent probability of being a Class II YSO.}

\edit{In our matched data set, we found a $P(\rm CII)>0.0079$ from \cite{2021AA...647A.116C} gave 182 Class II YSO candidates, equal to the number of candidates found with our $P_{\rm s}(\textsc{CII})>0.5$ threshold. These sets of candidates have 75 per cent of sources in common. The low threshold needed for an equal number of candidates from \cite{2021AA...647A.116C} causes some issues that we will address later.}

\edit{The common sources between our catalogues are limited to the NGC 2264 region, so a spatial comparison does not provide insight into the different classifications. Within the matched data set there are 138 labelled Class II YSOs by the simplified \cite{Gutermuth2009a} method. The \cite{2021AA...647A.116C} $P(\rm CII)>0.0079$ threshold recovers all 138 Class II YSOs, while our $P_{\rm s}(\textsc{CII})>0.5$ recovers 129. The 9 labelled Class II YSO not in our \textsc{CII} candidate list exhibit a lower level of \textit{Gaia} $G$-band variability. They also have little or no $(W1-W2)$, $(H-K)$ and $H\alpha$ excess. The isochronal age feature is consistent with young stars without giving a strong result. Hence, a classification of \textsc{Other} by our classifier is consistent with our features.}

\edit{There are 44 sources with $P(\rm CII)>0.0079$ from \cite{2021AA...647A.116C} that are not in their list of labelled Class II YSOs. By our classifier, 6 have a $P_{\rm s}(\textsc{CII}) > 0.8$ and 35 have $P_{\rm s}(\textsc{CII}) < 0.075$. This indicates that a few may be genuine Class II YSOs while the majority are probably contaminants. The implication is that we are incorrectly classifying them as \textsc{CII} by setting the \cite{2021AA...647A.116C} threshold too low. This is a weakness in our comparison approach, as we need to make up a similar number of candidate \textsc{CII} to our $P_{\rm s}(\textsc{CII})>0.5$ set, even though all the labelled \textsc{CII} have been found by this threshold.}

\edit{An alternative approach to the comparison presents itself for this survey. We can set the probability thresholds to return the number of labelled Class II YSOs in the matched set, that is about 138 candidates. A $P_{\rm s}(\textsc{CII})>0.94$ on our catalogue returns 137 candidate \textsc{CII}. Coincidentally the same threshold of $P(\rm CII)>0.94$ on the \cite{2021AA...647A.116C} catalogue returns 138 candidates. These sets share 88 per cent of their candidate \textsc{CII} in common. The number of \cite{2021AA...647A.116C} labelled Class II YSOs missed by our classifier doubles from 9 to 18, and goes from 0 to 5 for the \cite{2021AA...647A.116C} candidates. The number of misidentified \textsc{CII} in the \cite{2021AA...647A.116C} catalogue reduces from 44 to 5. Redoing the comparisons for the other literature catalogues using a threshold of $P_{\rm s}(\textsc{CII})>0.94$ does not materially alter those comparisons.}

\subsubsection{\edittwo{Comparison with \protect\cite{2022arXiv220606416E,2022arXiv221117238R,2022gdr3.reptE..10R}}}
\label{sec:gaiadr3_yso_variables}

\begin{figure*}
	\includegraphics[width=1\textwidth]{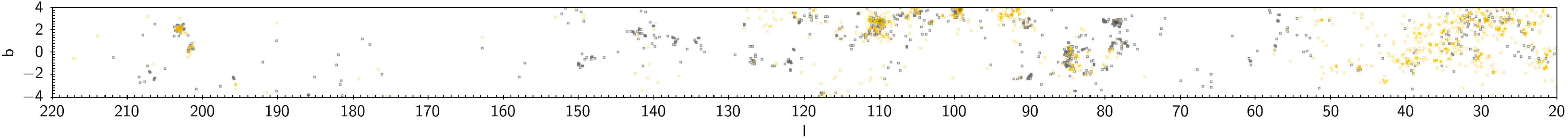}
	\includegraphics[width=1\textwidth]{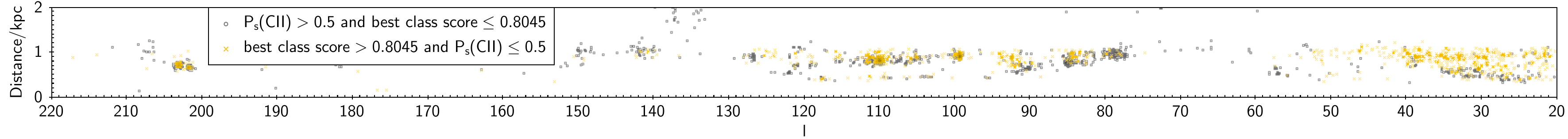}
	\caption{\edit{As Fig.~\ref{fig:Marton2019_l} but for the matches between our classifier and the \textit{Gaia} YSO candidates \protect\citep{2022arXiv220606416E,2022gdr3.reptE..10R}. Grey circles are our high confidence \textsc{CII} candidates not in the equivalent \textit{Gaia} high ranking YSO candidate set. Orange crosses are the \textit{Gaia} high ranking YSO candidates not in our candidate set.}}
	\label{fig:Eyer2022_l}
\end{figure*}

\edittwo{The variable source catalogue of \textit{Gaia} DR3 includes 79\,375 candidate YSOs \citep{2022arXiv221117238R}}. These variable sources were initially identified through the General Variability Detection path using an Extreme Gradient Boost classifier \citep{2022arXiv220606416E,2022gdr3.reptE..10R} applied to sources with over five field of view transits. A period search and Fourier modelling were then performed on these sources, followed by a set of Random Forest \citep{2001MachL..45....5B} and XGBoost \citep{2016arXiv160302754C} classifiers to determine the variability type \citep{2022gdr3.reptE..10R}. The results from the period search and modelling were input to the classifiers, along with other data including $G$, $G_{\rm BP}$ and $G_{\rm RP}$ time series, spectral shape information from the BP and RP spectro-photometers, and the parallax.

\edittwo{A contamination rate of <0.798 and a completeness of 0.091 were found by \citet[table 2]{2022arXiv221117238R}. The validation of the \textit{Gaia} DR3 YSO variable sources by \citet{2022arXiv220605796M} found a contamination rate <0.267 and a completeness rate of around a percent that varies significantly with evolutionary class.}

The \textit{Gaia} best class score is the combined median normalised ranks from up to 12 classifiers. To find the same number of \textit{Gaia} YSOs as \textsc{CII} from our $P_{\rm s}(\textsc{CII})>0.5$ set required a best class score greater than 0.8045, giving 1\,345 candidates in both sets. These sets are reasonably distinct, with 74 per cent of sources unique to each.

In order to establish whether the distinct subsets are likely to be genuine YSOs, we examine their spatial distribution in Fig.~\ref{fig:Eyer2022_l}. There are notable differences where we identify more candidates in the Cygnus-X region and the Gaia YSOs have more candidates to low Galactic longitude. However, both of these subsets show structure that closely matches known star forming regions including Cygnus, Cepheus, NGC 2264 and Mon R1. Hence, these subsets appear to be composed of genuine YSOs.

If both classifiers are working well, then we would expect a large proportion of our candidate \textsc{CII} to be \textit{Gaia} YSO candidates. Hence, that 74 per cent of our \textsc{CII} are not in the high ranking \textit{Gaia} YSOs might appear problematic. However, all of these sources are \textit{Gaia} YSO candidates, they are simply lower ranking with a best class score below 0.8045. Also, our \textsc{CII} should be a subset of the \textit{Gaia} YSOs as the \textit{Gaia} sample includes YSOs at other stages of evolution. Thus it is reasonable that our classifier does not consider 74 per cent of the high ranking \textit{Gaia} YSOs as \textsc{CII}.

\begin{figure}
	\includegraphics[width=\columnwidth]{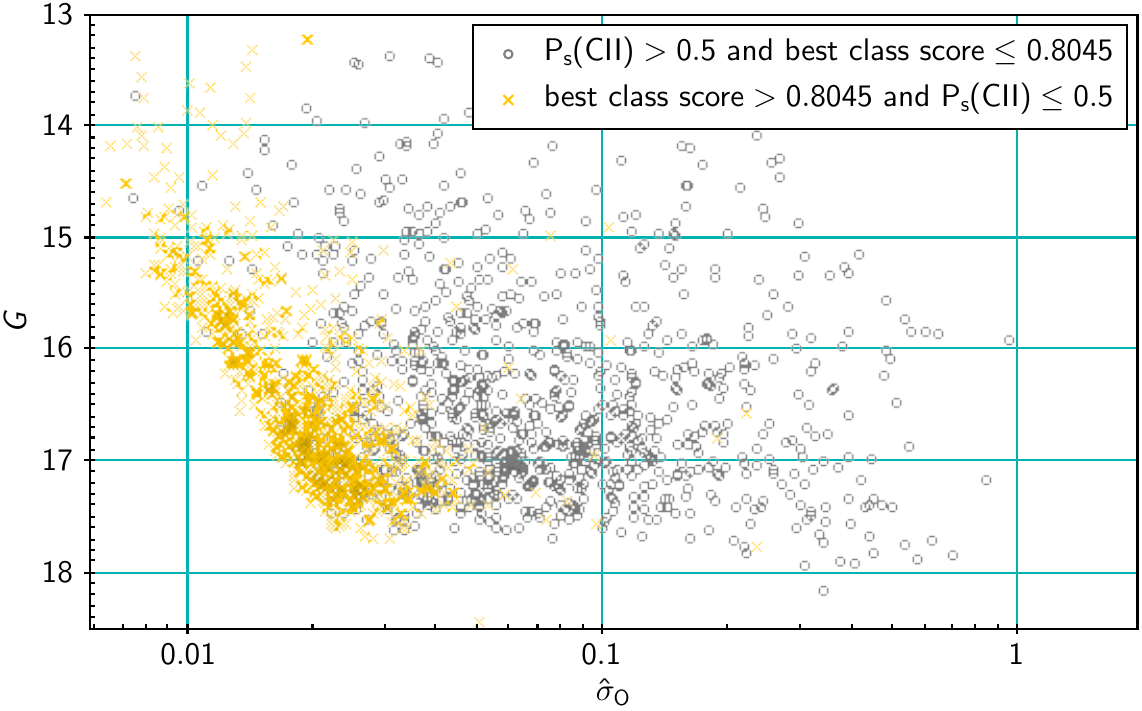}
	\caption{\edit{The \textit{Gaia} EDR3 $\hat{\sigma}_{\rm O}$ (fractional standard deviation of the flux) by $G$-band magnitude for the \textit{Gaia} DR3 variability candidate YSOs matched to our catalogue. The symbols have the same meaning as in Fig.~\ref{fig:Eyer2022_l}.}}
	\label{fig:Eyer2022_Gvar_G_vs_FSDF}
\end{figure}

\begin{figure}
	\includegraphics[width=\columnwidth]{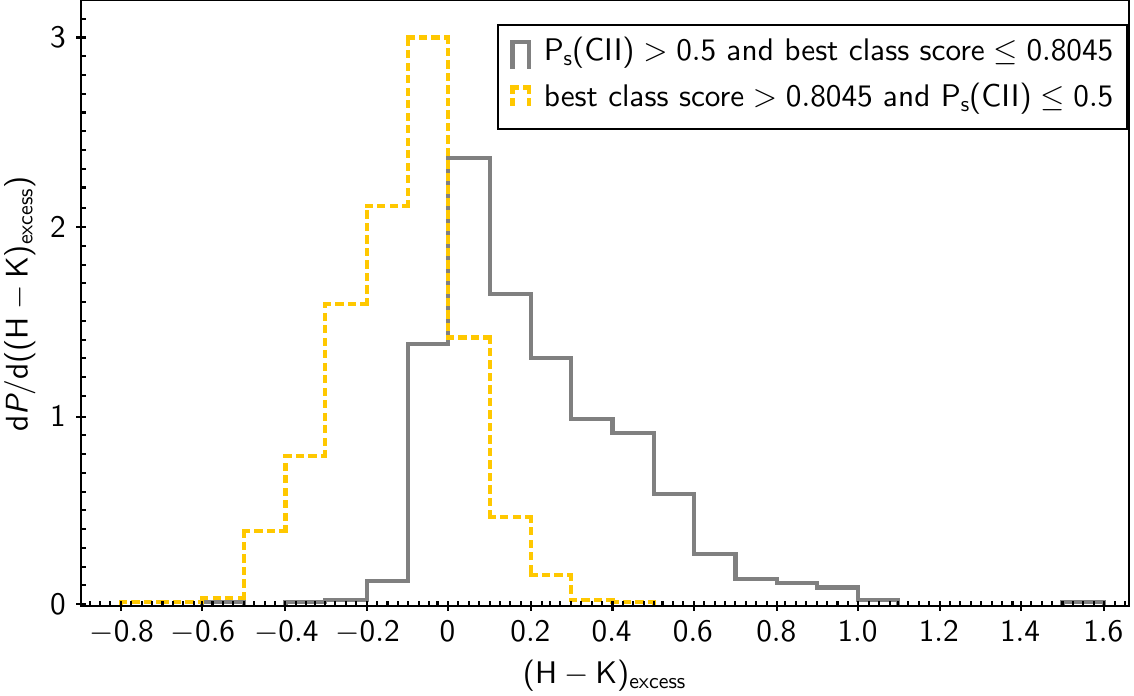}
	\caption{\edit{Histogram of the $(H-K)$ excess for the \textit{Gaia} YSOs matched to our catalogue. As UKIDSS and 2MASS have similar feature profiles in the training data they are shown together. The solid grey line is our high confidence \textsc{CII} candidates not in the equivalent \textit{Gaia} high ranking YSO candidate set. The orange dotted line is the \textit{Gaia} high ranking YSO candidates not in our \textsc{CII} candidate set.}}
	\label{fig:Eyer2022_HK_Hist}
\end{figure}

As our classifier provides feature information for a variety of stellar youth indicators, we can use them to investigate the properties of the distinct subsets. While this will simply confirm our \textsc{CII} classifications, the features can provide clues about the age of the \textit{Gaia} high ranking candidates not in our \textsc{CII} candidate list. Starting with our variability feature $\hat{\sigma}_{\rm O}$ in Fig.~\ref{fig:Eyer2022_Gvar_G_vs_FSDF}, the sources we consider high confidence \textsc{CII} that are lower ranking \textit{Gaia} YSOs show consistently higher $\hat{\sigma}_{\rm O}$ than the \textit{Gaia} high ranking YSO candidates not in our \textsc{CII} candidate set. This could be demonstrating that the \textit{Gaia} pipeline is able to discover lower variability YSOs than our basic approach. The \textit{WISE} $(W1-W2)$ mid-infrared, $(H-K)$ near-infrared and H$\alpha$ excess features all exhibit similar behaviour. We illustrate this with the $(H-K)$ histogram of Fig.~\ref{fig:Eyer2022_HK_Hist}. The high confidence \textsc{CII} from our classifier which are low ranking \textit{Gaia} YSOs show a strong infrared excess. This contrasts with the high ranking Gaia YSOs not in our \textsc{CII} set, where very few sources exhibit an infrared excess and where it is present it is weak. The isochronal age feature shows a slight tendency for our high confidence \textsc{CII} set to be more consistent with the young star age range than the high ranking \textit{Gaia} YSO candidates. However, it is weak, so we do not explore it in detail. Taking these results together, they confirm the high ranking \textit{Gaia} candidates not in our candidate list are unlikely to be younger than Class II, as we would expect an infrared excess and they would be difficult for Gaia to detect in the optical. They could be older YSOs or contaminants that are not young stars. As they cluster around known star forming regions this gives more credence to them being YSOs at a later stage of evolution, possibly Class III YSOs. \edittwo{A finding that is consistent with the conclusions of \citet{2022arXiv220605796M} that \textit{Gaia} is more sensitive to Class II/III YSOs.} This is an intriguing result as Class III YSOs are typically more difficult to identify.

\section{Potential improvements and known bias}
\label{sec:improvements_and_bias}

During the construction of the classifier, a variety of potential improvements presented themselves that we were unable to implement due to time and resource constraints. The catalogue we present is sufficient for testing the techniques, and carrying out initial science. In this Section we discuss these as potential future improvements that could enable additional science, and call out biases inherent to this work.

Our approach of constructing a 2-way classifier introduces a fundamental \edit{weakness}. Several types of object share the observational properties of Class II YSOs. These include young stars at different stages of evolution, as well very different objects such AGB stars \citep{Lee2021} and cataclysmic variables. Adding classifications for these types of object would improve the classifier by weakening the results for sources with feature properties common to a variety of object types. The effect would be to remove contaminants, making the classifier better at distinguishing genuine Class II YSOs. Additional classifications are introduced by training on the object types and a simple extension of the Bayes formula. The effort comes from compiling good quality training sets followed by training and optimisation. There is a cost in computer processing time, as likelihood ratios are needed for each pair of classes. The work to train the classifier grows as $(n^2-n)/2$ where $n$ is the number of classes. An alternative formulation is to construct a hierarchy of binary classifiers rather than a single n-way classifier. This would only require training on $2n-2$ likelihood pairs. So a hierarchical 3-way classifier would be constructed from two binary classifiers. For example, if we take the classes \textsc{CII}, \textsc{CIII} and \textsc{Other}. The first level binary classifier would classify into combined \textsc{CII/CIII} and \textsc{Other}. The second level would then distinguish \textsc{CII} from \textsc{CIII}.

Another area for potential improvement is our training/test sets. When creating the likelihoods we were frequently limited by the small number of \textsc{CII}s in our training set. \edit{Expanding our labelled \textsc{CII} set would allow finer structure and better accuracy in the likelihood ratios. The key limitations on the quantity of labelled \textsc{CII} were the quality cuts we imposed on the input data, along with the faint magnitude limit and the footprint we set for our catalogue.}

A critical factor in the classifier performance is the purity of the training sets. For the \textsc{Other} set we simply chose a bland region of the sky devoid of known star formation and significant reddening, our \textsc{Golden Rectangle}. This may have introduced a couple of problems. First, it is possible \edit{that} this data set is contaminated by a small quantity of Class II YSOs. Even a small level of contamination is a problem as we are trying to distinguish at a level of around one Class II YSO in a thousand sources. The second problem is this set may not be representative of the general population of other sources in our full survey region. As this region is bland with little extinction this means it is different to the bulk of the survey region, and this difference may introduce bias. Further, since it deliberately avoids other star forming regions, we should expect it to be devoid of young stars at all stages of evolution. For example, we have not trained our classifier to assign Class III YSO to the \textsc{Other} category. An alternative method for creating a training set of \textsc{Other} sources would be to take a random sample from the full data set, with known Class II YSOs removed. This introduces a different problem as we expect there to be unidentified Class II YSOs in the survey region.

A potential bias comes from the way we compiled our training set of known \textsc{CII}. We used data from a variety of papers providing catalogues of candidate Class II YSOs. Many of the source papers were based on the methods of \cite{Gutermuth2005,Gutermuth2008,Gutermuth2009a}. While we do not doubt their utility, this may have biased our classifier to find Class II YSOs matching the specific characteristics used by these methods.

\edit{Complete automation of the training would bring the benefits of fully reproducible likelihood ratios and a faster training process. For the four features where the likelihoods are based on histograms we manually tuned some of the bin edges to avoid unphysical results and large jumps in the likelihood ratio. These came about due to outliers and small quantities of training data in the critical regions where the likelihood ratio transitions from favouring \textsc{Other} to \textsc{CII}. A solution could be to use kernel density estimation to generate smoothed histograms of the training data. The kernel might need to vary with location to ensure we capture fine detail in the fast changing transition region without introducing unphysical fluctuations in the sparsely populated wings. The training of the variability feature is already semi-automated. We perform statistical fits of the training data to the model parameters, and then manually assess the best method to capture the behaviour of individual model parameters.}

The Isochronal Age feature, based on the overluminosity of young stars above the MS is the weakest of the features. \edit{Multiple star systems are known to be very common. For example \cite{Duchene2013} found multiplicity fraction from 22 per cent for low mass stars to 80 per cent for high mass stars.} However, for simplicity we only used isochrones for single stars and did not attempt to determine the multiplicity of sources. Unresolved binary and higher order multiple star systems will cause a source to have a higher luminosity than a single star of the same colour. Hence, we can expect a significant number of MS sources to lie above the MS line due to multiplicity rather than youth, and for some young stars to lie above the region occupied by single young stars. The selection criteria used to create our NGPn data set partially mitigated this problem, as only sources consistent with a single star model in the \textit{Gaia} EDR3 pipeline were included. This will have removed some but not all multiple star systems, those that are not resolved by Gaia but whose orbital motions degrade the parallax measurements. The problem is further mitigated as we can expect the training sets to contain multiple star systems. However, none of these deal with the underlying problem that only single star isochrones were used. An alternative approach might involve construction of multiple star isochrones weighted by the theoretical multiplicity by mass. Training would involve integrating theoretical probabilities to produce the likelihoods without use of the training data.

The \textit{WISE} photometry is affected by crowding in dense fields as the $W1$-band (3.4 $\micron$) and $W2$-band (4.6 $\micron$) have FWHM of 6.1$\arcsec$ and 6.2$\arcsec$ respectively \citep{Wright2010}. This is a particular problem at lower Galactic longitudes, resulting in fewer matches and a higher risk of contamination. The \textit{Spitzer} Space Telescope \citep{Werner2004} IRAC instrument has a FWHM of 1.6$\arcsec$ in similar band passes at 3.6 $\micron$ and 4.5 $\micron$ \citep{Fazio2004}. Hence, \textit{Spitzer} photometry is less affected by source confusion. However, unlike \textit{WISE}, \textit{Spitzer} did not observe the full sky and only covers patches of our NGPn footprint. We could potentially use \textit{Spitzer} with \textit{WISE} by taking a similar approach to our $(H-K)$ feature. We would use \textit{Spitzer} for a source where it is available, and otherwise \textit{WISE} where a reliable measurement exists.

The \textit{Gaia} archive provides uncertainties on the parallaxes and STILISM provides uncertainties on the reddening. There is the potential to weight feature results based on these uncertainties. This could be done by integrating across distance and reddening for each feature, weighted by the uncertainties. The computing time would be increased by the number of steps in the integration.

The construction of our NGPn catalogue requires that all sources have a detection in \textit{Gaia} EDR3 with a high quality parallax measurement and a $G_{\rm{RP}}<18$. \edit{This will bias our data selection}. First, young stars tend to emit strongly in the infrared with less in the optical where \textit{Gaia} is sensitive. There may be sources sufficiently bright in the infrared for our \textit{WISE} and $(H-K)$ features to give good results, but are excluded as they are faint in $G_{\rm{RP}}$. \edit{Second, the visibility in the optical will be affected by our viewing angle of the circumstellar disc. When a source is viewed edge on or the protostar is behind a flared disc, there will be a high level of optical extinction that could make it invisible to \textit{Gaia} or fainter than our $G_{\rm{RP}}<18$ limit. Hence, we should expect our input data and classifier to be biased towards Class II YSOs where the protostar is not obscured or only partially obscured by its disc. Third}, the parallax quality cut excludes many sources with good photometry. The flip side of this argument is the reliable parallax cut excludes galaxies and the majority of AGB stars, cleaning our master catalogue of potential young star interlopers.

By only selecting sources with good quality parallaxes from \textit{Gaia} EDR3, we are restricting the data to sources that are consistent with the single star model of the \textit{Gaia} pipeline. This will exclude multiple star systems where their orbital motion is detectable by \textit{Gaia} but they are not directly resolved into distinct sources. This selection is fundamental to our approach, so it is a bias we must accept. However, future \textit{Gaia} data releases will drop the single star assumption from the pipeline, and this may allow us to remove this bias.

We could extend our catalogue in Galactic longitude and latitude. The source surveys for our classifier provide data that would allow us to extend to $\pm$5 \degr in Galactic latitude. The survey could be extended in Galactic longitude by introducing the VPHAS+ survey \citep{Drew2014} the southern hemisphere equivalent of IGAPS, and VVV \citep{Lucas2008} where we are lacking UKDISS in the south. Extending to greater distance is more problematic. Beyond 1\,kpc the \textit{Gaia} parallax measurements start to be dominated by their uncertainties, and by our 2\,kpc limit this becomes a significant issue. Future \textit{Gaia} releases may allow us to push the limit further, and will otherwise improve our results to 2\,kpc. \edit{Any increases to the size of our catalogue would allow us to consider expansion of our training sets.} This would be of particular benefit to our \textsc{CII} training set since its small size limits the quality of the likelihood ratios.

It is mathematically straightforward to incorporate additional features to the naive Bayes classifier. The effort is taken with analysing features to determine what produces a good discriminator of \textsc{CII}, then training and optimisation. Examples of other known features of young stars are UV excess and X-ray emission.

\section{Summary and conclusions}

We created a naive Bayes classifier for identifying candidate Class II YSOs. In Section~\ref{sec:features} we derived five features from known observational properties of Class II YSOs: \textit{Gaia} EDR3 $G$-band variability, \textit{WISE} mid-infrared excess from $(W1-W2)$, near-infrared excess from $(H-K)$ using UKIDSS and 2MASS photometry, $H\alpha$ excess and Isochronal Age both using IGAPS photometry. In Section~\ref{sec:naive_bayes_classifier} we constructed the classifier based on these features.

The classifier was applied to the NGPn (Northern Galactic Plane) region we define $20<l<220$ and $|b|<4$ to a distance within 2\,kpc from inversion of \textit{Gaia} EDR3 parallaxes. In compiling the catalogue we prioritised data quality at the expense of completeness. This gives us confidence that the results reflect the performance of the classifier without being significantly influenced by poor quality data. As naive Bayes classifiers can cope with missing data, only a single feature is required to generate a result. Though the more features, the more robust the posterior. The posteriors are not true probabilities as we must rely on an estimated prior. We chose a representative prior of one Class II YSO in a thousand sources. Other values of the prior would lead to different posteriors and alter the scaling, though the order of results by posterior would always be preserved as long as a constant prior was used.

When using the catalogue, a posterior threshold for candidate Class II YSOs must be carefully chosen to match the task at hand. We chose $P_{\rm s}(\textsc{CII})>0.5$ to provide a data set of high purity Class II YSOs for analysis \edit{(see Section~\ref{sec:confusion})}. This classified 6\,504 sources as candidate \textsc{CII} with a false positive rate around 0.02 per cent and a true positive rate of approximately 87 per cent. In Section~\ref{sec:roc} we used a ROC curve to analyse the performance of the classifier across the full range of posteriors. The curve rises rapidly to a value close to one, with area under the ROC curve of around 0.998 or better.

In Section~\ref{sec:identified_star_forming_regions} we reviewed the $P_{\rm s}(\textsc{CII})>0.5$ set qualitatively by examining its distribution in Galactic coordinates and Galactic longitude with distance from the \textit{Gaia} parallax. This showed the high posterior \textsc{CII} sources are concentrated in regions of known star formation. For example significant groupings were found to coincide with Cep OB3b, the Rosette Nebula, NGC 2264, the Cygnus star forming complex, as well as many other locations. We also identified what may be three previously undiscovered young clusters or associations.

In Section~\ref{sec:comparisons_to_other_catalogues} we used the $P_{\rm s}(\textsc{CII})>0.5$ data set to assess the performance of our classifier against \edit{eight} other classifiers of young stars. The common theme across these comparisons was \edit{that} the matched sources shared a high probability young star classification at around one \edit{quarter to three quarters} of the time. The corollary to this finding is between one \edit{quarter and three quarters} of candidates are unique in these binary comparisons between classifiers. Examining the spatial distribution of these sources with opposing classifications showed they tend to coincide with known star forming regions. Thus a significant proportion of these candidates are likely to be genuine YSOs. This is telling us that none of the reviewed classifiers, including our classifier, is finding all of the young stars even when they assess the same sources.

The degree of agreement between our \textsc{CII} candidates and catalogues of candidate YSOs at a broader range of ages indicates our classifier is finding YSOs at other stages of evolution. The comparison to \cite{Kuhn2021} in Section~\ref{sec:kuhn2021_comparison} illustrates this well as their candidates are labelled by YSO evolutionary stage. While we match more Class II candidates from \cite{Kuhn2021} than other stages of evolution, we match the same proportion in the neighbouring evolutionary stages of Flat SED and Class III YSOs, with a lower proportion of the younger Class I sources. Such a result is not entirely unexpected, as Flat SED and Class III YSOs share some features with Class II YSOs. Further, as we do not explicitly train on other stages of YSO evolution, this may have given our classifier a tendency to classify these neighbouring evolutionary stages as weaker \textsc{CII} candidates rather than classifying them as \textsc{Other} sources.

In the coming years, spectra will be obtained for many of our candidate Class II YSOs using the WEAVE spectrograph on the WHT \citep{2012SPIE.8446E..0PD}, \editthree{acquired for the SCIP (Stellar, Circumstellar and Interstellar Physics) Science Team \citep{2022arXiv221203981J}}. This will enable a thorough assessment of the classifier results.

\section*{Acknowledgements}

\edit{The authors thank the anonymous \editthree{referees} for their constructive feedback that has improved this manuscript.} \editfour{The WISE cross matching was supported by Science and Technology Facilities Council (STFC) funding for UK participation in the Vera C. Rubin observatory (previously referred to as the Large Synoptic Survey Telescope, LSST), through grant ST/S 006117/1 supporting Tom J. Wilson and Tim Naylor. Ben Lakeland is supported by STFC studentship ST/V506679/1 and Tim Naylor for part of this work by a Leverhulme Research Project Grant.}

This work has made use of data from the European Space Agency (ESA) mission {\it Gaia} (\url{https://www.cosmos.esa.int/gaia}), processed by the {\it Gaia} Data Processing and Analysis Consortium (DPAC, \url{https://www.cosmos.esa.int/web/gaia/dpac/consortium}). Funding for the DPAC has been provided by national institutions, in particular the institutions participating in the {\it Gaia} Multilateral Agreement.

This publication makes use of data products from the Wide-field Infrared Survey Explorer, which is a joint project of the University of California, Los Angeles, and the Jet Propulsion Laboratory/California Institute of Technology, and NEOWISE, which is a project of the Jet Propulsion Laboratory/California Institute of Technology. \textit{WISE} and NEOWISE are funded by the National Aeronautics and Space Administration.

\editthree{This paper makes use of data obtained as part of the IGAPS merger of the IPHAS and UVEX surveys (\url{www.igapsimages.org}) carried out at the Isaac Newton Telescope (INT).} The INT is operated on the island of La Palma by the Isaac Newton Group in the Spanish Observatorio del Roque de los Muchachos of the Instituto de Astrofisica de Canarias. All IGAPS data were processed by the Cambridge Astronomical Survey Unit, at the Institute of Astronomy in Cambridge. The uniformly-calibrated bandmerged IGAPS catalogue was assembled using the high performance computing cluster via the Centre for Astrophysics Research, University of Hertfordshire.

This publication makes use of data products from the Two Micron All Sky Survey, which is a joint project of the University of Massachusetts and the Infrared Processing and Analysis Center/California Institute of Technology, funded by the National Aeronautics and Space Administration and the National Science Foundation.

The UKIDSS project is defined in \cite{Lawrence2007}. UKIDSS uses the UKIRT Wide Field Camera (WFCAM; \cite{Casali2007}) and a photometric system described in \cite{Hewett2006}. The pipeline processing and science archive are described in \cite{Hambly2008}. We have used data from UKIDSS DR11 plus, which is described in detail in \cite{Lucas2008}.

\edittwo{This publication makes use of the STILISM tridimensional dust maps described in \cite{Lallement2014,Capitanio2017,2018A&A...616A.132L}.}

The Pan-STARRS1 Surveys (PS1) and the PS1 public science archive have been made possible through contributions by the Institute for Astronomy, the University of Hawaii, the Pan-STARRS Project Office, the Max-Planck Society and its participating institutes, the Max Planck Institute for Astronomy, Heidelberg and the Max Planck Institute for Extraterrestrial Physics, Garching, The Johns Hopkins University, Durham University, the University of Edinburgh, the Queen's University Belfast, the Harvard-Smithsonian Center for Astrophysics, the Las Cumbres Observatory Global Telescope Network Incorporated, the National Central University of Taiwan, the Space Telescope Science Institute, the National Aeronautics and Space Administration under Grant No. NNX08AR22G issued through the Planetary Science Division of the NASA Science Mission Directorate, the National Science Foundation Grant No. AST-1238877, the University of Maryland, Eotvos Lorand University (ELTE), the Los Alamos National Laboratory, and the Gordon and Betty Moore Foundation.

This research made use of \textsc{astropy},\footnote{\url{http://www.astropy.org}} a community-developed core \textsc{python} package for Astronomy \citep{2013A&A...558A..33A, 2018AJ....156..123A}, \textsc{numpy} \citep{harris2020array}, TOPCAT \citep{Taylor2005} for analysis, and the SIMBAD database \citep{Wenger2000}, operated at CDS, Strasbourg, France.

%%%%%%%%%%%%%%%%%%%%%%%%%%%%%%%%%%%%%%%%%%%%%%%%%%
\section*{Data Availability}
\label{sec:data_availability}

\editthree{The catalogue of results is available at CDS via anonymous ftp to cdsarc.u-strasbg.fr (130.79.128.5) or via \url{http://cdsarc.unistra.fr/viz-bin/cat/VII/293}. A fits copy of the catalogue is available at \url{https://www.nyxanalytics.co.uk/catalogues.php}. We intend to update the catalogue if better versions become available. The \textsc{python} code is available on GitHub at \url{https://github.com/AndyJWil/bayescii}.}

\editthree{The input data were extracted from public archives. The Gaia EDR3 data was downloaded from the Gaia ESA Archive (\url{https://gea.esac.esa.int/archive/}). The ALLWISE, IGAPS, 2MASS and PanSTARRS DR1 data were downloaded from VizieR (\url{https://vizier.cds.unistra.fr/viz-bin/VizieR}) at the respective catalogue numbers II/328/allwise, V/165/igapsdr1, II/246/out and II/349/ps1. The UKIDSS DR11 Plus data was downloaded from the WFCAM Science Archive (\url{http://wsa.roe.ac.uk}). The Stilism dust map was accessed from \url{https://stilism.obspm.fr}.}

\editthree{The model atmospheres and interiors are available at \url{http://svo2.cab.inta-csic.es/theory/main/}, \url{http://stellar.dartmouth.edu/models/index.html} and \url{https://www.astro.ex.ac.uk/people/timn/isochrones/}.}



\bibliographystyle{mnras}
\bibliography{library_mnras} % if your bibtex file is called example.bib

%\begin{thebibliography}{99}
%\end{thebibliography}

%%%%%%%%%%%%%%%%%%%%%%%%%%%%%%%%%%%%%%%%%%%%%%%%%%

%%%%%%%%%%%%%%%%% APPENDICES %%%%%%%%%%%%%%%%%%%%%

\appendix

\section{Survey selection and construction of the base data set}
\label{sec:surveys}

A collection of surveys were identified to provide data for the features. Each survey provides good coverage of our NGPn footprint, defined $20<l<220$ and $|b|<4$. When selecting data from the surveys, quality was prioritised over completeness.

\subsection{Base data set}
\label{sec:base_data_set}

Rather than treating each survey in isolation, it was decided to define a base data set using \textit{Gaia} Early Data Release 3 \citep{2016A&A...595A...1G,2021A&A...649A...1G}, chosen for its high quality astrometry. Data from the other surveys are only retained where there exists a robust match to this base data set. Within the NGPn survey region \textit{Gaia} EDR3 contains 190\,016\,855 sources, though only 8\,080\,045 survive the quality cuts to make it into our base data set.

To be able to test the catalogue we need to be able to obtain good quality spectra, hence we imposed a magnitude limit of $r<18$. To a first approximation we implemented this via $G_{\rm{RP}}<18.0$ (\textit{Gaia} red prism photometer) as we can expect the majority of these sources to be brighter than $r=18$. This was useful in keeping data volumes to a size that was straight forward to process. A precise cut using PanSTARRS $r<18.0$ was imposed later in our pipeline.

Good quality parallax measurements were important for data selection and several features. In \textit{Gaia} EDR3 only sources with a 5 or 6-parameter astrometric solution have a reported parallax measurement \citep{2021A&A...649A...1G}. The 5-parameter solutions were derived using the effective wave number to incorporate chromaticism in the \edit{point spread function (PSF) and line spread function (LSF)}, while the 6-parameters solutions were derived using a default colour as the measured colour was of low quality. Thus only sources with 5 or 6-parameter solutions were selected for the base data set.

The \textit{Gaia} EDR3 processing pipeline assumes all sources are single stars \citep{Lindegren2021a}. Thus multiple star systems with small angular separations can give rise to spurious astrometry. The RUWE (renormalised unit weight error) \citep{GAIA-C3-TN-LU-LL-124-01} indicates whether a source is a good fit to the single-star model. We impose \edit{RUWE} $\leq 1.4$ \citep{GAIA-C3-TN-LU-LL-124-01,Fabricius2021} to ensure a good fit to the single star model.

The combined \edit{RUWE} $\leq 1.4$, $G_{\rm{RP}}<18.0$, and 5 or 6 astrometric parameters criteria reduced the number of sources to 53\,872\,361 within the NGPn survey region.

\subsubsection{Gaia EDR3 parallaxes}
\label{sec:gaia_parallaxes}

The \textit{Gaia} parallaxes are a significant improvement upon anything previously available. To maximize their potential we remove the best estimate of the bias from the parallaxes, and apply a correction factor to the uncertainties.

\cite{Lindegren2021b} found a parallax bias of a few tens of $\mu$as dependent on $G$, ecliptic latitude and effective wavelength for the 5-parameter astrometric solutions, and pseudocolour in place of effective wavelength for the 6-parameter solutions. This parallax bias was calculated using the \textsc{python} script provided on the \textit{Gaia} EDR3 website, and subsequently removed from the parallax measurements in this work.

Analysis by \cite{Fabricius2021} of \editfour{quasars (QSOs), clusters and Large Magellanic Cloud (LMC)} parallaxes indicate the formal parallax uncertainties in \textit{Gaia} EDR3 are slightly underestimated, defining the external error
\begin{equation}
	\sigma_{\rm{ext}} = \sqrt{ k^2 \sigma_{\rm \varpi}^2 + \sigma_{\rm s}^2 }
	\label{eq:adjusted_parallax_error}
\end{equation}
where the $k$ factor (unit-weight uncertainty) adjusts the formal uncertainties for their underestimation, $\sigma_{\rm \varpi}$ is the formal \textit{Gaia} error, and $\sigma_{\rm s}$ is the systematic uncertainty. We adopt their findings of an approximate $k$ factor of 1.05 for 5-parameter astrometric solutions, 1.22 for 6-parameter solutions, and a zero systematic uncertainty.

Giant stars and distant galaxies contaminate catalogues of YSOs as they can exhibit similar photometric properties. The accuracy of the \textit{Gaia} parallax measurements makes possible a new approach to cleaning out these contaminants. Extragalactic sources should have a parallax close to 0 mas. To select sources likely to lie within our Galaxy to a confidence level of 3 sigma we require
\begin{equation}
	\frac{\varpi}{\sigma_{\rm{ext}}} > 3.
	\label{eq:parallax_nonzero_criteria}
\end{equation}
To remove the majority of giant stars we select nearby sources with a parallax consistent with a distance within 2\,kpc
\begin{equation}
	\varpi > 0.5\,\rm{mas}.
	\label{eq:parallax_2kpc_criteria}
\end{equation}
While these parallax selections introduce bias, it is more important for our catalogue to be based on reliable nearby sources than a more complete and unbiased data set. These parallax selections reduce the number of sources within the NGPn survey region to 13\,358\,778.

\subsubsection{Faint magnitude limit}
\label{sec:faint_limit}

The base data set was matched to PanSTARRS DR1 \citep{Chambers2016,Magnier2016} using the official \textit{Gaia} EDR3 match table, yielding 8\,088\,045 sources with r<18. This constituted the NGPn base data set on which the classifier was run and training sets extracted.

\subsubsection{Check on spurious parallax measurements}
\label{sec:parallax_check}

\cite{Fabricius2021} defines spurious parallax measurements as close unresolved doubles or crowding, where the scan angle can result in a mix of single component or photocentre astrometric measurements. Where these are large, they can equally give rise to a spurious positive or negative parallax. They suggest a useful way to estimate the number of spurious parallax solutions is to examine an equivalent data set of negative parallaxes. We did this by imposing
\begin{equation}
	\frac{\varpi}{\sigma_{\rm{ext}}} < -3
	\label{eq:parallax_nonzero_criteria_neg}
\end{equation}
and
\begin{equation}
	\varpi < -0.5\,\rm{mas}
	\label{eq:parallax_2kpc_criteria_neg}
\end{equation}
along with our other base data set criteria and matching to PanSTARRS with the $r<18$. This gave 1\,989 sources, just 0.02 per cent the size of the positive parallax sample, giving high confidence in the lack of spurious parallaxes in the base data set.

\subsection{Gaia EDR3 photometry}
\label{sec:gaia_photometry}

Photometry from the \textit{Gaia} EDR3 $G$-band was used for the intrinsic variability feature. The $G$-band photometry is of a very high quality with uncertainties less than 0.01 magnitudes, so no quality cut was imposed. There are small known issues with some $G$-band magnitudes and fluxes, identified by \cite{Riello2021,Fabricius2021}. Of relevance to this work is a systematic error with the photometry associated with the 6-parameter astrometric solution, due to the default colour term used in place of the effective wavelength. The problem is window class and colour dependent, with larger corrections for redder sources and attaining values above 1 per cent for the reddest sources. We applied the official corrections of \cite{Riello2021} to the $G$-band magnitudes and fluxes using the \textsc{python} code provided on the \textit{Gaia} consortium website. This adjusted the photometry of 6 per cent of the NGPn base data set, with 1 per cent of the corrections above 0.01 magnitudes and a maximum correction of 0.026.

To ensure robust results from the variability feature, it was necessary to consider the total number of observations and the distribution of the observations in time. To avoid sources with a small number of observations, we followed the approach of \cite{Deason2017,Vioque2018} by requiring a minimum of 70 $G$-band observations for inclusion of the variability feature. The number of visibility periods is useful in gauging the temporal distribution of observations, where a visibility period is defined as a set of one or more measurements within 4 days. As all sources in the NGPn data set include parallax, and all parallaxes in EDR3 are based on at least 9 visibility periods. This ensured good temporal distribution of the measurements.

\subsection{WISE}
\label{sec:wise_survey}

Mid-infrared data was obtained from the Wide-field Infrared Survey Explorer (\textit{WISE}) space observatory \citep{Wright2010} ALLWISE source catalogue in the $W1$ (3.4 $\micron$) and $W2$ (4.6 $\micron$) pass bands. This combines single exposures from the 4-Band and 3-Band cryogenic phases with the NEOWISE \citep{Mainzer2011} post-cryogenic phase.

Quality limits were imposed by requiring magnitude uncertainties were below 0.1 magnitudes, and the contamination and confusion flags be zero to ensure sources were unaffected by artefacts such as diffraction spikes.

The data were matched to \textit{Gaia} EDR3 using an updated version of the cross-match table from \cite{Wilson2018a}. Good quality matches were selected by imposing a match probability greater than 50 per cent without a photometric match, as this could bias against young stars since they can be extremely red. This gave 2\,820\,933 matches to the NGPn base data set, about a 35 per cent match rate.

\subsection{IGAPS}
\label{sec:igaps}

IGAPS \citep{Monguio2020} is the merged IPHAS and UVEX surveys. The IPHAS (INT H$\alpha$ Survey of the Northern Galactic Plane) survey \citep{Drew2005} covers Galactic longitude from 30$^\circ$ to 215$^\circ$ and latitude -5$^\circ$ to +5$^\circ$. This encompasses the majority of the NGPn footprint, missing 10$^\circ$ at small longitude and 5$^\circ$ at large longitude.

The classifier makes use of the $r$, $i$ and H$\alpha$ AB photometry (we refer to IGAPS $r_{\rm I}$ as $r$ as we do not use $r_{\rm U}$). Quality criteria were applied to each filter, selecting only sources where the image was consistent with a stellar source ($Class=-1$) and good quality photometry by the saturation, vignetting, trailing, truncation and bad pixel flags all being false. We accepted deblended sources after our analysis showed the photometry was consistent between IGAPS and PanSTARRS DR1.

The IGAPS AB magnitudes were translated from INT photometry into the Pan-STARRS system by the introduction of colour terms \citep{Monguio2020}. We require photometry in the natural INT system calibrated by \cite{Bell2014}. We therefore undid the main effect of the translation by backing out the colour term from \citet[equation 5]{Monguio2020} on $i_{\rm AB}$
\begin{equation}
	i_{\rm INT} = i_{\rm AB} - 0.06 \times \left( r_{\rm AB} - i_{\rm AB} \right) .
	\label{eq:igaps_i_ab_int}
\end{equation}
The H$\alpha$ and $r_{\rm AB}$ have no colour terms in the \cite{Monguio2020} calibrations.

IGAPS was matched to the base data set using a 0.5$\arcsec$ radius in \textsc{TOPCAT}. This gave 6\,246\,318 matches, about 77 per cent of the NGPn base data set, with an estimated false match rate around 0.3 per cent.

\subsection{JHK photometry}
\label{sec:jhk}

Near-infrared JHK photometry was obtained from UKIDSS and 2MASS, with UKIDSS used in preference to 2MASS as it is of higher quality. 2MASS was used where no matching UKIDSS source was found, or UKIDSS was saturated as 2MASS remains reliable for brighter sources. The method of \cite{Lucas2008} was used to determine whether UKIDSS was saturated using the conditions $J < 13.25$, $H < 12.75$ and $K < 12.0$. Where it was available the 2MASS photometry was used to make this saturation determination since the UKIDSS would by definition be suspect for these sources. If no 2MASS photometry was available for saturated sources then an $(H-K)$ excess was not calculated.

\subsubsection{UKIDSS}
\label{sec:ukidss}

The UKIDSS GPS \citep[UKIRT Infrared Deep Sky Survey Galactic Plane Survey]{Lucas2008} provides data covering a large part of our NGPn footprint. There is a gap at approximately $107<l<141$ as this region of the Galactic plane lies above declination +60 \degr. The UKIRT design will not allow observation above this declination at the latitude of the observatory. Hence, only 2MASS observations were used for this region.

We used the UKIDSS DR11 Plus data set as our source. As we were prioritising reliability over completeness, we used the reliableGpsPointSource view designed for this purpose. This view ensures a seamless selection via the criteria priOrSec <= 0 or priOrSec = frameSetID. It contains high quality detections while allowing deblending, via ppErrBits between 0 and 31 in the $J$, $H$ and $K1$ filters. Only point source detections in all filters are included, by Class = -1. The positional tolerance between the detections in the 3 filters is assured to 0.5$\arcsec$ or better as Xi and Eta are limited to between -0.5 and +0.5.

As the data were used in $(J-H,H-K)$ diagrams, the uncertainties in the $J-H$ and $H-K$ colours were restricted to less than 0.1 magnitudes.

These criteria gave an initial data set containing 121\,494\,306 sources across the footprint. Matching these to the NGPn base data set using a 0.29$\arcsec$ match radius in \textsc{TOPCAT} gave 5\,510\,250 matches, about 68 per cent of the base data set with an estimated false match rate of 0.3 per cent.

\subsubsection{2MASS}
\label{sec:2mass}

The Two Micron All Sky Survey (2MASS) \citep{Skrutskie2006} provides $J$, $H$ and $K$ photometry to the classifier with full coverage of our NGPn footprint. To be consistent with the UKIDSS quality cuts of photometry to better than 0.1 magnitudes in $J-H$ and $H-K$, uncertainties in the 2MASS individual filters were limited to less than 0.05 magnitudes. Additionally a photometric quality flag of \lq A\rq\ (scan signal to noise ratio $\geq 10$ and the measurement errors $\leq 0.10857$) was required in all filters. Only sources where the contamination and confusion flag was zero were included, ensuring sources were unaffected by known artefacts.

These criteria yielded 11\,453\,173 sources across the NGPn footprint. Matching to the base data set with a 1.0$\arcsec$ radius in \textsc{TOPCAT} gave 2\,830\,365 sources, about 35 per cent of the data with an estimated 0.6 per cent false match rate.

\section{Interstellar extinction}
\label{sec:intersetllar_extinction}

\subsection{STILISM}
\label{sec:stilism}

An estimate of the reddening due to interstellar extinction is required for several of the classifier features. \edittwo{This work uses the STILISM 3-dimensional extinction map \citep{Lallement2014,Capitanio2017,2018A&A...616A.132L}.} This provides reddening estimates in $E(B-V)$ for the local neighbourhood out to 2\,kpc using parallaxes from \textit{Gaia} DR1. Where features needed reddening values, the $E(B-V)$ value was transformed into the relevant photometric system, as described in subsequent subsections.

We determined the approximate reddening to voxels in Galactic coordinates and distance, with the same reddening value applied to all the sources within a voxel. The voxels were a half a degree to a side in Galactic coordinates with a depth of 5\,pc. STILISM provided the $E(B-V)$ to the centre of the near and far sides of the voxels. The $E(B-V)$ of the voxel was set to the average of these two values.

\subsection{Suspect reddening values}
\label{sec:suspect_reddening}

A major contributor to the uncertainty in the reddening comes from the uncertainty on the \textit{Gaia} EDR3 parallaxes. As per equation~(\ref{eq:parallax_nonzero_criteria}), we require the parallax is greater than three times the parallax uncertainty. While this removes the lowest quality parallax measurements, it leaves many sources with a distance uncertainty that is a significant proportion of the distance estimate from the inverted parallax. Hence, even though STILISM provides good quality extinction estimates, some of our reddening values may have a large error.

A method of identifying suspect reddening values presented itself when we reviewed early iterations of the classifier results. We identified regions with concentrations of high posterior \textsc{CII} \edit{for comparing} against the locations of known star forming regions. A rectangular region bounded by $75.0<l<76.0$ and $-1.0<b<-0.5$ appeared anomalous as it was defined by straight edges and right angles. The majority of the higher $P(\textsc{CII})$ sources in this region had a very high $(r-H\alpha)$ excess likelihood ratio favouring \textsc{CII}. At this location there is a steep rise in the reddening by over 1.5 magnitudes in $E(B-V)$ at about 1.35\,kpc indicating a wall of dust. Small changes in the measured parallax of sources at this distance would place them in front or behind the dust, giving rise to large changes in reddening. The proportion of sources in our NGPn base data set with significant uncertainty in their parallax increases with distance. Closer than 1\,kpc only 4 per cent of sources have a relative parallax uncertainty greater than 10 per cent, compared with 45 per cent of sources at 1\,kpc and beyond. Hence, we should expect a significant number of sources to have a parallax measurement that places them on the wrong side of this wall. If a source is physically located in front of the dust but the parallax measurement places them behind the dust then this would cause the dereddening process to make them too blue. For the $(r-H\alpha)$ excess feature, this over dereddening would place sources on the blue side of empirical MS, giving the appearance of an ${r-H\alpha}$ excess.

The Isochronal Age feature gives us a robust way to identify sources which have been significantly over dereddened. The unreddened young stellar isochrone have a blueward $(r-i)$ limit of about -0.3. Hence, we consider sources with an $(r-i) \leq -0.3$ to have suspect reddening. The main drawback with this approach is it only works for sources where we have IGAPS photometry. This resulted in 1\,355\,639 sources with an $(H-K)$ excess where we were unable to check for suspect reddening.

The features affected by bad reddening are $(r-H\alpha)$ excess, Isochronal Age and $(H-K)$ excess. For each of these features, we calculate the feature value for all sources, but assign likelihood values of unity to all classes for sources with suspect reddening. This has the effect of removing the feature from the Bayes calculation.

\subsection{Dereddening UKIDSS and 2MASS J, H and K}
\label{sec:dereddening_jhk}

Before calculating an $(H-K)$ excess it was necessary to deredden the observations in $(J-H)$ and $(H-K)$. Transformations from $E(B-V)$ to $E(J-H)$ and $E(H-K)$ were derived by plotting the \edit{differences between an $E(B-V)$ of zero and one using 1\,Myr isochrones (see Appendix~\ref{sec:isochrones})}. The differences between different temperatures and gravities covered just over a thousandth in $E(J-H)$ and $E(H-K)$, so the linear transformations were simply taken as the approximate centre of these plots. For UKIDSS
\begin{equation}
	E(J-H) = 0.2747 \times E(B-V)
	\label{eq:JH_UKIDSS_trans}
\end{equation}
and
\begin{equation}
	E(H-K) = 0.1815 \times E(B-V).
	\label{eq:HK_UKIDSS_trans}
\end{equation}
For 2MASS
\begin{equation}
	E(J-H) = 0.2963 \times E(B-V)
	\label{eq:JH_2MASS_trans}
\end{equation}
and
\begin{equation}
	E(H-K) = 0.1668 \times E(B-V).
	\label{eq:HK_2MASS_trans}
\end{equation}

\subsection{Dereddening \edit{the} IGAPS \edit{H-$\alpha$ feature}}
\label{sec:dereddening_igaps}

\edit{To explore the extinction in the range $0 \leq E(B-V) \leq 6$, we folded model atmospheres through filter responses using the process described in \cite{Bell2014} and Appendix \ref{sec:isochrones} with the \citet{2007ApJ...663..320F} extinction curve.}
This showed that below 4\,000\,K the extinction in $r-H\alpha$ varies steeply with temperature, and there is no good, single number which can be used.
However, the extinction in $r-H\alpha$ is small, at about 0.1 mag per magnitude in $E(B-V)$, and so the answer is not too critical.
We therefore fitted the extinction for $T_{\rm eff}=10\,000$\,K and $\log(g)$=4.5 with a quadratic \edit{function}, which gave
\begin{equation}
	A(r-H\alpha) = 0.0065 + 0.2448 \times E(B-V) - 0.0224 \times E(B-V)^2.
	\label{eq:EBV_to_ArHa_trans}
\end{equation}
We chose 10\,000\,K as this is where nominal $E(B-V)$ is defined \citep[see][]{Bell2014}.

The extinction in $r-i$ has a much weaker dependency on temperature, and so we fitted the average of the extinctions for 10\,000\,K and 3\,500\,K and $\log(g)$=4.5 to give 
\begin{equation}
	A(r-i) = 0.0048 + 0.7333 \times E(B-V) - 0.0096 \times E(B-V)^2.
	\label{eq:EBV_to_Ari_trans}
\end{equation}

\section{Model star isochrones}
\label{sec:isochrones}

We used model isochrones to create the main-sequence in the $J$-$H$ vs $H$-$K$ CCD (Section \ref{sec:hk_feature}) and to obtain stellar ages from the $r$ vs $r$-$i$ CMD (Section \ref{sec:isochronal_age_feature}).  For $J$-$H$ vs $H$-$K$ we used BT-Settl AGSS model atmospheres \citep{2009ARA&A..47..481A,Allard2011} with \citet{Baraffe2015} interiors, folded through the UKIDSS and 2MASS filter bands.  The situation is more complicated for $r$ vs $r$-$i$ because standard isochrones over-estimate the temperature of M-dwarfs for a given luminosity (e.g. \citet{Jackson2018,2019MNRAS.489.2615M}). We therefore used the isochrones of \citet{Bell2014} which have an empirical correction for this calculated in the same system as we correct IGAPS to (see Appendix \ref{sec:igaps}).  These isochrones are based on the \citet{Dotter2008a} z = 0.018 interiors combined with BT-Settl AGSS model atmospheres tuned by \citet{Bell2014}. \editfive{We reddened the isochrones using the \citet{2007ApJ...663..320F} law.}

\section{Variability model}
\label{sec:variability_model}

The variability model for $\hat{\sigma}_{\rm O}$ was composed from a $\chi^2$ distribution for the instrumental noise, an exponential function for the variable stars and a $\delta$-function for the non-variable stars.
For a star of constant luminosity, the standard deviation of the observed flux $F$ would solely depend on the measurement scatter due to the instrument (including photon noise)
\begin{equation}
	\sigma_{\rm O}^2 = \langle (\bar{F} - F_i)^2 \rangle.
	\label{eq:sigma_flux}
\end{equation}
The generic relation between the fractional standard deviation and the standard deviation is defined
\begin{equation}
	\hat{\sigma}_{\rm X} = \frac{\sigma_{\rm X}}{\bar{F}} ,
	\label{eq:sigma_varamp}
\end{equation}
where the subscript $\rm X$ may take values of $\rm O$ for observed, $\rm I$ for instrumental, or $\rm A$ for astrophysical. Hence the observed standard deviation can be replaced with
\begin{equation}
	\bar{F}^2 \hat{\sigma}_{\rm O}^2 = \langle (\bar{F} - F_i)^2 \rangle.
	\label{eq:sigma_flux_sigma_fig}
\end{equation}

For a given star the measured value of $\hat{\sigma}_{\rm O}$ will deviate from this theoretical value, as finite data sets are composed from a random sample that will introduce scatter around the theoretical value. 
Hence our sum now becomes over a finite number of samples, and with a little re-arrangement of equation~(\ref{eq:sigma_flux_sigma_fig}) \edit{we} obtain
\begin{equation}
	\frac{n \hat{\sigma}_{\rm O}^2}{\hat{\sigma}_{\rm I}^2} = \sum_{i=1}^n \frac {(\bar{F} - F_i)^2} {\sigma_{\rm I}^2}.
	\label{eq:sigma_fig_sq_obs_inst}
\end{equation}
The right-hand side is the $\chi^2$ of the flux measurements. Hence if we were to observe many stars it would follow the $\chi^2$ distribution. As we calculated $\bar{F}$ from the same data used to calculate $\sigma$, so it will be a $\chi^2$ distribution with $n-1$ degrees of freedom
\begin{equation}
	\frac{\mathrm d P}{\mathrm d \left( n \hat{\sigma}_{\rm O}^2 / \hat{\sigma}_{\rm I}^2 \right) } = 
	\chi^2 \left( \frac{n \hat{\sigma}_{\rm O}^2}{\hat{\sigma}_{\rm I}^2} \middle| n-1 \right) .
	\label{eq:chisq_sigma_fig}
\end{equation}
This can be re-written
\begin{equation}
	\frac{\mathrm d P}{\mathrm d \hat{\sigma}_{\rm O}} = 
	2 \hat{\sigma}_{\rm O}
	\chi^2 \left( \frac{n \hat{\sigma}_{\rm O}^2}{\hat{\sigma}_{\rm I}^2} \middle| n-1 \right)  \frac{n}{\hat{\sigma}_{\rm I}^2} .
	\label{eq:chisq_sigma_fig_three}
\end{equation}
For large $n$, the $\chi^2$ distribution tends to a Gaussian distribution
\begin{equation}
	\chi^2 ( x | n ) \rightarrow \mathcal{N} ( x | n , 2n ) .
	\label{eq:chi_to_gauss}
\end{equation}
Hence
\begin{equation}
	\begin{split}
		\frac{\mathrm d P}{\mathrm d \hat{\sigma}_{\rm O}} &= 
		2 \hat{\sigma}_{\rm O}
		\mathcal{N} \left( \frac{n \hat{\sigma}_{\rm O}^2}{\hat{\sigma}_{\rm I}^2} \middle| n-1 , 2 \left( n-1 \right) \right) 
		\frac{n}{\hat{\sigma}_{\rm I}^2} \\
		&= P_{\rm C} .
	\end{split}
	\label{eq:gauss_sigma_fig_three}
\end{equation}
Here ${\rm d} P$ is the fraction of stars with no measurable intrinsic variability that have an observed fractional standard deviation from $\hat{\sigma}_{\rm O}$ to $\hat{\sigma}_{\rm O}+ \mathrm d\hat{\sigma}_{\rm O}$. In other words, this provides the model component for the constant stars that have no \edit{intrinsic variability measurable} by \textit{Gaia}.

Now we introduce the astrophysical variability $\hat{\sigma}_{\rm A}$. As discussed in Section~\ref{sec:variability_likelihoods}, there is a decreasing trend in the proportion of variable stars with increasing variability amplitude. This is modelled by the exponential function
\begin{equation}
	\begin{split}
		\frac{\mathrm d P}{\mathrm d \hat{\sigma}_{\rm A}} &= \frac{1}{C_{\rm{scale}}} \times e^{- \frac{\hat{\sigma}_{\rm A}}{C_{\rm{scale}}}} \\
		&= A \left( \hat{\sigma}_{\rm A}\right)
	\end{split}
	\label{eq:var_mag_exp}
\end{equation}
where $C_{\rm{scale}}$ is the exponential scale factor.

To find $\hat{\sigma}_{\rm O}$ for a set of stars with a given $\hat{\sigma}_{\rm A}$,  we replace $\hat{\sigma}_{\rm I}$ in the $\chi^2$  distribution of equation \ref{eq:gauss_sigma_fig_three} with the sum in quadrature of $\hat{\sigma}_{\rm I}$ and $\hat{\sigma}_{\rm A}$.  We can then see that the distribution of $\hat{\sigma}_{\rm O}$ integrated over all values of variable star amplitudes, $A(\hat{\sigma}_{\rm A})$, is
\begin{equation}
	\frac{\mathrm d P}{\mathrm{d \hat{\sigma}_{\rm O}} } = 
	\int 2 \hat{\sigma}_{\rm O}\chi^2 \left( \frac{n \hat{\sigma}_{\rm O}^2}{\hat{\sigma}_{\rm I}^2 + \hat{\sigma}_{\rm A}^2} \middle| n-1 \right) \left( \frac{n}{\hat{\sigma}_{\rm I}^2 + \hat{\sigma}_{\rm A}^2} \right) A \left( \hat{\sigma}_{\rm A}\right) \mathrm{d} \hat{\sigma}_{\rm A}.
	\label{eq:chi_frac_inst_astro}
\end{equation}
Again for large $n$, the $\chi^2$ tends to a Gaussian distribution $\mathcal{N} ( x | n , 2n )$
\begin{equation}
	\begin{split}
		\frac{\mathrm d P}{\mathrm{d \hat{\sigma}_{\rm O}} } &= \int_0^\infty
		2 \hat{\sigma}_{\rm O}\mathcal{N} \left( \frac{n \hat{\sigma}_{\rm O}^2}{\hat{\sigma}_{\rm I}^2 + \hat{\sigma}_{\rm A}^2} \middle| n-1 , 2 \left( n-1 \right) \right)
		\left( \frac{n}{\hat{\sigma}_{\rm I}^2 + \hat{\sigma}_{\rm A}^2} \right) \\
		&\qquad A \left( \hat{\sigma}_{\rm A}\right)
		\mathrm d \hat{\sigma}_{\rm A}\\
		&= P_{\rm A} .
	\end{split}
	\label{eq:gauss_frac_inst_astro_integ}
\end{equation}
In the implementation of the model, the upper limit of the integral is set to $4 C_{\rm{scale}}$.

$P_{\rm A}$ provides the component due to variable stars and must be combined with equation~(\ref{eq:gauss_sigma_fig_three}) for the non-variable stars to give the overall likelihood. Both equation~(\ref{eq:gauss_sigma_fig_three}) and (\ref{eq:gauss_frac_inst_astro_integ}) must be combined in proportion to the fraction of variable and non-variable stars, determined from the training data
\begin{equation}
	P_{\rm T} = h P_{\rm C} + \left( 1 - h \right) P_{\rm A}
	\label{eq:pdf_variability}
\end{equation}
where $h$ is the fraction of non-variable stars.

\subsection{Approximation for large fractional standard deviation of the flux}

When $\hat{\sigma}_{\rm O}\gg \hat{\sigma}_{\rm I}$ the effect of the instrumental noise becomes negligible. 
We could then reasonably expect the model in equation \ref{eq:gauss_frac_inst_astro_integ} to be simplified to a version of the exponential function for the astrophysical variability alone, which would be faster to evaluate than the integral.
We can show this is the case by first replacing $\hat{\sigma}_{\rm A}$ as the variable of integration in equation~(\ref{eq:gauss_frac_inst_astro_integ}) with
\begin{equation}
	y = \frac{n \hat{\sigma}_{\rm O}^2}{\hat{\sigma}_{\rm I}^2 + \hat{\sigma}_{\rm A}^2}.
	\label{eq:y}
\end{equation}
If at the same time we assume that $n\gg1$ then equation~(\ref{eq:gauss_frac_inst_astro_integ}) becomes
\begin{equation}
	\begin{split}
		&\frac{\mathrm d P}{\mathrm{d \hat{\sigma}_{\rm O}} } = \\
		&\int_0^{n(\hat{\sigma}_{\rm O}/\hat{\sigma}_{\rm I})^2}{
			{\mathcal{N}} \left( y \middle| n , 2 n \right)
			\frac{n\hat{\sigma}_{\rm O}}{\sqrt{yn\hat{\sigma}_{\rm O}^2 - y^2\hat{\sigma}_{\rm I}^2}}
			A \left( \sqrt{{n\over y}\hat{\sigma}_{\rm O}^2- \hat{\sigma}_{\rm I}^2}\ \right)
			\mathrm d y }.
	\end{split}
	\label{eq:gauss_frac_inst_astro_integ_y}
\end{equation}
We can then remove the integral if we can replace $\mathcal{N}$ with a $\delta$-function centered on $y=n$. 
This we can do provided the remaining integrand varies slowly over the width of the normal function ($\sqrt{2n}$).
In fact the fastest variation is imposed by the upper limit of the integral (which corresponds to 
$\hat{\sigma}_{\rm A}=0$) which effectively sets the integrand to zero.
The condition for the center of the normal to be far removed from this limit is
\begin{equation}
	n(\hat{\sigma}_{\rm O}/\hat{\sigma}_{\rm I})^2 - n \gg \sqrt{2n}.
	\label{eq:long_approx}
\end{equation}
Rather than trying to solve this inequality, we used the slightly different condition 
\begin{equation}
	\hat{\sigma}_{\rm O}> \hat{\sigma}_{\rm I}+ \frac {10 \hat{\sigma}_{\rm I}} {\sqrt{ 2 \left( n-1 \right) } } .
	\label{eq:used_approx}
\end{equation}
\edit{We can show that if the condition for \ref{eq:used_approx} is satisfied so is that for \ref{eq:long_approx}, by substituting for $\hat{\sigma}_{\rm O}/\hat{\sigma}_{\rm I}$ in \ref{eq:long_approx} using \ref{eq:used_approx}.} 
If equation \ref{eq:long_approx} is satisfied then equation \ref{eq:gauss_frac_inst_astro_integ_y} becomes the simplified version of \ref{eq:gauss_frac_inst_astro_integ} we were searching for,
\begin{equation}
	\frac{\mathrm d P}{\mathrm{d \hat{\sigma}_{\rm O}} } \simeq 
	\frac{\hat{\sigma}_{\rm O}}{\sqrt{\hat{\sigma}_{\rm O}^2 - \hat{\sigma}_{\rm I}^2}} 
	A \left( \sqrt{\hat{\sigma}_{\rm O}^2 - \hat{\sigma}_{\rm I}^2} \right).
	\label{eq:gauss_frac_inst_astro_integ_approx}
\end{equation}

%%%%%%%%%%%%%%%%%%%%%%%%%%%%%%%%%%%%%%%%%%%%%%%%%%

%%%% Supplementary Material

\section{Symbols}

The symbols used throughout the paper are listed in Table~\ref{tab:symbol_table}.

\begin{table*}
	\centering
	\caption{Table of symbols used throughout this paper.}
	\label{tab:symbol_table}
	\begin{tabular*}{\textwidth}{ll} % four columns, alignment for each
	\hline
	Symbol & Description \\
	\hline
	RUWE & \textit{Gaia} renormalised unit weight error \\
	$F$ & \textit{Gaia} $G$-band flux measurement \\
	$\bar{F}$ & \textit{Gaia} $G$-band mean flux \\
	$S_{\rm O}$ & Standard error of the \textit{Gaia} $G$-band mean flux \\
	$\sigma_{\rm O}$ & Standard deviation of the \textit{Gaia} $G$-band flux \\
	$\hat{\sigma}_{\rm O}$ & Fractional standard deviation of the \textit{Gaia} $G$-band flux \\
	$\sigma_{\rm I}$ & Standard deviation due to instrumental noise in the variability model \\
	$\hat{\sigma}_{\rm I}$ & Fractional standard deviation due to instrumental noise in the variability model \\
	$\hat{\sigma}_{\rm A}$ & Fractional standard deviation due to astrophysical variability in the variabilty model \\
	$C_{\rm{scale}}$ & Exponential scale factor in the variabilty model \\
	$N_{\rm{obs}}$ & Number of \textit{Gaia} observations (CCD transits) contributing to the $G$-band photometry \\
	$N_{\rm{eff}}$ & Effective number of \textit{Gaia} observations used in the variability model \\
	$n$ & Number of observations in the variability model \\
	$h$ & Fraction of non-variable stars in the variability model \\
	$A$ & Exponential function from the variabilty model \\
	$P_{\rm A}$ & Astrophysical variability probability density function \\
	$P_{\rm C}$ & Non-variable star probability density function \\
	$P_{\rm T}$ & Combined variability probability density function \\
	$\varpi$ & \textit{Gaia} parallax with bias removed \\
	$\sigma_{\rm \varpi}$ & \textit{Gaia} archive parallax uncertaintiy \\
	$\sigma_{\rm{ext}}$ & External error on \textit{Gaia} parallaxes \\
	$\sigma_{\rm s}$ & \textit{Gaia} systematic uncertainty \\
	$k$ & k factor on the \textit{Gaia} formal parallax uncertainties \\
	\textsc{CII} & The Class II YSO classification including training and verification sets \\
	\textsc{Other} & The Other classification including training and verification sets \\
	$C_x$ & Classifier class $x$ \\
	$D_y$ & The data for feature $y$ \\
	$L_f$ & The likelihood for feature $f$ \\
	$L(\textsc{CII})$ & The \textsc{CII} likelihood for an unspecified feature in the classifier \\
	$L(\textsc{Other})$ & The \textsc{Other} likelihood for an unspecified feature in the classifier \\
	$P_{\rm r}(\textsc{CII})$ & The \textsc{CII} prior \\
	$P_{\rm r}(\textsc{Other})$ & The \textsc{Other} prior \\
	$P_{\rm s}(\textsc{CII})$ & The \textsc{CII} posterior \\
	$P_{\rm s}(\textsc{Other})$ & The \textsc{Other} posterior \\
	$T$ & The number of Class II YSOs we set as the binary result true \\
	$F$ & The number of \textsc{Other} objects we set as the binary result false \\
	$TP$ & True Positive, the number of objects correctly accepted by the classifier as Class II YSOs \\
	$TN$ & True Negative, the number of objects correctly rejected by the classifier as \textsc{Other} \\
	$FP$ & False Positive, the number of objects falsely accepted by the classifier as Class II YSOs \\
	$FN$ & False Negative, the number of objects falsely rejected by the classifier as \textsc{Other} \\
	$TPR$ & True Positive Rate, the fraction of Class II YSOs the classifier correctly accepts as Class II YSOs \\
	$FPR$ & False Positive Rate, the fraction of \textsc{Other} objects the classifier correctly rejects as \textsc{Other} \\
	ROC curve & Receive Operating Characteristic curve (plots TPR vs FPR) \\
	AUC & Area under the ROC curve \\
	\hline
	\end{tabular*}
\end{table*}

\section{Online catalogue data dictionary}
\label{sec:data_dictionary}

A data dictionary for the published catalogue of naive Bayes classifier results is given in Table~\ref{tab:data_dictionary}.

\begin{table*}
	\caption{Data dictionary for the catalogue of naive Bayes classifier results.}
	\centering
	\label{tab:data_dictionary}
\begin{tabular*}{\textwidth}{lccl} % four columns, alignment for each
	
	\hline
	Field name & Data type & Unit & Description \\
	\hline
	\edittwo{GaiaEDR3SourceId} & 64-bit integer &  & Gaia EDR3 unique source designation \edittwo{(EDR3 and DR3 source ids are identical)}. \\
	RAdeg & 64-bit floating point & deg & Right ascension from Gaia EDR3. \\
	DEdeg & 64-bit floating point & deg & Declination from Gaia EDR3. \\
	GLON & 64-bit floating point & deg & Galactic longitude from Gaia EDR3. \\
	GLAT & 64-bit floating point & deg & Galactic latitude from Gaia EDR3. \\
	plx & 64-bit floating point & mas & Gaia EDR3 parallax adjusted for zero-point correction using the Python script from the Gaia EDR3 website. \\
	EBmV & 32-bit floating point & mag & E(B-V) derived from STILISM. \\
	SuspectReddening & boolean &  & Suspect reddening flag. \\
	GmagCorr & 32-bit floating point & mag & Gaia EDR3 $G$-band magnitude with corrections to sources with 6-parameter astrometric solution using the Python code from the Gaia EDR3 website. \\
	rmagParCorrDered & 32-bit floating point & mag & \editfour{Parallax corrected dereddened IGAPS $r$-band magnitude using Ar=2.38E(B-V).} \\
	rmiDered & 32-bit floating point & mag & IGAPS $(r-i)$ dereddened using transformed STILISM reddening values. \\
	FeaturesCalc & 16-bit integer &  & Number of features used in the Bayes calculation. \\
	PriorCII & 32-bit floating point &  & CII prior. \\
	PriorOther & 32-bit floating point &  & Other prior. \\
	PosteriorCII & 64-bit floating point &  & Naive Bayes CII posterior. \\
	PosteriorOther & 64-bit floating point &  & Naive Bayes Other posterior. \\
	GFracStdDevFlux & 64-bit floating point &  & Gaia EDR3 $G$-band feature, the observed fractional standard devation of the flux. \\
	GvarCIILike & 64-bit floating point &  & Variability feature CII likelihood. \\
	GvarOtherLike & 64-bit floating point &  & Variability feature Other likelihood. \\
	GvarLikeRatio & 64-bit floating point &  & Variability feature CII likelihood divided by Other likelihood. \\
	GvarCalc & boolean &  & Variability feature likelihood calculated\edittwo{?} \\
	GvarCIICapped & boolean &  & Variability feature CII likelihood capped\edittwo{?} \\
	GvarOtherCapped & boolean &  & Variability feature Other likelihood capped\edittwo{?} \\
	W1mW2 & 32-bit floating point & mag & AllWISE $(W1-W2)$ feature. \\
	W1mW2CIILike & 64-bit floating point &  & $(W1-W2)$ feature CII likelihood. \\
	W1mW2OtherLike & 64-bit floating point &  & $(W1-W2)$ feature Other likelihood. \\
	W1mW2LikeRatio & 64-bit floating point &  & $(W1-W2)$ feature CII likelihood divided by Other likelihood. \\
	W1mW2Calc & boolean &  & $(W1-W2)$ feature likelihood calculated\edittwo{?} \\
	W1mW2CIICapped & boolean &  & $(W1-W2)$ feature CII likelihood capped\edittwo{?} \\
	W1mW2OtherCapped & boolean &  & $(W1-W2)$ feature Other likelihood capped\edittwo{?} \\
	HmKExSource & 7 character string &  & $(H-K)$ feature photometry source, UKIDSS or 2MASS. \\
	HmKEx & 32-bit floating point & mag & $(H-K)$ feature offset. \\
	HmKExCIILike & 64-bit floating point &  & $(H-K)$ feature CII likelihood. \\
	HmKExOtherLike & 64-bit floating point &  & $(H-K)$ feature Other likelihood. \\
	HmKExLikeRatio & 64-bit floating point &  & $(H-K)$ feature CII likelihood divided by Other likelihood. \\
	HmKExCalc & boolean &  & $(H-K)$ feature likelihood calculated\edittwo{?} \\
	HmKExCIICapped & boolean &  & $(H-K)$ feature CII likelihood capped\edittwo{?} \\
	HmKExOtherCapped & boolean &  & $(H-K)$ feature Other likelihood capped\edittwo{?} \\
	rmHaEx & 32-bit floating point & mag & H$\alpha$ feature IGAPS $(r-H\alpha)$ offset. \\
	rmHaExCIILike & 64-bit floating point &  & H$\alpha$ feature CII likelihood. \\
	rmHaExOtherLike & 64-bit floating point &  & H$\alpha$ feature Other likelihood. \\
	rmHaExLikeRatio & 64-bit floating point &  & H$\alpha$ feature CII likelihood divided by Other likelihood. \\
	rmHaExCalc & boolean &  & H$\alpha$ feature likelihood calculated\edittwo{?} \\
	rmHaExCIICapped & boolean &  & H$\alpha$ feature CII likelihood capped\edittwo{?} \\
	rmHaExOtherCapped & boolean &  & H$\alpha$ feature Other likelihood capped\edittwo{?} \\
	log10Age & 32-bit floating point &  & Isochronal age feature $log_{10}(Age)$ in Myr derived from the isochrone interpolation. \\
	OlderMaxIsoAge & boolean &  & Isochronal age feature flag for older than isochrones. \\
	YoungerMinIsoAge & boolean &  & Isochronal age feature flag for younger than isochrones. \\
	GoodClassIIYSOFit & boolean &  & Isochronal age feature flag for good fit to YSO isochrones. \\
	IsoAgeCIILike & 64-bit floating point &  & Isochronal age feature CII likelihood. \\
	IsoAgeOtherLike & 64-bit floating point &  & Isochronal age feature Other likelihood. \\
	IsoAgeLikeRatio & 64-bit floating point &  & Isochronal age feature CII likelihood divided by Other likelihood. \\
	IsoAgeCalc & boolean &  & Isochronal age feature calculated\edittwo{?} \\
	IsoAgeCIICapped & boolean &  & Isochronal age feature CII likelihood capped\edittwo{?} \\
	IsoAgeOtherCapped & boolean &  & Isochronal age feature Other likelihood capped\edittwo{?} \\
	\hline
	
\end{tabular*}
\end{table*}

%%%%%%%%%%%%%%%%%%%%%%%%%%%%%%%%%%%%%%%%%%%%%%%%%%

% Don't change these lines
\bsp	% typesetting comment
\label{lastpage}
\end{document}